\documentclass{article}
\usepackage[english]{babel}
\usepackage[utf8]{inputenc}
\usepackage{amsmath}
\usepackage{amsfonts}
\usepackage{amssymb}
\usepackage{graphicx}
\usepackage[style=apa, citestyle=apa, backend=biber, uniquename=false, mincitenames=1, maxcitenames=2, citetracker, maxbibnames=99, doi=false, isbn=false, url=false, eprint=false]{biblatex}
\usepackage{setspace}
\usepackage{wrapfig}
\usepackage{fancyhdr}
\usepackage{tabularx}
\usepackage{geometry}
\usepackage[right]{eurosym}
\usepackage[printonlyused]{acronym}
\usepackage{subcaption}
\usepackage{floatflt}
\usepackage[usenames,dvipsnames]{color}
\usepackage{colortbl}
\usepackage{paralist}
\usepackage{array}
\usepackage{titlesec}
\usepackage{parskip}
\usepackage{longtable}
\usepackage[bottom,flushmargin]{footmisc}
\usepackage{hyperref}
\hypersetup{colorlinks=true,linkcolor=black,urlcolor=black,citecolor=black,
	pdftitle={Meta-Learners for Estimation of Causal Effects: Finite Sample Cross-Fit Performance}, 	pdfauthor={Gabriel Okasa},
	pdfsubject={Econometrics},
	pdfkeywords={meta-learners, causal machine learning, heterogeneous treatment effects, simulation},
	pdfstartpage=1,
	pdfpagelabels=true}
\usepackage{csquotes}
\usepackage{mathtools}
\usepackage{booktabs,caption}
\usepackage[flushleft]{threeparttable}
\usepackage{multicol}
\usepackage{multirow}
\usepackage{pdflscape}
\usepackage{titlesec}
\usepackage{changepage}% http://ctan.org/pkg/changepage
\usepackage{lipsum}% http://ctan.org/pkg/lipsum
\usepackage{xcolor, soul}
\sethlcolor{white}
\setulcolor{white}
\setstcolor{white}
\usepackage[inline]{enumitem}
\usepackage{booktabs,caption}
\usepackage{pdflscape}
\usepackage{titlesec}
\usepackage{pifont}
\usepackage{threeparttable}
\usepackage{threeparttablex}
\usepackage{float}
\usepackage{appendix}
\usepackage{chngcntr}
\usepackage{etoolbox}
\usepackage{tablefootnote}
\usepackage{comment}
\usepackage{everypage}
\usepackage{tikz}
\usetikzlibrary{positioning}
\usetikzlibrary{shapes,arrows,arrows,positioning,fit}
\usetikzlibrary{calc}
\usepackage[ruled,vlined]{algorithm2e}
\usepackage{dsfont}
\usepackage{calligra}
\usepackage{indentfirst}
\usepackage[object=vectorian]{pgfornament}

\titleformat
{\chapter} % command
[display] % shape
{\Huge\bfseries} % format
{Chapter \ \thechapter} % label
{0.5cm} % sep
{
	\LARGE
    \rule{\textwidth}{0.75pt}
    \vspace{0.025ex}
    \centering
} % before-code
[
\vspace{-0.5cm}%
\rule{\textwidth}{0.75pt}
] % after-code
 
\titlespacing*{\section}     {0pt}{5.25ex plus 1ex minus .2ex}{3.45ex plus .2ex}
\titlespacing*{\subsection}   {0pt}{4.875ex plus 1ex minus .2ex}{2.25ex plus .2ex}
\titlespacing*{\subsubsection}{0pt}{4.875ex plus 1ex minus .2ex}{2.25ex plus .2ex}

\newcommand\independent{\protect\mathpalette{\protect\independenT}{\perp}}
\def\independenT#1#2{\mathrel{\rlap{$#1#2$}\mkern2mu{#1#2}}}

\providecommand{\keywords}[1]
{
  \noindent
  \small	
  Keywords: #1
}

\providecommand{\jelclass}[1]
{
  \noindent
  \small	
  JEL classification: #1
}

\makeatletter
   {\par}
\makeatother

\newcolumntype{H}{>{\setbox0=\hbox\bgroup}c<{\egroup}@{}}

\AtBeginEnvironment{subappendices}{%
\chapter*{Appendix}
\addcontentsline{toc}{chapter}{Appendices}
\counterwithin{figure}{section}
\counterwithin{table}{section}
}

\definecolor{darkgreen}{cmyk}{1,0,0.93,0.5}
\definecolor{lightgreen}{cmyk}{0.8,0,0.95,0.25}
\definecolor{lightlightgreen}{cmyk}{0.2,0,0.235,0.065}
\definecolor{femalegreen}{rgb}{0.91, 0.45, 0.32}
\definecolor{malegreen}{rgb}{0.0, 0.55, 0.55}

\def\plusplus{{\nolinebreak[4]\hspace{-.05em}\raisebox{.4ex}{\tiny\bf ++ }}}

\newcommand{\Lpagenumber}{\ifdim\textwidth=\linewidth\else\bgroup
  \dimendef\margin=0 %use \margin instead of \dimen0
  \ifodd\value{page}\margin=\oddsidemargin
  \else\margin=\evensidemargin
  \fi
  \raisebox{\dimexpr -\topmargin-\headheight-\headsep-0.5\linewidth}[0pt][0pt]{%
    \rlap{\hspace{\dimexpr \margin+\textheight+\footskip}%
    \llap{\rotatebox{90}{\thepage}}}}%
\egroup\fi}
\AddEverypageHook{\Lpagenumber}%

\newtheorem{assumption}{Assumption}

\DeclareMathOperator\supp{supp}

\makeatletter
\AtEveryCitekey{%
  \ifciteseen
    {}
    {\defcounter{maxnames}{6}}%
}
\makeatother

\numberwithin{equation}{section}

\title{Meta-Learners for Estimation of Causal Effects:\\
Finite Sample Cross-Fit Performance\\
~\\}
\author{Gabriel Okasa\footnote{A previous version of this paper was presented at
  research seminars of the University of St.Gallen, EPFL and the GESIS
  Spring Seminar in Cologne. We thank participants, in particular
  Michael Lechner, for helpful comments and suggestions. We also thank
  Francesco Audrino, Daniele Ballinari, Martin Biewen, Jonathan Chassot,
  Daniel Goller, Sandro Heiniger, Daniel Jacob, Michael Knaus, Jana Mareckova,
  Matthias Roesti, Kenneth Younge and Michael Zimmert for their useful feedback.
  The usual disclaimer applies. %\newline
  Email: \href{mailto:gabriel.okasa@epfl.ch}{gabriel.okasa@epfl.ch}.
  Website: \href{https://okasag.github.io/}{okasag.github.io}.}\\
~\\
TIS-EPFL\\
Chair for Technology and Innovation Strategy\\
Swiss Federal Institute of Technology in Lausanne, Switzerland\\
~\\
}
\date{\today}
 
\begin{document}
%\onehalfspacing
\maketitle

\vspace{-1cm}
%\begin{center}
%\Large{Gabriel Okasa}
%\end{center}
\vspace{2cm}
%\begin{adjustwidth}{2cm}{2cm}
\begin{abstract}
Estimation of causal effects using machine learning methods has become
an active research field in econometrics. \ul{In this paper,}
we study the finite sample performance of meta-learners for
estimation of heterogeneous treatment effects \ul{under} the usage of
sample-splitting and cross-fitting to reduce the overfitting bias.
In both synthetic and semi-synthetic simulations we find that the
performance of the meta-learners in finite samples greatly depends on
the estimation procedure. The results imply that sample-splitting and
cross-fitting are beneficial in large samples for bias reduction and
efficiency of the meta-learners, respectively, whereas full-sample
estimation is preferable in small samples. Furthermore, we derive
practical recommendations for \ul{application} of specific meta-learners in
empirical studies depending on particular data characteristics such
as treatment shares and sample size.\\

\keywords{Meta-learners, causal machine learning, heterogeneous treatment effects, Monte Carlo simulation, sample-splitting, cross-fitting.}\\

\jelclass{C15, C18, C31.}
\end{abstract}

\thispagestyle{empty}

\pagebreak

\setcounter{page}{1}

\section{Introduction}

In recent years there has been a growing interest in the estimation of
causal effects using machine learning algorithms, particularly in the field of economics
\parencite{Athey2018}. The newly emerging synthesis of machine learning
methods with causal inference has a large potential for a more
\ul{comprehensive} estimation of causal effects \parencite{Lechner2019}. On the one
hand, it enables a more flexible estimation of average effects which are
of main interest in microeconometrics \parencite{Imbens2009}. On the other
hand, it advances the estimation beyond the average effects and allows
for a systematic analysis of effect heterogeneity \parencite{Imbens2017}.
Both of these aspects contribute to a better description of the causal mechanisms
and thus to a possibly more efficient treatment allocation
\parencites{Zhao2012}{Kitagawa2018}{Athey2021}{Nie2021}. Hence, applied empirical researchers
can greatly benefit from the usage of machine learning methods ranging
from evaluation of public policies and business decisions to designing
\ul{personalized interventions} \parencites{Andini2018}{Bansak2018}.

Machine learning estimators as such are, however, primarily designed \ul{for}
prediction problems and thus cannot be used \ul{directly} for causal inference.
Therefore, new approaches for the estimation of causal parameters 
using machine learning emerged \parencite[see][for an overview]{Athey2019a}. In particular, 
the development of the so-called meta-learners have received considerable attention
\parencites(see e.g.)(){Kunzel2019}{Kennedy2020}[or][]{Nie2021a}. Meta-learners decompose the causal problem into separate prediction problems that can be solved by standard machine learning algorithms and subsequently combined to estimate the causal parameters of interest \parencite{Curth2021}. Such an approach is advantageous for several reasons. First, the meta-learners do not modify the objective function of the machine learning methods but rather combine their predictions in order to estimate the causal effect. This enables to directly leverage the superior prediction power of machine learning estimators. Second, the
meta-learners are generic algorithms refraining from a specific usage of
any particular machine learning method. This allows to apply any
suitable supervised learning method for the particular prediction
problem at hand. Third, the meta-learners are attractive due to the ease of
implementation using standard statistical software. This permits
researchers to apply the meta-learners without any potential
restrictions due to limited availability in software packages and
enables tailored implementation for particular types of data. Despite
the attractive features of the meta-learners, there is little guidance
for applied empirical researchers on how to choose from a variety of the
meta-learners proposed in the literature, with lack of unifying
simulation evidence for an assessment of the performance of the
meta-learners in applied empirical settings.

The complexity of the meta-learners \ul{for the estimation of the causal parameters}
varies widely and often hinges on \ul{a prior} estimation of the conditional means
of the outcome and the treatment, \ul{also referred to as the nuisance functions,
that are not of primary interest} \parencite{Chernozhukov2018c}. 
Due to the machine learning, \ul{or in general flexible} estimation of such nuisance
functions the meta-learners are prone to the overfitting bias, i.e. own observation bias,
\ul{which stems from fitting the data too well such that the prediction performance gets compromised} \parencite{Hastie2009}. \ul{This bias then pollutes the estimation of the
causal parameters, when the machine learning estimators of the nuisance functions are
directly plugged into the estimation of the causal effect using the same data} \parencite{Newey2018}. Therefore, sample-splitting
has been proposed in the literature to reduce the overfitting
bias by using one part of the sample for estimation of the nuisance
functions and the other part of the sample for estimation of the causal effect.
\ul{In order to regain the full sample size efficiency of the estimator
cross-fitting repeats the estimation by swapping the samples and averaging
the estimated causal effects} \parencite{Chernozhukov2018c}. 
\ul{However, the usage of sample-splitting and cross-fitting is not well understood
in practice and the specific definitions
of meta-learners differ substantially in their implementation of
these procedures.} Despite the ambiguous
definitions, there is a lack of simulation evidence concerned with the
usage of sample-splitting and cross-fitting within the meta-learning
framework and thus limited guidance for or against specific
implementations. Moreover, there appears to be limited knowledge about
how the asymptotic arguments translate into finite sample properties of
the meta-learners.

In this paper, we address both of the above issues and study the finite
sample properties of the \ul{machine learning based} meta-learners for the
estimation of causal effects based on the specific implementations using the full-sample,
sample-splitting and cross-fitting procedures for varying sample sizes.
We focus on \ul{evaluating the estimation} of heterogeneous treatment
effects \ul{as these provide the most detailed description of the underlying
causal mechanisms and thus allow for a better assessment of the individualized
impacts of an intervention}.
For this purpose, we review the most widely used meta-learning algorithms
together with their \ul{assumptions} with respect to
sample-splitting and cross-fitting and identify their strengths and
weaknesses. We conduct both synthetic and empirically grounded semi-synthetic simulations
comparing the performance of the meta-learners in various 
settings featuring unequal treatment shares, non-linear functional forms
and large-dimensional feature sets. \ul{Importantly}, within the simulations
we explicitly study the
convergence performance of the meta-learners based on growing sample
sizes \ul{up to $32'000$ observations.} \ul{Furthermore}, we derive practical
recommendations on the choice of specific meta-learners and the
respective estimation procedures for applied empirical work.

The results of our simulation experiments reveal that the choice of the
estimation procedure has a \ul{large} impact on the performance of the
\ul{machine learning based} meta-learners in finite samples. For
sufficiently large samples we provide evidence for the theoretical
arguments of bias reduction via sample-splitting and cross-fitting,
while for smaller samples we observe adverse effects of these procedures
\ul{when using machine learning}. The results show that, \ul{if computation time
is not a constraint,} cross-fitting is always preferable to
sample-splitting as it keeps the bias low, while successfully reducing
the variance of the estimators even in small samples. \ul{However}, the
results imply heterogeneous impacts of the estimation procedures on the
performance of the different meta-learners. \ul{We detect meta-learners for which the
performance is stable regardless of the estimation procedure and for
which the performance is more sensitive to the choice thereof. This holds
not only with regard to bias and variance, but also with regard to convergence rates.}
\ul{Beyond the impacts of the estimation procedures, the results reveal
clear patterns for the performance of particular meta-learners based on
data characteristics. As such, we identify meta-learners suitable for
empirical settings with highly unbalanced treatment shares, irrespective
of the sample size, as well as meta-learners that are unstable in small
samples but have superior performance once larger samples become available.
Finally, we set apart meta-learners with undesirable statistical
properties that should be avoided for estimation of causal effects.}

This paper contributes to the causal machine learning literature in several ways.
First, we provide unifying simulation evidence of meta-learning
algorithms for the estimation of heterogeneous causal effects in large-dimensional and
highly non-linear settings based on synthetic and semi-synthetic simulations.
Second, we explicitly study the meta-learners under the full-sample,
sample-splitting and cross-fitting implementations, respectively and
thereby provide evidence on the contrast between the asymptotic
arguments and finite sample properties. Third, we empirically
investigate the convergence \ul{performance} of the meta-learners by repeating
the simulation experiments with growing sample sizes. Finally, we derive relevant
practical recommendations for applied empirical work which are based on
the particular observable data characteristics.

\ul{The rest of} this paper is organized as follows. We briefly discuss the related
literature in Section \ref{c1sec:lit}. Section \ref{c1sec:notation} introduces
the notation, the parameters of interest and their identification.
Section \ref{c1sec:learners} reviews the considered meta-learners and the
estimation procedures. Section \ref{c1sec:sim} describes the
synthetic as well as semi-synthetic simulations and presents the corresponding results.
The main findings of the study are discussed in Section \ref{c1sec:disc}.
Section \ref{c1sec:end} concludes.
Further details including descriptive statistics, an exhaustive summary of the main and supplementary results as well as a computation time analysis are provided in Appendices \ref{c1app:desc}, \ref{appendix-b-simulation-results} and \ref{appendix-d-computation-time}, respectively.

\subsection{Related Work}\label{c1sec:lit}

Within the fast developing causal machine learning literature,
many different meta-learning algorithms of different complexities
have been proposed. Following the naming convention of meta-learners
based on the capital letters, the less complex meta-learners
include the S-learner and the T-learner \parencites[see][and]{Lo2002}[][for early ideas of these approaches]{Hansotia2002}[and][for applications in the marketing domain]{Kane2014}\footnote{In the marketing literature the so-called uplift modelling
is concerned with the same target of estimating causal effects and developed in
parallel to the classical econometric and causal machine learning literature. For a comprehensive
overview of the literature on uplift modelling see \textcite{Devriendt2018}. \textcite{Gutierrez2016} and \textcite{Zhang2022} provide unified surveys of uplift
modelling harmonized with the econometric and causal machine learning literature,
respectively.}
which besides the treatment effect function do not require estimation of any additional
nuisance functions. However, the most prominent and widely used
meta-learners in the literature consist of the more complex 
X-learner \parencite{Kunzel2019}, the DR-learner \parencite{Kennedy2020},
and the R-learner \parencite{Nie2021a}, which all require prior estimation and
combination of several nuisance functions, such as the conditional means
of the outcome and the treatment, to estimate the causal effect.
We define the meta-learners formally and discuss them in more detail
in Section \ref{c1sec:meta}.  In this paper, we focus on the above-listed meta-learners in
order to provide a contrast between less and more complex algorithms for
the estimation of causal effects. Moreover, these particular meta-learners
have been extensively studied theoretically as well as applied in various empirical
settings, including economics \parencites{Knaus2020b}{Jacob2021}{Sallin2021}{Valente2022}, public policy \parencites{Kristjanpoller2021}{Shah2021}, marketing \parencites{Gubela2020}{Gubela2021},
medicine \parencites{Lu2018}{Duan2019} or sports \parencite{Goller2021b}.
Some further examples of meta-learners proposed in the
literature consist of the U-learner and Y-learner \parencite{Stadie2018}, or
more recently the IF-learner \parencite{Curth2020} and RA-learner \parencite{Curth2021},
which are, however, beyond the scope of this paper.

Besides the meta-learning framework, \ul{there has been also a substantial
development of specific causal estimators based on direct modifications of
particular machine learning algorithms. Especially, the tree-based estimators
have been studied extensively in this respect. These include
Causal Trees} \parencite{Athey2016} \ul{as well as} Causal Boosting \parencite{Powers2018}
\ul{and Causal Forests} \parencite{Wager2018} \ul{with the extensions of the Modified Causal Forests} \parencite{Lechner2019} \ul{and the Generalized Random Forests} \parencite{Athey2019}.
These methods are based on the underlying predictive algorithms of
Regression Trees \parencite{Breiman1984}, Boosted Trees \parencite{Friedman2001}
and Random Forests \parencite{Breiman2001}, respectively. Furthemore,
Bayesian versions of Regression Trees\ul{, the so-called BART}
\parencite{Chipman1998} have been adapted for estimation of causal effects as well
\parencites{Hill2011}{Taddy2016}{Hahn2020}. Besides the estimators based on
recursive partitioning, important causal adjustments have been applied
in respect to regularization based estimators such as the Lasso
\parencites{Qian2011}{Belloni2013a}{Tian2014} or Lasso-augmented Support
Vector Machines \parencite{Imai2013}. Additionally, further machine learning
algorithms such as the Nearest Neighbours \parencite{Fan2018} or Neural
Networks \parencites{Johansson2016}{Shalit2017}{Schwab2018}{Shi2019} have been
transformed towards causal inference as well. For a comprehensive
overview of many of these estimators, we refer the interested reader to
\textcite{Athey2019a}, \textcite{Kreif2019} and \textcite{Jacob2021},
or to \textcite{Baiardi2021} for numerous empirical examples. While many of
these methods have well-established theoretical properties, they restrict the
researcher in the choice of the machine learning method. In this paper, although we
focus on the machine learning estimation of causal effects, we refrain from
an analysis of these methods due to major conceptual differences to the
meta-learning framework and the lack of comparability in terms of the
usage of sample-splitting and cross-fitting procedures.

In general, the literature on the finite sample properties of causal
machine learning estimators under unified framework seems to be rather
scarce. One of the exceptions in the econometric literature\footnote{\textcite{Wendling2018} conduct similar empirically grounded simulation
study in medical context, while \textcite{McConnell2019} use synthetic data.} is
\textcite{Knaus2021} who study a wide range of estimators for heterogeneous
as well as (group) average treatment effects, including direct
estimators as well as some meta-learners in an Empirical Monte Carlo
Study as developed in \textcite{Huber2013} and \textcite{Lechner2013}.
\textcite{Knaus2021} find no estimator to perform uniformly best, but
notice that estimators which model both the outcome as well as the
treatment process are substantially more robust throughout all data
generating processes considered, which can be observed in our
simulations as well. Among the meta-learners considered, the DR-learner
and the R-learner perform especially well in terms of the root mean
squared error. Moreover, using the Random Forest as a base learner turns
out to be more stable with better statistical properties in contrast to
using the Lasso, particularly in smaller samples, which also motivates
the usage of the Random Forest in our simulations. \ul{However, although}
both meta-learners are implemented with cross-fitting, an explicit consideration of different
sample-splitting or cross-fitting schemes is missing. A recent work by \textcite{Naghi2021}
also focuses on a variety of causal machine learning estimators, inclusive of
selected meta-learning algorithms, in an empirically grounded simulation design.
\textcite{Naghi2021} find Bayesian estimators, including the DR-learner with BART
as a base learner, to perform best in estimating the heterogeneous treatment effects.
In addition to \textcite{Knaus2021}, the simulations of \textcite{Naghi2021}
provide results on statistical inference in terms of coverage rates and length of
the confidence intervals for the estimated causal effects. Nevertheless, albeit \textcite{Naghi2021}, similarly as \textcite{Knaus2021}, implement the
relevant estimators using cross-fitting, a devoted assessment of this procedure
is omitted. \textcite{Curth2021} \ul{focus directly on meta-learning
algorithms for estimation of heterogeneous treatment effects,
but refrain from studying sample-splitting and cross-fitting procedures and rely fully
on the full-sample estimation.} In this regard, \textcite{Zivich2021} study the
performance of treatment effect estimators based on cross-fitting,
including some meta-learners as well. Similarly to \textcite{Knaus2021}
they find the DR-learner with an ensemble machine learning base learners
together with cross-fitting to perform the best among all considered
estimators, both in comparison to \ul{cases without} cross-fitting and
to parametric base learners. However, \textcite{Zivich2021} study \ul{exclusively} the
estimation of average effects without examining convergence performance
of the estimators, considering only a single sample size of $3'000$ observations.\footnote{Numerous other recent studies in the epidemiology literature focus on the estimation
of average effects using the DR-learner based on cross-fitting, typically in small samples.
For details see \textcite{Balzer2021}, \textcite{Meng2021}, \textcite{Naimi2021},
\textcite{Zhong2021} and \textcite{ConzueloRodriguez2022}.}
Recently, \textcite{Jacob2020} focuses on the
estimation of heterogeneous treatment effects under various
cross-fitting schemes for selected meta-learning algorithms. Also, in
this simulation study the DR-learner together with the R-learner achieve
consistently the best results. \textcite{Jacob2020}
stresses the \ul{heterogeneous impacts} of the particular sample-splitting and
cross-fitting procedures on each meta-learner, which is documented in our simulations as well.
\ul{Nevertheless, even though considering varying sample sizes within the simulation experiments, the considered sample sizes are limited to $2'000$ observations.}
Overall, none of the above studies focuses directly on the convergence
performance of the meta-learners under various estimation procedures
which still remains an open question. To the best of our knowledge, this
is the first paper that \ul{empirically} studies the convergence properties of the
meta-learners under full-sample, sample-splitting and cross-fitting
implementations with growing sample sizes up to several thousands of
observations, \ul{reaching $32'000$ in our simulations.}

\section{Framework and Identification}\label{c1sec:notation}

In order to describe the effects of interest and their corresponding
identification assumptions we rely on the potential outcome framework
\parencite{Rubin1974}. We assume a population \(\mathcal{P}\) \ul{from which a realization of} \(N\) \textit{i.i.d.} random variables is given consisting of a random sample
\(\{Y_i(1), Y_i(0),W_i,X_i\} \sim \mathcal{P} \). Here, we consider a binary treatment
variable \(W_i\) that is equal to \(1\) for the treated group and equal
to \(0\) for the control group, respectively. According to the treatment
status we define the potential outcome \(Y_i(1)\) under treatment for
the case when \(W_i=1\) and correspondingly the potential outcome
\(Y_i(0)\) under control for \(W_i=0\). Additionally, we define a
\(p\)-dimensional vector of exogenous pre-treatment covariates such that
\(X_i \in \mathbb{R}^p\). Given this definition we can characterize the
\textit{Individual Treatment Effect} (ITE) as follows:
\[\xi_i=Y_i(1)-Y_i(0).\]
However, the fundamental problem of causal inference is that we never
observe both potential outcomes at the same time
\parencite{Holland1986}. \ul{Hence, the observed outcomes are defined
according to the observational rule as} \(Y_i=Y_i(W_i)\). \ul{The observed data
then consists of the triple} \(\{Y_i,W_i,X_i\}_{1\leq i\leq N}\).
Nevertheless, it is still possible to identify the expectation of
\(\xi_i\) under additional assumptions \parencites(compare)(){Rubin1974}[or][]{Imbens2015}.
Thus, we shift the effect of interest towards the
\textit{Conditional Average Treatment Effect} (CATE) which
takes the expectation of \(\xi_i\), conditional on covariates \(X_i\)
and is given as:
\[\tau(x)=\mathbb{E}\big[\xi_i \mid X_i=x\big]=\mathbb{E}\big[Y_i(1)-Y_i(0) \mid X_i=x\big]=\mu_1(x)-\mu_0(x)\]
where \(\mu_1(x)=\mathbb{E}[Y_i(1) \mid X_i=x]\) and \(\mu_0(x)=\mathbb{E}[Y_i(0) \mid X_i=x]\) are the response functions \ul{for potential outcomes}
under treatment and under control, \ul{respectively}. \ul{In this paper we always refer to
the CATE with conditioning on all observed exogeneous covariates and thus focusing on the finest level
of heterogeneity} \parencite[see e.g.][]{Knaus2021}.\footnote{In general,
  the term CATE describes conditional average treatment effects on
  various aggregation levels \parencite{Lechner2019}. \ul{In our case, the CATE corresponds
  to the \textit{Individualized Average Treatment Effect} (IATE).} Additionally,
  researchers and especially policy makers might be interested in a \ul{low-dimensional
  heterogeneity} level based on some pre-specified heterogeneity
  covariates of interest, which are referred to as the \textit{Group Average Treatment Effects}
  (GATEs). Such effects are, however, beyond the scope of our study and
  the interested reader is referred to \textcite{Zimmert2019}, \textcite{Jacob2019}
  and \textcite{Semenova2021} for a theoretical analysis
  and to \textcite{Knaus2021} for simulation based results or to
  \textcite{Cockx2019}, \textcite{Knaus2020a}, \textcite{Hodler2020} and \textcite{Goller2021} for
  empirical applications estimating policy relevant GATEs.} \textcite{Kunzel2019}
point out that the best estimator for \(\tau(x)\) is also the best estimator
for \(\xi_i\) in terms of the mean squared error (MSE).

In order to identify the effects of interest, we
need a set of identification assumptions. We operate under the
selection-on-observables strategy\footnote{For estimation of heterogeneous treatment effects under different
identification strategies see e.g. \textcite{Athey2019}, \textcite{BargagliStoffi2020} and \textcite{Biewen2021} for the case of instrumental variables and \textcite{Zimmert2018}, \textcite{SantAnna2020} and \textcite{Wager2021} for the case of difference-in-differences.} \parencite[see e.g.][]{Imbens2015} and
assume that we observe all relevant confounders,
i.e.~all covariates \(X_i\) that \textit{jointly} influence
both the treatment \(W_i\) and the potential outcomes, \(Y_i(0)\) and \(Y_i(1)\).
We state the following identification assumptions:

\vspace{0.25cm}

\begin{assumption}[Conditional Independence]\label{as:cia} 
$\big(Y_i(0),Y_i(1)\big) \independent W_i \mid X_i=x, \forall x \in \supp(X_i)$.
\end{assumption}
%\vspace{0.25cm}
\begin{assumption}[Common Support]\label{as:cs} 
$ 0 < \mathbb{P}\big[W_i=1 \mid X_i=x\big] < 1, \forall x \in \supp(X_i)$.
\end{assumption}
%\vspace{0.25cm}
\begin{assumption}[SUTVA]\label{as:sutva} 
$ Y_i=W_i \cdot Y_i(1)+(1-W_i) \cdot Y_i(0)$.
\end{assumption}
%\vspace{0.25cm}
\begin{assumption}[Exogeneity]\label{as:exog} 
$ X_i(0)=X_i(1)$.
\end{assumption}

\vspace{0.25cm}

According to Assumption \ref{as:cia}, also referred to as the
conditional ignorability or unconfoudedness assumption, we assume that
the potential outcomes are independent of the treatment assignment once
conditioned on the covariates, i.e. we assume that there are no hidden
confounders. Assumption \ref{as:cs}, also known as the overlap
assumption, states that the conditional treatment probability is bounded
away from \(0\) and \(1\) and thus
it is \ul{possible} to observe treated as well as control units for each
realization of \(X_i=x\). Assumption \ref{as:sutva} is known as the
stable unit treatment value assumption and indicates that the observed
treatment value for a unit is independent of the treatment exposure for
other units, which rules out any general equilibrium or spillover
effects between treated and controls. Lastly, Assumption \ref{as:exog}
specifies that the covariates are not influenced by the treatment.\footnote{\ul{Analogously
to the definition of potential outcomes, we denote potential covariates
under control and under treatment as $X_i(0)$ and $X_i(1)$, respectively.}} Under
these assumptions it follows that
\begin{align}
\tau(x)&=\mathbb{E}\big[Y_i(1)-Y_i(0) \mid X_i=x\big]\\
&=\mathbb{E}\big[Y_i(1)\mid X_i=x\big] - \mathbb{E}\big[Y_i(0)\mid X_i=x\big]\\
&=\mathbb{E}\big[Y_i(1)\mid X_i=x, W_i=1\big] - \mathbb{E}\big[Y_i(0)\mid X_i=x, W_i=0\big]\\
&=\mathbb{E}\big[Y_i\mid X_i=x, W_i=1\big] - \mathbb{E}\big[Y_i\mid X_i=x, W_i=0\big]\label{eq:identification}
%&=\mu_1(x)-\mu_0(x)
\end{align}
and thus the CATE can be nonparametrically identified from observable
data \parencite{Hurwicz1950}.

\section{Meta-Learning Algorithms and Estimation Procedures}\label{c1sec:learners}

\ul{In the machine learning literature meta-learning represents algorithms
that exploit knowledge about learning to improve the algorithm's performance,
as generally defined by} \textcite{Vilalta2002}. \ul{These include various algorithms
that learn to solve new task from prior learning experience, i.e.} \textit{learning to learn} \parencites{Schmidhuber1987}{Thrun1998}, \ul{algorithms that learn from multiple related tasks, i.e.
\textit{multi-task learning}} \parencite{Caruana1997}, \ul{or algorithms that learn from
multiple models solving identical task, i.e.} \textit{ensemble learning} \parencite{Dietterich2000}.\footnote{\ul{For a recent survey on meta-learning, see} \textcite{Vanschoren2019}.}
\ul{Recently, the meta-learning framework has been adopted within the causal machine
learning literature for learning causal effects from multiple prediction models} 
\parencite[see for example][]{Kunzel2019}, \ul{which could be termed accordingly as
\textit{causal learning}.} 

At a high level the meta-learners for estimation of heterogeneous
causal effects are two-step algorithms. In the first step they define regression
functions, in the causal machine learning literature often denoted as the 
nuisance functions \parencites{Chernozhukov2018c}{Kennedy2020}, which can
be estimated by any \ul{supervised learning method fulfilling suitable
regularity conditions}, i.e. the base learner.\footnote{The base learners can be in
general any set of black-box methods as long as they are consistent estimators of the
nuisance functions and sufficiently minimize the prediction error in terms of
the MSE \parencites{Kunzel2019}{Kennedy2020}{Nie2021a}. However, in order to provide
explicit error bounds and achieve specific convergence rates on the estimation of the
causal effect, the base-learners must fulfill a set of regularity conditions such as
smoothness or sparsity \parencite{Alaa2018}. We discuss the particular conditions
required for each meta-learner in Section \ref{c1sec:meta}.} In the second step they use the
estimated nuisance functions to construct an estimator for the causal
effect, \ul{i.e. the meta-learner}. Various meta-learners then differ in the definitions of the
nuisance functions and their subsequent usage to obtain the final
estimator for the causal effects. Depending on the algorithm complexity, some meta-learners require
estimation of only one single model whereas others require estimation of
multiple models. This raises the question of data usage within the
estimation procedure and thus the possible need for sample-splitting and
cross-fitting, respectively.\footnote{\ul{Recently, related issue of data usage
of the meta-learning algorithms with respect to splitting into training and
validation set for the learning to learn domain has been discussed by}
\textcite{Bai2020} and \textcite{Saunshi2021}.}

In general, the nuisance functions are defined as conditional
expectations of various types. The most common types are the
propensity score function and the response function. First, the propensity
score is defined as the conditional \ul{probability} of a binary treatment
\(W_i\) given the covariates \(X_i\) as follows:
\[e(x)=\mathbb{P}\big[W_i=1 \mid X_i=x\big].\]
In the causal
inference literature the propensity score plays a central role \parencite{Rosenbaum1983} in many
matching and reweighting methods to balance the distributions of treated
and controls \parencites(see)(){Hahn1998}[and][among others]{Huber2013}.
Second, the response function is broadly defined as the conditional
expectation of an outcome variable \(Y_i\) given a conditioning set of
explanatory variables. The particular definitions of the response
function then differ in the specification of the conditioning set and
the subset of the data used. For the meta-learners studied in this
paper, the following definitions of the response function are of
interest:
\begin{align}
\mu(x,w)&=\mathbb{E}[Y_i \mid X_i=x, W_i=w] \label{eq:full}\\
%\mu_1(x)&=\mathbb{E}[Y_i \mid X_i=x, W_i=1] \label{eq:2}\\
%\mu_0(x)&=\mathbb{E}[Y_i \mid X_i=x, W_i=0] \label{eq:3}\\
\mu(x)&=\mathbb{E}[Y_i \mid X_i=x] \label{eq:without}
\end{align}
where Equation \ref{eq:full} defines the response function with
conditioning on both the covariates \(X_i\) as well as the treatment
indicator \(W_i\), while \(\mu(x,1)\) and \(\mu(x,0)\) describe the
response functions with conditioning on the covariates \(X_i\) in the
\ul{subpopulation} under treatment \(W_i=1\) and under control \(W_i=0\),
accordingly. Similarly, Equation \ref{eq:without} defines the response function
with conditioning only on covariates. The meta-learners then use selected
nuisance functions together with the available data as inputs for the
estimation of the CATE function which can be generally denoted as
follows:
\[\tau(x)=\zeta\big(W_i,X_i,Y_i,e(x),\mu(x,w),\mu(x)\big)\]
\ul{where $\zeta(\cdot)$ is a function of the respective inputs, which}
is detailed for each particular meta-learning algorithm in Section
\ref{c1sec:meta}. The problem arises when estimating the nuisance
functions using flexible machine learning methods as these are prone to
the overfitting bias, i.e.~the `own observation bias'. \ul{The overfitting bias
emerges when the in-sample data is fitted too well such that the out-of-sample
performance is compromised }\parencite[see e.g.][for a general discussion of the overfitting issue in machine learning]{Hastie2009}. \ul{Hence, a single observation $i$ can have a
large influence on the predictions for covariates $X_i$ as pointed out by} \textcite{Athey2019a}.
\textcite{Chernozhukov2018c} and \textcite{Newey2018} thus
propose sample-splitting procedures that allow for elimination of such
overfitting biases.\footnote{Original ideas of using sample-splitting
  procedures to eliminate own observation bias stem from the literature
  on density estimation going back to \textcite{Bickel1982},
  \textcite{Bickel1988} and \textcite{Powell1989} among others.}

\subsection{Sample-Splitting and Cross-Fitting}\label{c1sec:cross}

Theoretical arguments express the need for sample-splitting when the
causal estimator involves several estimation steps such as the
estimation of nuisance functions. Within the meta-learning framework the
nuisance functions are typically highly complex and potentially
high-dimensional functions estimated by supervised machine learning
methods such as penalized regression, tree-based methods, neural
networks, etc. Using the same data sample for machine learning
estimation of the nuisance function as well as for estimation of the
causal effect leads to overfitting which induces a bias in the CATE
estimator. On a high level, the bias of the CATE estimator can \ul{generally
be} decomposed into an estimation error of learning the CATE
function itself, and the estimation error in learning the nuisance functions,
encompassing the overfitting bias \parencite[see e.g.][]{Kennedy2020}.\footnote{\ul{The
machine learning estimation of the nuisance functions might additionally induce so-called
regularization bias. This is due to the effective MSE minimization that restricts
the variance of the estimator, but introduces a bias, which might not decay fast enough.
For a discussion of the regularization bias in treatment effect estimation, see}
\textcite{Chernozhukov2018c}.} \textcite{Chernozhukov2018c} show that \ul{for the ATE estimation} the \ul{overfitting bias} can be controlled by using sample-splitting, \ul{while}
\textcite{Kennedy2020} and \textcite{Nie2021a} \ul{extend this concept for the CATE
estimation}. In that case one part of the sample is used to estimate the
nuisance functions and the other part is used to estimate the causal effect.\footnote{Sample-splitting procedures are frequently used in causal machine
  learning literature including Double Machine Learning
  \parencite{Chernozhukov2018c}, Causal Forests
  \parencites{Wager2018}{Lechner2019} or the here-discussed meta-learners
  \parencites{Kennedy2020}{Nie2021a}.} As a result, \ul{the bias term stemming from overfitting}
can be shown to be bounded and to converge to zero \ul{sufficiently fast}. Building upon this
result, \textcite{Newey2018} propose \ul{an advanced} sample-splitting scheme
called \textit{double} sample-splitting. In this case, not only the
nuisance functions are estimated together on a separate part of the sample
but each single nuisance function is estimated on an own separate part
of the sample. In practice, the training data is split into
\(M+1\) equally sized parts, with \(M\) being the number of nuisance
functions to estimate and the remaining part of the data serves for
estimation of the causal effect. \textcite{Newey2018} show that under the
double sample-splitting the bias term converges to
zero at a faster rate \ul{compared to standard sample-splitting where all nuisances are
estimated on the same sample}.\footnote{\ul{The intuition for this result comes from the
observation that for estimators using multiple nuisance functions, such as the doubly robust estimators as e.g. the herein discussed DR-learner,
the estimation error involves a product of the biases from the estimation of
the} \(M\) \ul{nuisance functions.
This induces additional nonlinearity bias if all} \(M\) \ul{nuisance functions
are estimated using the same data, which gets effectively removed by
using separate samples for estimation of each of the} \(M\) functions. For more details
see \textcite{Newey2018} and \textcite{Kennedy2020}.} The double sample-splitting procedure
has also been recently implemented by \textcite{Kennedy2020} in the context of the DR-learner.

In general, the overfitting bias could also be controlled for by restricting
the complexity of the nuisance functions which would, however, prevent
high-dimensional settings as well as usage of a variety of machine
learning estimators or ensembles of those.\footnote{For results in the
  context of the Lasso estimation under sparsity see
  \textcite{Belloni2017}.} Hence, the advantage of using sample-splitting
is to allow for a \ul{high degree of} complexity of the nuisance functions
estimated by a wide class of machine learning estimators \parencite{Kennedy2020}.

It follows that, \ul{theoretically}, sample-splitting prevents overfitting and
thus reduces the bias in the final causal estimator
\parencites{Chernozhukov2018c}{Wager2018}. At the same time, however,
the variance of the estimator increases as less data is
effectively used \ul{for estimation}. Cross-fitting \parencite{Chernozhukov2018c} and respectively \textit{double} \ul{cross-fitting} \parencite{Newey2018}
have been proposed in the literature in
order to reduce the variance loss induced by sample-splitting. 
\ul{In this procedure}, the roles of the data parts get
switched such that each part has been used for both the estimation of
nuisances as well as the causal effect estimation. The final CATE estimator is
then an average of the separate effect estimators produced. This
\ul{method} can be further extended to use more than \(M+1\) splits
denoted as \(K\)-fold cross-fitting \parencite{Chernozhukov2018c} with the
final CATE estimator given as:
\[\hat{\tau}(x)=\frac{1}{K} \sum^{K}_{k=1} \hat{\tau}_k(x)\] where
\(\hat{\tau}_k(x)\) is the CATE estimator based on the \(k\)-th
fold.\footnote{Increasing the efficiency of a sample-splitting based
  estimator by swapping the roles of the data samples and averaging the
  resulting estimates goes back to \textcite{Schick1986} in the context of
  estimation of semi-parametric models.}

\begin{figure}[ht]
\centering
\caption{CATE distributions under full-sample, sample-splitting and cross-fitting estimation.}\label{fig:example}
\includegraphics[scale=0.52]{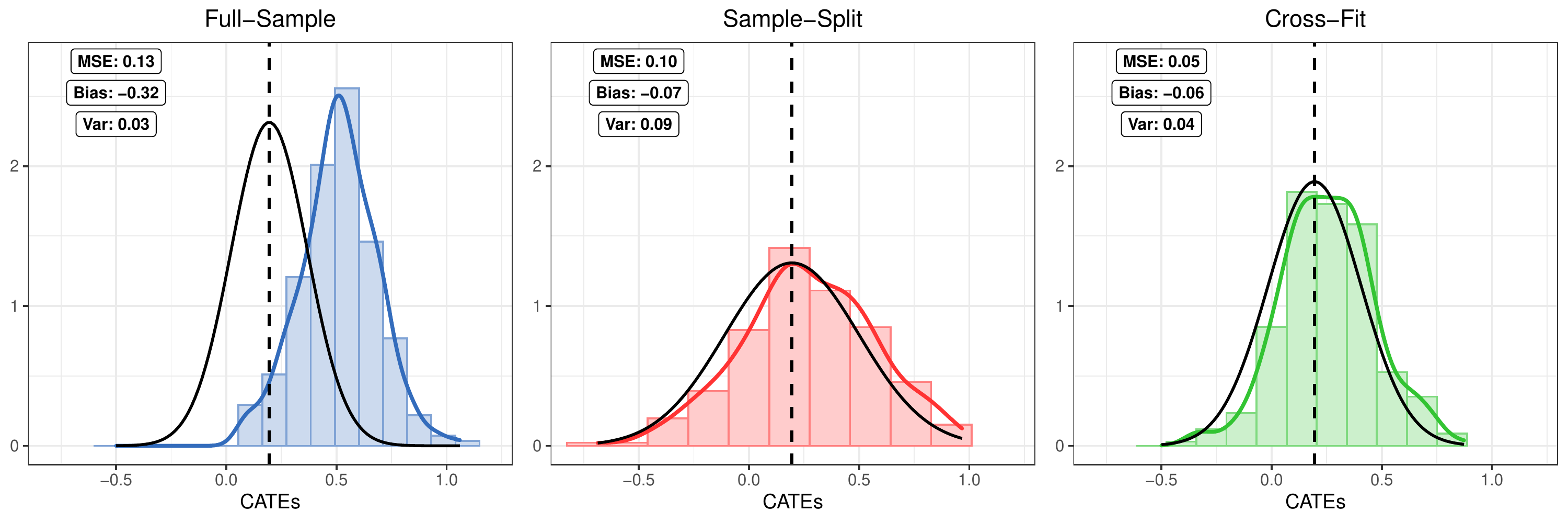}
    \vspace{0.25cm}
\caption*{\footnotesize{\textit{Note:} Distributions of the CATE parameter under full-sample estimation (blue), double sample-splitting (red) and double cross-fitting (green) as a result of a Monte Carlo simulation. The CATE distributions are smoothed with the Gaussian kernel using the Silverman's bandwidth. The dashed black line defines the true value of the CATE while the solid black line plots the normal distribution around the true parameter with variance of the estimated CATE distribution. The CATEs are estimated by the DR-learner based on a training sample of $N^T=32'000$ observations with $250$ simulation replications and predicted out-of-sample. Detailed description of the simulation design is given in Section \ref{c1sec:sim}, while a detailed description of the DR-learner is given in Section \ref{c1sec:meta}.}}
\end{figure}

The above theoretical arguments have a direct impact on the
implementation of various meta-learning algorithms. \ul{Under the double
sample-splitting} the more models have to be estimated within the
meta-learning algorithm, the more data splits are being implicitly induced,
\ul{while} the impact \ul{thereof} in finite samples is not
clear \textit{a priori} as pointed out by \textcite{Newey2018}.
\ul{As such, the} researcher faces a typical
bias-variance trade-off with respect to sample-splitting. In order to
illustrate the issue it is instructive to \ul{decompose} the MSE of a
CATE estimator \(\hat{\tau}(x)\):
\[MSE\bigg(\hat{\tau}(x)\bigg) = Var\bigg(\hat{\tau}(x)\bigg) + \bigg(Bias\big(\hat{\tau}(x)\big)\bigg)^2.\]
Naively using the full data sample for estimation of both the nuisance
functions as well as the CATE function leads to a higher bias due to
overfitting but at the same time to lower variance as all available data is used
for estimation. Using sample-splitting eliminates the overfitting bias
but results in higher variance due to less data being used for
estimation. In contrast, cross-fitting both removes the overfitting bias
and reduces the variance by effectively using \ul{all the available information
from the data} for estimation. Figure \ref{fig:example} illustrates this theoretical
argument by contrasting the distributions of the CATE parameter
under full-sample estimation, double sample-splitting and double
cross-fitting, \ul{resulting from a Monte Carlo simulation} based
on a large training sample of \(32'000\)
observations (further details on the meta-learner and the simulation
design are provided in Sections \ref{c1sec:meta} and \ref{c1sec:sim},
respectively). We observe that the \ul{theoretical} arguments can be
documented in finite samples too. As such, the full sample version
exhibits substantial bias due to overfitting as its distribution is
shifted away from the true value of the CATE parameter, but with a
rather low variance. On the contrary, the double sample-splitting
version successfully eliminates the overfitting bias as the simulated
distribution is centered around the true value of the CATE, however with
much larger variance. Finally, the double cross-fitting version keeps
the reduction in bias whilst having a much lower variance in comparison
to the double sample-splitting version as the spread of the
CATE distribution comes close to the full sample version, indicating the
gain in efficiency of this procedure.

Apart from the illustrative example above, the empirical question
remains the precise quantification of this bias-variance trade-off for
various meta-learners and to what degree this might vary with different
sample sizes. Different meta-learners use different nuisance
functions in different ways which might have an influence on the
performance under the particular estimation procedures. Even
though sample-splitting and cross-fitting help to eliminate the
overfitting bias, in finite samples less data available for estimation
might even lead to higher bias due to errors in learning the CATE
function itself, especially for small sample sizes. In this paper we
address this open question via Monte Carlo simulations and compare the
performance of various meta-learners under full-sample, double
sample-splitting and double cross-fitting procedure for several
different sample sizes to shed more light onto the finite sample
properties. We follow \textcite{Newey2018} and choose the double
sample-splitting, respectively double cross-fitting procedure due to its
theoretically faster convergence rates. Furthermore, we opt for the
setting with equally sized \(K=M+1\) folds as suggested by
\textcite{Kennedy2020}. Additionally, we always distinguish between the
training and validation data. We use the training data for
learning the nuisance function and the CATE function, including the
double sample-splitting and double cross-fitting procedure, while we
evaluate the CATEs on a set of new validation data. An illustration of
the data usage under full-sample estimation, double sample-splitting and
double cross-fitting procedure is provided in Figure \ref{fig:diagram}.

\begin{figure}[ht]
    \caption{Illustration of the full-sample, sample-splitting and cross-fitting procedure.}
    \label{fig:diagram}
\small
\centering
    \tikzstyle{decision} = [diamond, draw, fill=blue!20, 
    text width=5.5em, text badly centered, inner sep=0pt]
    \tikzstyle{block} = [rectangle, draw, fill=gray!10, 
    text width=11.5em, text centered, rounded corners, minimum height=2em]
    \tikzstyle{fold} = [rectangle, draw, fill=black!20, 
    text width=2.75em, text centered, rounded corners, minimum height=2em]
    \tikzstyle{blocksmall} = [rectangle, draw, fill=blue!20, 
    text width=2.75em, text centered, rounded corners, minimum height=2em]
    \tikzstyle{line} = [draw, -latex']
    \tikzstyle{line2} = [draw]
    \tikzstyle{cloud} = [draw, ellipse,fill=red!20,
    minimum height=4em]
    \tikzstyle{decisionn} = [diamond, draw, fill=red!20, 
    text width=4em, text badly centered, node distance=4cm, inner sep=0pt]
    \tikzstyle{blockk} = [rectangle, draw, fill=blue!20, 
    text width=8em, text centered, rounded corners, minimum height=2em]
    \tikzstyle{blockkk} = [rectangle, draw, fill=red!20, 
    text width=6em, text centered, rounded corners, minimum height=2em]
    
    \begin{tikzpicture}[node distance = 5cm, auto]\label{ams1}
    % Place nodes
    \node [block] (Training1) {Training Data\\ $W_i,X_i,Y_i$};
    \node [block, right of=Training1] (Training2) {Training Data\\ $W_i,X_i,Y_i$};
    \node [block, right of=Training2] (Training3) {Training Data\\ $W_i,X_i,Y_i$};
    % full sample
    \node [block, below of=Training1, fill=green!20, yshift=2cm] (Nuisances) {$\hat{e}(x)$};
    \node [block, below of=Nuisances, fill=green!20, yshift=3.5cm] (Nuisancemu) {$\hat{\mu}(x)$};
    \node [block, below of=Nuisancemu, fill=green!20, yshift=3.5cm] (CATE) {$\hat{\tau}(x)$};
    \node [block, below of=CATE, yshift=3.5cm] (Validation) {Validation Data\\ $X_i$};
    \node [block, below of=Validation, yshift=3.5cm] (Final1) {CATE\\ $\hat{\tau}_i^{F}=\hat{\tau}(X_i)$};
    % Draw edges
    \path [line] (Training1) -- (Nuisances);
    \path [line2] (Nuisances) -- (Nuisancemu);
    \path [line] (Nuisancemu) -- (CATE);
    \path [line] (CATE) -- (Validation);
    \path [line] (Validation) -- (Final1);
    
    % sample split
    \node [fold, below of=Training2, yshift=3.5cm, xshift=-1.5cm] (fold1) {Fold 1};
    \node [fold, below of=Training2, yshift=3.5cm] (fold2) {Fold 2};
    \node [fold, below of=Training2, yshift=3.5cm, xshift=1.5cm] (fold3) {Fold 3};
    \node [blocksmall, right of=Nuisances, xshift=-1.5cm, fill=green!20] (pscore) {$\hat{e}_1(x)$};
    \node [blocksmall, right of=Nuisancemu, xshift=0cm, yshift=0cm, fill=green!20] (mu) {$\hat{\mu}_2(x)$};
    \node [blocksmall, right of=CATE, xshift=1.5cm, fill=green!20] (CATE2) {$\hat{\tau}_3(x)$};
    \node [block, below of=CATE2, xshift=-1.5cm, yshift=3.5cm] (Validation2) {Validation Data\\ $X_i$};
    \node [block, below of=Validation2, yshift=3.5cm] (Final2) {CATE\\ $\hat{\tau}_i^{S}=\hat{\tau}_3(X_i)$};
    % Draw edges
    \path [line] (Training2) -- (fold1);
    \path [line] (Training2) -- (fold2);
    \path [line] (Training2) -- (fold3);
    \path [line] (fold1) -- (pscore);
    \path [line] (fold2) -- (mu);
    \path [line] (fold3) -- (CATE2);
    \path [line2] (pscore) -- (mu);
    \path [line] (mu) -- (CATE2);
    \path [line] (CATE2) -- (Validation2);
    \path [line] (Validation2) -- (Final2);
    
    % cross fit
    \node [fold, below of=Training3, yshift=3.5cm, xshift=-1.5cm] (fold11) {Fold 1};
    \node [fold, below of=Training3, yshift=3.5cm] (fold22) {Fold 2};
    \node [fold, below of=Training3, yshift=3.5cm, xshift=1.5cm] (fold333) {Fold 3};
    \node [blocksmall, right of=pscore, fill=green!20] (pscore) {$\hat{e}_1(x)$};
    \node [blocksmall, right of=pscore, xshift=-3.5cm, fill=red!20] (e2) {$\hat{e}_2(x)$};
    \node [blocksmall, right of=pscore, xshift=-2cm] (e3) {$\hat{e}_3(x)$};
    \node [blocksmall, right of=CATE2, fill=green!20] (CATE3) {$\hat{\tau}_3(x)$};
    \node [blocksmall, right of=CATE2, xshift=-3cm, fill=red!20] (CATE31) {$\hat{\tau}_1(x)$};
    \node [blocksmall, right of=CATE2, xshift=-1.5cm] (CATE32) {$\hat{\tau}_2(x)$};
    \node [blocksmall, right of=mu, xshift=-1.5cm, yshift=0cm] (mu31) {$\hat{\mu}_1(x)$};
    \node [blocksmall, right of=mu, xshift=0cm, yshift=0cm, fill=green!20] (mu32) {$\hat{\mu}_2(x)$};
    \node [blocksmall, right of=mu, yshift=0cm,  xshift=1.5cm, fill=red!20] (mu33) {$\hat{\mu}_3(x)$};
    \node [block, below of=CATE3, xshift=-1.5cm, yshift=3.5cm] (Validation3) {Validation Data\\ $X_i$};
    \node [block, below of=Validation3, yshift=3.5cm] (Final3) {CATE\\ $\hat{\tau}_i^{C}=\frac{1}{3}\sum^3_{k=1}\hat{\tau}_k(X_i)$};
    % Draw edges
    \path [line] (Training3) -- (fold11);
    \path [line] (Training3) -- (fold22);
    \path [line] (Training3) -- (fold333);
    \path [line] (fold11) -- (pscore);
    \path [line] (fold22) -- (e2);
    \path [line] (fold333) -- (e3);
    \path [line2] (pscore) -- (mu32);
    \path [line] (mu32) -- (CATE3);
    \path [line2] (e2) -- (mu33);
    \path [line] (mu33) -- (CATE31);
    \path [line2] (e3) -- (mu31);
    \path [line] (mu31) -- (CATE32);
    \path [line] (CATE31) -- (Validation3);
    \path [line] (CATE32) -- (Validation3);
    \path [line] (CATE3) -- (Validation3);
    \path [line] (Validation3) -- (Final3);
    
    % draw boxes around
    \draw[black,thick] ($(Validation.north west)+(-0.3,-2.7)$)  rectangle ($(Training1.south east)+(0.3,1.2)$);
    \node[above, yshift=0.95cm] (rec1) {Full-Sample};
    \draw[black,thick] ($(Validation2.north west)+(-0.3,-2.7)$)  rectangle ($(Training2.south east)+(0.3,1.2)$);
    \node[right of=rec1, yshift=0cm] (rec2) {Sample-Split};
    \draw[black,thick] ($(Validation3.north west)+(-0.3,-2.7)$)  rectangle ($(Training3.south east)+(0.3,1.2)$);
    \node[right of=rec2, yshift=0cm] (rec3) {Cross-Fit};

    \end{tikzpicture}
    \vspace{0.5cm}
\caption*{\footnotesize{\textit{Note:} Illustration of the full-sample (left), double sample-splitting (middle) and double cross-fitting (right) procedures with $K=3$ folds. The propensity score function is defined by $e(x)$, the response functions in general are denoted by $\mu(x)$ and the CATE function is characterized by $\tau(x)$. Subscripts for the nuisance functions and the CATE function correspond to the fold used for estimation, while the colors indicate the combination of the estimated functions across different folds.}}
 \end{figure}

\ul{Further} motivation for the usage of sample-splitting and cross-fitting
stems from the theoretical arguments for conducting statistical
inference about the causal parameters of interest. As such,
sample-splitting plays a crucial role in obtaining estimators that are
not only approximately unbiased but also normally distributed which in
turn allows for a valid construction of confidence intervals. In this
vein, \textcite{Chernozhukov2018c} provide results for the estimators of
average treatment effect (ATE) that rely on sample-splitting and
cross-fitting procedures. \textcite{Semenova2021} and \textcite{Zimmert2019}
extend this analysis for parametric and nonparametric estimators of
group average treatment effects (GATEs), respectively. In the context of
Causal Forests, \textcite{Wager2018}, \textcite{Lechner2019}, and
\textcite{Athey2019} also rely on sample-splitting procedures termed
`honesty' to provide inference for causal effects on various levels of
aggregation. Nonetheless, in the context of meta-learning estimation of
causal effects, there \ul{appears to be lack of} unifying model-free theory for conducting
statistical inference so far. One exception is the study by
\textcite{Kunzel2019} that analyses various versions of bootstrapping for
estimation of standard errors for the CATEs. Recently, \textcite{Jacob2021}
makes use of such bootstrapping procedures to construct confidence
intervals in an empirical application. Besides the computational burden,
however, none of the bootstrapping procedures studied by
\textcite{Kunzel2019} seems to reliably provide accurate coverage rates.
However, the meta-learners analyzed in \textcite{Kunzel2019} do not make
use of sample-splitting, which could potentially improve the performance
of the bootstrapping for estimation of standard errors, given the
insights from the related literature. While we do study the properties
of the distribution of the CATEs within the simulation
experiments in Section \ref{c1sec:sim}, we do not further analyse the
estimation of standard errors mainly due to computational reasons and
focus primarily on the point estimators. However, apart from the
computational aspects, we note that combining sample-splitting and
cross-fitting with bootstrapping for statistical inference about causal
effects within the meta-learning framework might be a promising avenue
for future research.

\subsection{Meta-Learners}\label{c1sec:meta}

\ul{In the following, we review the meta-learning algorithms for estimation of
heterogeneous treatment effects and discuss their advantages
and disadvantages in particular empirical settings.}

\subsubsection{S-learner}

The first meta-learning algorithm we investigate is the S-learner as denoted by \textcite{Kunzel2019}. The early ideas of the S-learner stem from the marketing
literature on uplift modelling \parencite{Lo2002} and the epidemiology
literature, where the S-learner is also being referred to as
\textit{g}-computation \parencites{Robins1986}{Snowden2011}. More recently,
the S-learner has been proposed for the estimation of heterogeneous treatment
effects based on the Regression Trees \parencite{Athey2015} and Random
Forests \parencite{Lu2018} as well as their
Bayesian versions, i.e. BART \parencites{Hill2011}{Green2012}. According to the naming
convention of \textcite{Kunzel2019}, \textit{S-} stands for \textit{Single} as this
meta-learner involves only one single model, namely the response
function, \(\mu(x,w)\), that needs to be estimated. The final causal
effect is, in this case, obtained as a difference \ul{between} predictions of the
response function with setting the treatment indicator to $W_i=1$ and
$W_i=0$, respectively.\footnote{Similar notion of predicted difference in the scope of model interpretation has been proposed by \textcite{King2000}.} The algorithm can be described as follows:\footnote{As a matter of notation, we refer to the training data used for
  model estimation with superscript \(^T\) as \(\{(X_i,Y_i,W_i)\}^T\)
  and the validation data used for effect prediction with superscript
  \(^V\) as \(\{(X_i)\}^V\).}\\
~\\
\begin{algorithm}[H]
\SetAlgoLined
\KwIn{Training Data: $\{(X_i,Y_i,W_i)\}^T$, Validation Data: $\{(X_i)\}^V$}
\KwOut{CATE: $\hat{\tau}_S(x)=\hat{E}[Y_i(1)-Y_i(0) \mid X_i=x]$}
 \Begin{
 \textsc{Response Function}\;
  estimate: $\mu(x,w)=E[Y_i \mid X_i=x, W_i=w]$ in $\{(X_i,Y_i,W_i)\}^T$\;
  %predict: $\hat{\mu}(x,1)=\hat{E}[Y_i \mid X_i=x, W_i=1]$ for $\{(X_i,W_i^1)\}^V$\;
  %predict: $\hat{\mu}(x,0)=\hat{E}[Y_i \mid X_i=x, W_i=0]$ for $\{(X_i,W_i^0)\}^V$\;
  \vspace{0.2cm}
  \textsc{CATE Function}\;
  define: $\hat{\tau}_S(x) = \hat{\mu}(x,1) - \hat{\mu}(x,0)$\; %in $\{(X_i)\}^V$\;
  %\vspace{0.2cm}
  predict:  $\hat{\tau}_S(X_i) = \hat{\mu}(X_i,1) - \hat{\mu}(X_i,0)$ in $\{(X_i)\}^V$
  %predict: $\hat{\tau}_{S} = \frac{1}{N^V}\sum^{N^V}_{i=1}\hat{\tau}_S(X_i)$ in $\{(X_i)\}^V$
  }
\caption{\textsc{S-learner}}
\label{Slearner}
\end{algorithm}
~\\
As can be seen from Algorithm \ref{Slearner}, the S-learner does not
assign any special role to the treatment indicator \(W_i\) within the
estimation procedure and uses it only \textit{post hoc} in the
computation of the causal effect. Thus, if the treatment indicator is
not strongly predictive for the outcome the S-learner will tend to estimate
a zero treatment effect.\footnote{\textcite{Kunzel2019} argue that the
  S-learner is actually biased towards zero.} Nevertheless, the
S-learner will perform particularly well if the true CATE function is
indeed zero, i.e.~if \(\tau(x)=0\), which has also been documented in
the simulation experiments of \textcite{Kunzel2019}. For the forest based
S-learner, \textcite{Kunzel2019diss} proposes a modification of the
algorithm such that it shrinks towards the ATE instead of zero by
performing a Ridge regression in the final leaves of the trees within
the forest.\footnote{For a detailed explanation of this procedure see
  \textcite{Kunzel2019diss}.} In our simulations, we study a simpler
modification of the forest based S-learner by always including the
treatment indicator in the random subset of covariates when determining
the splits. By doing so, we always give the S-learner the chance to split
on the treatment indicator which might potentially alleviate the
zero-bias issue. We will henceforth denote such learner as the
SW-learner, where the \textit{W} reflects the enforcement of the
treatment indicator into the splitting set of covariates.
We discuss the behaviour of the SW-learner more
closely throughout the simulation results in
Section \ref{c1sec:simresults}. Furthermore, notice that the
Algorithm \ref{Slearner} consists of only one nuisance function that
needs to be estimated and thus does not require any sample-splitting or
cross-fitting within the training sample induced by multiple nuisance
functions, hence it always has access to the full sample of the training
data.\footnote{Nevertheless, an optional additional sample-splitting or
cross-fitting could potentially improve the performance of the
S-learner by reducing the possible overfitting of the base learner as
such. This is, however, beyond the scope of our analysis and is left
for future research.} However, from a theoretical perspective, \textcite{Alaa2018}
show that modelling the CATE as a single response function as in the S-learner is not optimal in terms of achieving the fastest convergence rate as this depends on the complexity, defined as sparsity-to-smoothness ratio, of the response functions in the treated, \(\mu(x,1)\), and the control sample, \(\mu(x,0)\), and pooling these into a single response function enforces the complexity for treated and controls to be the same.

\subsubsection{T-learner}

The T-learner is another common and widely used meta-learner that we
investigate in our study. Similarly to the S-learner, its early applications
emerged in the marketing literature on the uplift modelling \parencites{Hansotia2002}{Radcliffe2007}, while more recently it has been suggested for estimating
heterogeneous treatment effects in the fields of medicine \parencite{Foster2011}
and econometrics \parencite{Athey2015}.
In the \ul{econometric} literature it is sometimes also called
as the \textit{basic} \parencite{Lechner2019}, \textit{plug-in} \parencite{Kennedy2020} or \textit{naive} \parencite{Nie2021a} CATE estimator.
According to \textcite{Kunzel2019}, \textit{T-} stands for \textit{Two} as
this meta-learner involves two models that need to be estimated,
defined by the treatment indicator \(W_i\). These are namely the
response function in the treated sample, \(\mu(x,1)\), and the response
function in the control sample, \(\mu(x,0)\). This is in contrast to the
above S-learner which pools the two response functions into a single
one. However, similarly to the S-learner the causal effect is computed
as a difference in predictions of the two response functions, \ul{which
is motivated by the identification result as in Equation} \eqref{eq:identification}.
The algorithm can be summarized as follows:\footnote{Notationwise, we refer
  to a subset of the data defined by a specific value of the variable as
  for example \(W_i=1\) by a subscript as \(\{(X_i,Y_i)\}_{W_i=1}\).}\\
~\\
\begin{algorithm}[H]
\SetAlgoLined
\KwIn{Training Data: $\{(X_i,Y_i,W_i)\}^T$, Validation Data: $\{(X_i)\}^V$}
\KwOut{CATE: $\hat{\tau}_T(X_i)=\hat{E}[Y_i(1)-Y_i(0) \mid X_i=x]$}
 \Begin{
 \textsc{Response Functions}\;
  estimate: $\mu(x,1)=E[Y_i \mid X_i=x, W_i=1]$ in $\{(X_i,Y_i)\}^T_{W_i=1}$\;
  %predict: $\hat{\mu}_1(x)=\hat{E}[Y_i \mid X_i=x]$ for $\{(X_i)\}^V$\;
  %\vspace{0.2cm}
  estimate: $\mu(x,0)=E[Y_i \mid X_i=x, W_i=0]$ in $\{(X_i,Y_i)\}^T_{W_i=0}$\;
  %predict: $\hat{\mu}_0(x)=\hat{E}[Y_i \mid X_i=x]$ for $\{(X_i)\}^V$\;
  \vspace{0.2cm}
  \textsc{CATE Function}\;
  define: $\hat{\tau}_T(x) = \hat{\mu}(x,1) - \hat{\mu}(x,0)$\; %in $\{(X_i)\}^V$\;
 % \vspace{0.2cm}
  predict: $\hat{\tau}_T(X_i) = \hat{\mu}(X_i,1) - \hat{\mu}(X_i,0)$ in $\{(X_i)\}^V$
  %predict: $\hat{\tau}_{T} = \frac{1}{N^V}\sum^{N^V}_{i=1}\hat{\tau}_T(X_i)$ in $\{(X_i)\}^V$
  }
\caption{\textsc{T-learner}}
\label{Tlearner}
\end{algorithm}
~\\
Hence the T-learner uses the treatment indicator to split the estimation
of the response function into two parts. This procedure is expected to
work particularly well if the CATE function is complicated and there are
no common trends in the response functions. This
phenomenon finds supportive evidence in several simulation studies
\parencites(see for example)(){Kunzel2019}{Jacob2020}{Curth2021}[or][]{Nie2021a}.
Nonetheless, it is expected to work rather poorly if the CATE function is simple, as
the response functions are not trained jointly and thus their difference
might be unstable \parencites{Lechner2019}{Kennedy2020}{Nie2021a}. In terms of
the estimation of the nuisance functions, the T-learner behaves
similarly to the S-learner, as only the response functions need to be
estimated to compute the CATE. As such no additional sample-splitting
induced by multiple nuisance functions is required as the response
functions are themselves estimated on separate samples defined by
treated and control.\footnote{Again, this does not preclude that an
optional sample-splitting or cross-fitting might be beneficial for the
same reason as in the case of the S-learner \parencite[][provides some results on this issue for the T-learner]{Jacob2020}. Using an honest forest as
a base learner would also add an implicit sample-splitting procedure,
however, this is not analysed herein.} Theoretically, the convergence rate
of the base learners for estimating the response functions directly translates
into the rate for estimating the CATE \parencites{Kunzel2019}{Kennedy2020} and thus
depends on the complexity, i.e. the dimension and the smoothness, of the response
functions \parencite{Alaa2018}.\footnote{For a detailed analysis of the optimal
minimax rates for the T-learner, we refer to \textcite{Alaa2018} and \textcite{Kunzel2019}.}

\subsubsection{X-learner}

The above mentioned problems of the T-learner \ul{are aggravated} if
the treatment assignment is \ul{highly} unbalanced, meaning that the vast majority
of observations in the sample belongs to only one treatment status.
\textcite{Kunzel2019} therefore propose the X-learner which
addresses this issue. The X-learner builds on the T-learner and, as such,
first estimates the two response functions \(\mu(x,1)\) and
\(\mu(x,0)\). It then uses these estimates to impute the unobserved
individual treatment effects \ul{for the treated}, \(\tilde{\xi}_i^1\), and
the control, \(\tilde{\xi}_i^0\).
\ul{The imputed effects} are in turn used as pseudo-outcomes to
estimate the treatment effects in the treated sample, \(\tau(x,1)\), and
the control sample, \(\tau(x,0)\), respectively. The final CATE estimate
\(\tau(x)\) is then a weighted average of these treatment effect
estimates weighted by the propensity score, \(e(x)\).\footnote{In the
  original definition of the X-learner, the estimation of the propensity
  score is not exactly specified as it could be any weighting function
  in general. However, in practice the estimation of the propensity
  score is recommended \parencite{Kunzel2019}.} Thus the X-learner
additionally uses the information from the treated to learn about the
controls and vice-versa in a \textit{Cross} regression style, hence the
\textit{X} term in its naming label. The learning algorithm can be
detailed as follows:\\
~\\
\begin{algorithm}[H]
\SetAlgoLined
\KwIn{Training Data: $\{(X_i,Y_i,W_i)\}^T$, Validation Data: $\{(X_i)\}^V$}
\KwOut{CATE: $\hat{\tau}_X(X_i)=\hat{E}[Y_i(1)-Y_i(0) \mid X_i=x]$}
 \Begin{
 \textsc{Response Functions}\;
  estimate: $\mu(x,1)=E[Y_i \mid X_i=x, W_i=1]$ in $\{(X_i,Y_i)\}^T_{W_i=1}$\;
  %predict: $\hat{\mu}_1(x^0)=\hat{E}[Y_i \mid X_i=x]$ for $\{(X_i)\}^T_{W_i=0}$\;
  %\vspace{0.2cm}
  estimate: $\mu(x,0)=E[Y_i \mid X_i=x, W_i=0]$ in $\{(X_i,Y_i)\}^T_{W_i=0}$\;
  %predict: $\hat{\mu}_0(x^1)=\hat{E}[Y_i \mid X_i=x]$ for $\{(X_i)\}^T_{W_i=1}$\;
  \vspace{0.2cm}
  \textsc{Imputed Effects}\;
  predict: $\tilde{\xi}_i^1=Y_i-\hat{\mu}(X_i,0)$ in $\{(X_i,Y_i)\}^T_{W_i=1}$\;
  predict: $\tilde{\xi}_i^0=Y_i-\hat{\mu}(X_i,1)$ in $\{(X_i,Y_i)\}^T_{W_i=0}$\;
  \vspace{0.2cm}
  \textsc{Treatment Effects}\;
  estimate: $\tau(x,1)=E[\tilde{\xi}_i^1 \mid X_i=x, W_i=1]$ in $\{(X_i,Y_i)\}^T_{W_i=1}$\;
  %predict: $\hat{\tau}_1(x)=\hat{E}[\tilde{D}_i^1 \mid X_i=x]$ for $\{(X_i)\}^V$\;
  %\vspace{0.2cm}
  estimate: $\tau(x,0)=E[\tilde{\xi}_i^0 \mid X_i=x, W_i=0]$ in $\{(X_i,Y_i)\}^T_{W_i=0}$\;
  %predict: $\hat{\tau}_0(x)=\hat{E}[\tilde{D}_i^0 \mid X_i=x]$ for $\{(X_i)\}^V$\;
  \vspace{0.2cm}
  \textsc{Propensity Score}\;
  estimate: $e(x)=P[W_i=1 \mid X_i=x]$ in $\{(X_i,W_i)\}^T$\;
  %predict: $\hat{e}(x)=\hat{E}[W_i \mid X_i=x]$ for $\{(X_i)\}^V$\;
  \vspace{0.2cm}
  \textsc{CATE Function}\;
  define: $\hat{\tau}_X(x) = \hat{e}(x) \cdot \hat{\tau}(x,0) + \big(1-\hat{e}(x)\big) \cdot \hat{\tau}(x,1)$\; %in $\{(X_i)\}^V$\;
  %\vspace{0.2cm}
  predict: $\hat{\tau}_X(X_i) = \hat{e}(X_i) \cdot \hat{\tau}(X_i,0) + \big(1-\hat{e}(X_i)\big) \cdot \hat{\tau}(X_i,1)$ in $\{(X_i)\}^V$
  %predict: $\hat{\tau}_{X} = \frac{1}{N^V}\sum^{N^V}_{i=1}\hat{\tau}_X(X_i)$ in $\{(X_i)\}^V$
  }
\caption{\textsc{X-learner}}
\label{Xlearner}
\end{algorithm}
~\\
According to Algorithm \ref{Xlearner}, the X-learner, in contrast to the
T-learner, firstly uses the response functions for imputing the
unobserved individual treatment effects instead of directly estimating
the CATE. Secondly, these imputed individual treatment effects are used
for estimating the CATE and reweighted by the propensity scores. The
reweighting helps to put more weight on the treatment effects which have
been estimated more precisely, i.e.~the ones coming from the larger
treated or control sample, respectively. For this reason, the X-learner is
expected to work particularly well in unbalanced settings, which is often
the case in practice as the share of treated might be restricted
financially or otherwise
\parencites(see)(){Arceneaux2006}{Gerber2008}{Broockman2016}[or][for such unbalanced empirical settings]{Goller2020}. Furthermore, by
directly estimating the treatment effects in the second step it enables
the estimator to learn structural properties of the CATE function from
the data and is thus expected to work well if the CATE function is
approximately linear or sparse \parencite{Kunzel2019}. Theoretically,
\textcite{Kunzel2019} indeed prove that the X-learner achieves a faster convergence
rate than the T-learner in cases where the treatment assignment is highly unbalanced
or when the CATE function is linear. Even if no assumptions about the CATE function
are imposed, the X-learner can be proven to achieve the same rate as the T-learner.
This hinges on regularity conditions such as the Lipschitz continuity of the response
functions.\footnote{More specifically, \textcite{Kunzel2019} prove that the X-learner
can indeed achieve the parametric $\sqrt{N}$ rate if the CATE function is estimated by OLS,
while the nuisance functions can be estimated at any nonparametric rate. Furthermore,
they prove that the X-learner achieves the minimax optimal rate if both the nuisance
functions and the CATE function are estimated via $k$-NN regression.} In simulations of
\textcite{Kunzel2019} the X-learner performs
reasonably well even in other non-favourable settings. Notice further
that Algorithm \ref{Xlearner} requires more estimation steps than the
previous two meta-learners. Additionally to the estimation of the
response functions, the X-learner requires the estimation of the
treatment effect functions as well as the propensity score function.
This raises the question of possible overfitting and hence the need for
sample-splitting and cross-fitting, respectively. However, there is
theoretically no explicit requirement for sample-splitting in the case of
the X-learner when estimating the nuisance functions, apart from
training and validation data split \parencite{Kunzel2019}. Yet, it might well be that the
sample-splitting and further cross-fitting have a non-negligible
influence on the performance of the learner in finite samples. We
address this issue by implementing the double sample-splitting and
double cross-fitting version of the X-learner in the simulation study.
For the case of the full-sample estimation we use the
out-of-bag predictions of the underlying forest as estimates of the
nuisance functions \ul{as recommended by} \textcite{Lu2018} and
\textcite{Athey2019a}. The out-of-bag predictions are based on the
observations that have been left `out of the bag' when drawing
bootstrap samples to estimate the trees of the forest
\parencite{Hastie2009}. Such observations, however, randomly appear both as
training as well as validation observations \ul{across the trees of the forest}
and thus such out-of-bag predictions are neither the classical in-sample fitted
values nor proper out-of-sample predictions \ul{and should not be confused with
the `honest' predictions}. We use the out-of-bag predictions for all
meta-learners within our analysis.

\subsubsection{DR-learner}

Although the X-learner makes use of the estimation of multiple nuisance
functions, it does not provide the double robustness property which
exploits the fact that the estimator remains consistent if either the
response function or the propensity score function is misspecified
\parencites{Kennedy2017}{Lee2017}. Recently, \textcite{Kennedy2020} proposed the
DR-learner where \textit{DR} symbolizes the
\textit{Double Robustness} property of the learner. The DR-learner
constructs a doubly robust score in the first estimation stage and
estimates the CATE in the second stage. There have been many other
versions of the DR-learner proposed in the literature, but these were
restricted to a particular estimator used in the second stage and are
thus not part of the meta-learning framework. For example,
\textcite{Semenova2021} propose a linear estimation of the CATE function,
whereas a local-constant estimation is proposed by \textcite{Zimmert2019}
and \textcite{Fan2020}, which works well for the estimation of GATEs,
i.e.~for low-dimensional conditioning set. The main advantage of the
DR-learner in comparison to the other versions lies in the general
model-free second stage \ul{estimation of the CATE function}. However,
common to all versions in the literature is the estimation of the doubly
robust score\footnote{Also called efficient score or efficient
  influence function in the literature \parencites{Robins1995}{Hahn1998}.} by
machine learning methods in the first stage also known as Double
Machine Learning \parencite{Chernozhukov2018c}. For a comprehensive overview
of the CATE estimators building on the doubly robust score see
\textcite{Knaus2020b}. The specific algorithm for the DR-learner is then
defined as follows:\\
~\\
\begin{algorithm}[H]
\SetAlgoLined
\KwIn{Training Data: $\{(X_i,Y_i,W_i)\}^T$, Validation Data: $\{(X_i)\}^V$}
\KwOut{CATE: $\hat{\tau}_{DR}(x)=\hat{E}[Y_i(1)-Y_i(0) \mid X_i=x]$}
 \Begin{
 \textsc{Response Functions}\;
  estimate: $\mu(x,1)=E[Y_i \mid X_i=x, W_i=1]$ in $\{(X_i,Y_i)\}^T_{W_i=1}$\;
  %predict: $\hat{\mu}_1(x)=\hat{E}[Y_i \mid X_i=x]$ for $\{(X_i)\}^T$\;
  %\vspace{0.2cm}
  estimate: $\mu(x,0)=E[Y_i \mid X_i=x, W_i=0]$ in $\{(X_i,Y_i)\}^T_{W_i=0}$\;
  %predict: $\hat{\mu}_0(x)=\hat{E}[Y_i \mid X_i=x]$ for $\{(X_i)\}^T$\;
  \vspace{0.2cm}
  \textsc{Propensity Score}\;
  estimate: $e(x)=P[W_i=1 \mid X_i=x]$ in $\{(X_i,W_i)\}^T$\;
  %predict: $\hat{e}(x)=\hat{E}[W_i \mid X_i=x]$ for $\{(X_i)\}^T$\;
  \vspace{0.2cm}
  \textsc{Pseudo Outcome}\;
  predict: $\hat{\psi}_i=\frac{W_i\big(Y_i-\hat{\mu}(X_i,1)\big)}{\hat{e}(X_i)}-\frac{(1-W_i)\big(Y_i-\hat{\mu}(X_i,0)\big)}{1-\hat{e}(X_i)} + \hat{\mu}(X_i,1) - \hat{\mu}(X_i,0)$ in $\{(X_i,Y_i,W_i)\}^T$\;
  \vspace{0.2cm}
  \textsc{CATE Function}\;
  estimate: $\tau_{DR}(x) = E[\hat{\psi}_i \mid X_i=x]$ in $\{(X_i,Y_i,W_i)\}^T$\;
  %\vspace{0.2cm}
  predict: $\hat{\tau}_{DR}(X_i) =  \hat{E}[\hat{\psi}_i \mid X_i=x]$ in $\{(X_i)\}^V$
  %predict: $\hat{\tau}_{DR} = \frac{1}{N^V}\sum^{N^V}_{i=1}\hat{\tau}_{DR}(X_i)$ in $\{(X_i)\}^V$
  }
\caption{\textsc{DR-learner}}
\label{DRlearner}
\end{algorithm}
~\\
As can be seen in Algorithm \ref{DRlearner} above, the DR-learner
estimates the very same nuisance functions, \(\mu(x,0), \mu(x,1)\) and
\(e(x)\), as the X-learner but uses them in a completely different
manner. It combines the nuisance functions \ul{as well as} the outcome and treatment
data in a doubly robust way to construct the pseudo-outcome \(\psi_i\),
i.e.~the doubly robust score. The score is then regressed on the
covariates to estimate the final CATE function. Therefore, the
DR-learner can also adapt to structural properties of the CATE such as
smoothness or sparsity. For this reason the DR-learner is expected to
work well in similar situations as the X-learner with a more balanced
treatment assignment, as too extreme propensity scores might possibly
yield the estimator unstable \parencites{Huber2013}{Powers2018}, especially in
high dimensions \parencite{DAmour2021}. Moreover, it
should have an additional advantage over the X-learner thanks to its
double robustness property. The theoretical analysis of \textcite{Kennedy2020} uses the
double sample-splitting procedure in order to derive a sharp error bound that
rests only on stability conditions for the estimation of the CATE function.
The theoretical results then exploit the rate double robustness which allows
for faster error rates for the second-stage CATE estimation under weaker rate
conditions for the first-stage estimation of the nuisance functions.
The simulations of \textcite{Kennedy2020} also suggest a faster convergence
rate of the DR-learner in comparison to the X- and T-learner.
In order to achieve the optimal rates\footnote{\textcite{Kennedy2020} shows that
the DR-learner achieves the minimax optimal rate under smoothness or sparsity
conditions for the nuisance and the CATE functions.} the DR-learner
explicitly requires the double sample-splitting as defined by
\textcite{Newey2018}, while the double cross-fitting procedure remains
optional. Theoretically it is not clear how important the role of the
optional cross-fitting is for the DR-learner in finite samples and how much
of the efficiency loss due to sample-splitting can be thereby regained.
In order to shed light on this issue we investigate the
implementations of the DR-learner with double sample-splitting, double
cross-fitting, as well as a version with full-sample estimation.

\subsubsection{R-learner}

Yet another approach of first estimating nuisance functions and then
using them to learn the treatment effects stems from the literature on
\ul{partially linear model} originally developed by \textcite{Robinson1988}.
\textcite{Nie2021a} build on these ideas to flexibly estimate heterogeneous
treatment effects and develop the R-learner, where the
\textit{R} stands for the \ul{recognition of the contribution of} \textcite{Robinson1988}
as well as for the \textit{Residualization} \ul{approach}. In the first step,
the R-learner estimates the response function, \(\mu(x)\),
similarly to the S-learner but without conditioning on the treatment
indicator, as well as the propensity score function \(e(x)\). It then
residualizes the outcome and the treatment by the predictions of the
response and the propensity score function, respectively, to construct a
modified outcome. In the second step, the R-learner regresses the
modified outcome on the covariates, weighted by the squared residualized
treatment\footnote{An estimation procedure without the weighting step
  is in literature referred to as the U-learner
  \parencites{Stadie2018}{Kunzel2019}{Nie2021a}. However, such estimator
  turned out to be quite unstable in the simulation experiments in
  \textcite{Nie2021a} as well as in those of \textcite{Kunzel2019} and will
  thus not be considered further in our analysis.}, i.e.
\(\big(W_i-\hat{e}(X_i)\big)^2\), to estimate the CATE function
\parencite{Schuler2018}. Analogous transformation of the outcome is also
used by the Causal Forest of \textcite{Athey2019} termed local
centering, or in the \textit{G}-estimation for sequential trials by
\textcite{Robins2004}. The full estimation procedure of the R-learner can
be summarized as follows:\\
~\\
\begin{algorithm}[H]
\SetAlgoLined
\KwIn{Training Data: $\{(X_i,Y_i,W_i)\}^T$, Validation Data: $\{(X_i)\}^V$}
\KwOut{CATE: $\hat{\tau}_{R}(x)=\hat{E}[Y_i(1)-Y_i(0) \mid X_i=x]$}
 \Begin{
 \textsc{Response Function}\;
  estimate: $\mu(x)=E[Y_i \mid X_i=x]$ in $\{(X_i,Y_i)\}^T$\;
  %predict: $\hat{\mu}(x,w)=\hat{E}[Y_i \mid X_i=x, W_i=w]$ for $\{(X_i,W_i)\}^T$\;
  \vspace{0.2cm}
  \textsc{Propensity Score}\;
  estimate: $e(x)=P[W_i=1 \mid X_i=x]$ in $\{(X_i,W_i)\}^T$\;
  %predict: $\hat{e}(x)=\hat{E}[W_i \mid X_i=x]$ for $\{(X_i)\}^T$\;
  \vspace{0.2cm}
  \textsc{Modified Outcome}\;
  predict: $\hat{\phi}_i=\frac{\big(Y_i-\hat{\mu}(X_i)\big)}{\big(W_i-\hat{e}(X_i)\big)}$ in $\{(X_i,Y_i,W_i)\}^T$\;
  \vspace{0.2cm}
  \textsc{CATE Function}\;
  estimate: $\tau_{R}(x) = E[\hat{\phi}_i \mid X_i=x]$ weighted by $\big(W_i-\hat{e}(X_i)\big)^2$ in $\{(X_i,Y_i,W_i)\}^T$\;
  %\vspace{0.2cm}
  predict: $\hat{\tau}_{R}(X_i) =  \hat{E}[\hat{\phi}_i \mid X_i=x]$ in $\{(X_i)\}^V$
  %predict: $\hat{\tau}_{R} = \frac{1}{N^V}\sum^{N^V}_{i=1}\hat{\tau}_{R}(X_i)$ in $\{(X_i)\}^V$
  }
\caption{\textsc{R-learner}}
\label{Rlearner}
\end{algorithm}
~\\
As follows from Algorithm \ref{Rlearner}, the R-learner separates the
estimation into two steps. First, it eliminates the spurious
correlations between the response function \(\mu(x)\) and the propensity
score function \(e(x)\) and second, it optimizes the CATE function
\(\tau_R(x)\). From this standpoint the R-learner follows a related
estimation scheme as the DR-learner and is expected to work well in
similar settings where the nuisance functions as well as the CATE
function might have \ul{a high degree of} complexity. A possible advantage of the
R-learner over the DR-learner might stem from the additional weighting
which reduces the impact of extreme propensity scores as pointed out by
\textcite{Jacob2021}. In their simulation experiments, \textcite{Nie2021a} show
good performance of the R-learner in settings with complicated nuisance
functions and rather simple CATE function. The theoretical analysis of
\textcite{Nie2021a} shows that the convergence rate of the R-learner
depends on the complexity of the CATE function and is not affected by the
complexity of the nuisance functions as long as they are estimated at sufficiently
fast rates.\footnote{For the case of estimating the CATE function via penalized kernel
regression, \textcite{Nie2021a} prove that the R-learner achieves the minimax optimal
rate and additionally show that the X-learner does not achieve the optimal rate in
general unless the treatment assignment is not highly unbalanced.} Furthermore,
for these theoretical results \textcite{Nie2021a} explicitly require sample-splitting
and cross-fitting, respectively. In particular, they advocate for a 5-
or 10-fold cross-fitting procedure as defined by
\textcite{Chernozhukov2018c}. In order to examine the importance of the
cross-fitting in finite samples we compare the performance of the
R-learner as in the above cases with full-sample estimation, double
sample-splitting and double cross-fitting, respectively.

\section{Simulation Study}\label{c1sec:sim}

We study the finite sample performance of meta-learners for estimation
of heterogeneous treatment effects based on Random Forests
\parencites()(){Breiman2001}[see also][for a comprehensive introduction]{Biau2016}.
The focus of the Monte Carlo study lies in an
assessment of the influence of sample-splitting and cross-fitting in the
causal effect estimation. For this purpose we compare the above
discussed meta-learners estimated with full-sample, double
sample-splitting, and double cross-fitting.

We rely on the Random Forest
as the base learner for all meta-learners for several reasons. First,
different meta-learners require estimation of different nuisance
functions which involve different types of outcome variables. As such,
the response functions mostly involve a continuous outcome variable
whereas the propensity score function includes a binary outcome. Hence,
when using Random Forests no additional adjustments need to be done in
terms of estimation as it automatically estimates probabilities in case
of binary outcome and expected values in case of continuous outcomes,
respectively. This is in contrast to linear learners such as the Lasso
\parencite{Tibshirani1996}, Ridge \parencite{Hoerl1970} or Elastic Net
\parencite{Zou2005} where the estimator needs to be modified using
appropriate link function for proper probability estimation \parencite[see
for example][for the Logit-Lasso]{Hastie2009}. Second, Random Forest is
a local nonparametric method which does not need any data pre-processing
to flexibly learn the underlying functional form from the data
\parencite{Hastie2009}. Thus, Random Forest is able to approximate any
function with different degrees of complexity which is often the case in
treatment effect estimation where the nuisance functions tend to be
rather difficult complex functions while the CATE function itself is
often \ul{argued to be} simple and sparse \parencites{Kunzel2019}{Kennedy2020}{Sekhon2021}.
This is again an advantage in comparison to the linear learners mentioned above which
become more flexible once an augmented covariate set consisting of
polynomials and interactions is created and thus can be regarded as
global nonparametric methods \parencite{Hastie2009}. Third, in contrast to
other flexible state-of-the-art machine learners such as Neural Networks
the theoretical properties of Random Forests are better understood which
makes it less of a black-box method \parencite[see][ for a discussion of statistical
properties of Random Forests]{Meinshausen2006, Biau2012, Wager2014, Wager2014a,
Scornet2015, Mentch2016, Wager2018, Athey2019}. In particular, \textcite{Scornet2015}
prove the consistency of the original Random Forest algorithm as developed by \textcite{Breiman2001} that we employ in our simulations, without relying on the
`honesty' condition.\footnote{The consistency (in MSE) result holds under the condition that
the number of leaves is smaller than the number of observations. In our simulations,
we ensure this by growing trees with minimum leaf size bigger than 1 (see Table \ref{simtable}).
We refrain from the honesty feature to avoid additional sample-splitting
that would further reduce the
effective sample size.} \ul{This is important as the consistency of the base learner
is a fundamental condition shared across all considered meta-learners.}
\textcite{Scornet2015} \ul{also show that Random Forests can effectively
adapt to sparsity of the underlying model, which we explicitly make use of within
our simulation design.} Additionally, another reason why we do not use the Lasso
and linear learners as such is due to a substantial increase in variance as
they are prone to outliers as documented in the simulation studies of
\textcite{Jacob2020} as well as \textcite{Knaus2021}. Furthermore,
Random Forests are a popular choice as a base-learner in empirical
studies using meta-learners too \parencites{Duan2019}{Knaus2020b}.
Lastly, from the
practical standpoint there is a vast variety of fast and reliable
software implementations of Random Forests which makes it easy to use
for practitioners.\footnote{In our simulations we use the \textsf{R}-package
  \textsf{ranger} which provides a fast C\plusplus implementation of Random
  Forests, particularly suited for high-dimensional data
  \parencite{Wright2017}. Further options in the \textsf{R} language
  \parencite{rstats} include the \textsf{grf} package
  written by \textcite{grf}, the \textsf{forestry} package by
  \textcite{forestry} or the \textsf{randomForest} package by
  \textcite{randomForest}.}

In order to objectively evaluate the performance and the robustness of
different meta-learners in estimating heterogeneous
treatment effects with regard to the double sample-splitting and double cross-fitting,
we design several simulation scenarios. On the one hand, for each
meta-learner we construct such a data generating process (DGP) that
\ul{suits the particular advantages} of the \ul{respective}
meta-learner, i.e.~we design a
simulation scenario where each meta-learner is expected to work best.
Hence, we are able to check if the particular meta-learner outperforms
the others and how big the performance discrepancies are for the other
meta-learners in comparison to the expected best performing
meta-learner. On the other hand, we design \ul{a simulation scenario with a complex DGP}
where none of the meta-learners has \textit{a priori} an explicit \ul{advantage},
\ul{which serves as our main simulation design of interest}.
Thus we can compare the performance of the meta-learners in an objective
manner and quantify the deviations to their respective best performance
cases. Furthermore, common to all DGPs is the observational study
design, i.e.~there is always selection into treatment and thus all
considered meta-learners have to deal with confounding and not only with
modelling the treatment effect itself. Moreover, in contrast to many
simulation studies where the nuisances are simple low-dimensional
functions \parencites{Wager2018}{Kunzel2019}{Kennedy2020}, we model
all nuisance functions as highly non-linear but sparse functions with
large-dimensional covariate space to test the potential of the
machine learning methods, though still largely obeying the theory
induced limitations. For other \ul{complex} simulation designs see also
\textcite{Jacob2020} or \textcite{Zivich2021} as well as \textcite{Lechner2019}
and \textcite{Knaus2021} for the Empirical Monte Carlo Studies.
\ul{Importantly,} in order to study the \ul{approximate convergence rates} of the
meta-learners we repeat each simulation scenario several times with
increasing training sample sizes using \(N^{T}=\{ 500, 2'000, 8'000, 32'000 \}\).
\ul{We emphasize that the considered sample sizes substantially exceed the 
ones from previous simulation studies devoted to the analysis of sample-splitting methods,
which were limited to} $2'000$ \parencite{Jacob2020} and $3'000$ \parencite{Zivich2021}
\ul{observations, respectively. Furthermore, the large samples enable us to study
the performance of the meta-learners in settings in which the application
of machine learning methods is arguably more relevant.} We choose to always
quadruple the sample size, which allows us to easily benchmark
the results with the parametric \(\sqrt{N}\) rate, \ul{in which case the
estimation error is expected to halve with each increase of the sample size.}
We then evaluate the performance measures on a validation set with sample size of
\(N^V=10'000\) to reduce the prediction noise as is usual in many
Monte Carlo studies \parencites{Janitza2016}{Hornung2017}{Okasa2019}{Jacob2020}{Knaus2021}.
Lastly, in terms of the tuning parameters for the Random Forest
base-learner we stick to the default, in the literature commonly used
settings, corresponding to \(1'000\) trees, the number of randomly
chosen split variables set to the square root of number of features,
and the minimum leaf size equal to \(5\).\footnote{We
  refrain from cross-validation or other tuning
  parameter optimization procedures due to computational constraints. We
  recommend such optimization in the applied work as it might
  considerably improve the performance of the estimator \parencite[see][for an evidence based on Neural Networks]{Curth2021}, however, for the
  purposes of the simulation study it would not change the relative
  ranking of the meta-learners as each of them uses the very same base
  learner.} Finally, for
each DGP we simulate the training data \(R=\{2'000, 1'000, 500, 250\}\)
times in total, where we use \(2'000\) replications for the smallest
sample size and decrease the number of replications down to \(250\) for
the largest sample size, due to computational reasons.\footnote{Notice, however,
  that we only halve the number of replications while quadrupling the
  sample size and as such we may \ul{limit} a possible deterioration of the
  performance in terms of the simulation error. A similar strategy
  for balancing the precision and the computational burden has been used
  in the simulations by \textcite{Lechner2019}, \textcite{Lu2018} or \textcite{Knaus2021}.
  Detailed results on the simulation error are provided in Appendix \ref{appendix-e-additional-results}.}

\subsection{Performance Measures}\label{c1sec:performance-measures}

For the evaluation of the performance of the considered meta-learners
with regard to the sample-splitting and cross-fitting in detail, we
employ several evaluation measures. First, to assess the overall
estimator performance we compute the root mean squared error
for each observation \(i\) from the validation sample over
the \(R\) simulation replications:\footnote{We
  take the square root of the MSE to have the same \ul{scale} as for the other
  performance measures, i.e.~the absolute bias and the standard
  deviation.}
\[RMSE \big(\hat{\tau}(X_i)\big)= \sqrt{\frac{1}{R} \sum^R_{r=1} \big( \tau(X_i) - \hat{\tau}^r(X_i) \big)^2}.\]
Next, we decompose the root mean squared error and evaluate the
bias and variance component separately to contrast the theoretically
expected asymptotic behaviour of sample-splitting and cross-fitting with
the finite sample properties. Hence, we additionally compute the mean
absolute bias:
\[\mid BIAS \big(\hat{\tau}(X_i)\big)\mid= \frac{1}{R} \sum^R_{r=1} \mid \tau(X_i) - \hat{\tau}^r(X_i) \mid\]
as well as the standard deviation of the treatment effects:
\[SD \big(\hat{\tau}(X_i)\big)= \sqrt{\frac{1}{R} \sum^R_{r=1} \bigg( \hat{\tau}^r(X_i) - \frac{1}{R} \sum^R_{r=1} \hat{\tau}^r(X_i) \bigg)^2}.\]
Furthermore, inspired by the simulation study of \textcite{Knaus2021} we
also compute the Jarque-Bera statistic \parencites{Jarque1980}{Bera1981} to test for the \ul{normality}
of the treatment effect predictions:\footnote{See \textcite{Thadewald2007} \ul{for
a discussion of the Jarque-Bera test and its comparison to other tests for normality.}}
\[JB \big(\hat{\tau}(X_i)\big)= \frac{R}{6}\bigg( S\big(\hat{\tau}(X_i)\big)^2 + \frac{1}{4} (K\big(\hat{\tau}(X_i)\big)-3)^2 \bigg)\]
where \(S\big(\hat{\tau}(X_i)\big)\) and \(K\big(\hat{\tau}(X_i)\big)\) is the skewness and the kurtosis of the \(R\) treatment effect predictions \ul{for observation} \(i\), respectively.
As a matter of presentation for CATEs,
we report the mean values of the RMSE, absolute bias, standard deviation
and the Jarque-Bera statistic over \(N^V\) validation observations.\footnote{\ul{As such,
we define the average RMSE as} \(\overline{RMSE}=\frac{1}{N^V}\sum^{N^V}_{i=1} RMSE \big(\hat{\tau}(X_i)\big)\) \ul{and analogously for the remaining performance measures. Additionally, for the Jarque-Bera statistic we report also the share of CATEs from the
validation sample for which the normality gets rejected at the 5\% confidence level.
For details, see Appendix} \ref{appendix-e-additional-results}.}
\ul{Additionally, we provide evaluation of further performance measures in Appendix} \ref{appendix-e-additional-results}.

\subsection{Simulation Design}\label{c1sec:simulation-design}

In the general simulation design we follow \textcite{Kunzel2019} and
specify the response functions \ul{for potential outcomes} under treatment, \(\mu_1(x)\), and
control, \(\mu_0(x)\), the propensity score, \(e(x)\), and the treatment
effect function, \(\tau(x)\), respectively. First, we simulate a
\(p\)-dimensional matrix of covariates \(X_i \in \mathbb{R}^p\) drawing
from the uniform distribution, as previously used in simulations of
\textcite{Wager2018}, \textcite{Kunzel2019} or \textcite{Nie2021a} among others,
such that:
\[X_i \sim \mathcal{U}\big([0,1]^{n\times p}\big)\] and defining the
correlation structure according to \textcite{Falk1999} using a random
correlation matrix \(\Sigma_p\) generated by the method of
\textcite{Joe2006}.\footnote{For a detailed correlation heat map of the
  covariates and descriptive statistics of the simulated
  datasets see Appendix \ref{c1app:desc}.} Second, we specify the response
functions and simulate the potential outcomes as:
\[Y_i(0)=\mu_0(X_i)+\epsilon_i(0)\] \[Y_i(1)=\mu_1(X_i)+\epsilon_i(1)\]
with errors \(\epsilon_i(0), \epsilon_i(1) \stackrel{iid}{\sim} \mathcal{N}(0,1)\)
that are independent of the covariates \(X_i\). Third, we define the
propensity score function and simulate the treatment assignment
according to:
\[W_i \sim Bern\big(e(X_i)\big)\] and use the observational rule to set
the observed outcomes such that:
\[Y_i = W_i \cdot Y_i(1) + (1-W_i) \cdot Y_i(0)\] to complete the
observable triple \(\{(X_i,Y_i,W_i)\}\). The subsequent simulation
designs then differ only with respect to how the corresponding
functions, namely \(\mu_0(x), \mu_1(x), e(x)\) and \(\tau(x)\) are
specified. For all of our simulations we define the response function
under non-treatment according to the well-known Friedman function
(\cite*{Friedman1991}) to create a difficult yet
standardized setting, which has been used also in the simulations of
\textcite{Biau2012} and \textcite{Nie2021a}, as follows:
\begin{equation}\label{eq:friedman}
\mu_0(x)= sin\big(\pi \cdot x_1 \cdot  x_2\big) + 2 \cdot \big(x_3-\frac{1}{2}\big)^2 + x_4 + \frac{1}{2} \cdot x_5
\end{equation}
hence effectively creating a highly non-linear but sparse response
function which is difficult to estimate on its own.\footnote{Note that $\pi$ refers
to the mathematical constant, i.e. $\pi \approx 3.14$.} The response
function under treatment is then defined simply as:
\[\mu_1(x)= \mu_0(x)+\tau(x)\] while we vary the specification of the
treatment effect function \(\tau(x)\) throughout our simulation designs.
Lastly, we model the propensity score function similarly to
\textcite{Wager2018} and \textcite{Kunzel2019} using the \(\beta\)
distribution with parameters \(2\) and \(4\) such that:
\begin{equation}\label{eq:ps}
e(x)=\alpha\bigg(1+\beta_{2,4}\big(f(x)\big)\bigg)
\end{equation}
while the scale parameter \(\alpha\) controls the share of treated in
the sample and at the same time helps to bound the resulting
probabilities away from \(0\) and \(1\) and thus to avoid extreme
propensity scores which might yield some meta-learners using such
propensities for reweighting unstable \parencites{Huber2013}{Powers2018}. We
additionally make the propensity score dependent on features \(X_i\) of
dimension \(p^e\) in a non-linear fashion using the functional form
specification of \textcite{Nie2021a} and set:
\[f(x)=sin(\pi\cdot x_1 \cdot x_2 \cdot x_3 \cdot x_4)\] which creates a
non-linear setting that is hard to model as opposed to,
e.g.~polynomial transformations alone. Similarly, such non-linear
transformations for the propensity scores using the sine function have
been used also in simulations by \textcite{Lechner2019} and
\textcite{Knaus2021}.

\begin{table}[ht]
\centering
\caption{\label{simtable}Overview of the Simulation Study} 
\begin{tabular}{lr}
  \toprule
  \multicolumn{2}{c}{General Settings} \\
  \midrule
 Number of DGPs & 6\\
 Number of Replications $R$ &  $\{2'000, 1'000, 500, 250\}$ \\
 Training Sample $N^T$ & $\{500, 2'000, 8'000, 32'000\}$\\
 Validation Sample $N^V$ & $10'000$ \\
 \midrule
 \multicolumn{2}{c}{DGP Settings} \\
  \midrule
 Covariate Space Dimension $p$ & $100$ \\
 Signal Covariates in Response Function $p^{\mu}$ & $5$ \\
 Signal Covariates in Propensity Score Function $p^{e}$ & $4$ \\
 Signal Covariates in Treatment Function $p^{\tau}$ & $\{0,1,2,3\}$ \\
 \midrule
  \multicolumn{2}{c}{Forest Settings} \\
  \midrule
 Number of Trees & $1'000$ \\
 Random Subset of Split Covariates & $\sqrt{p}$ \\
 Minimum Leaf Size & $5$ \\
 \bottomrule
\end{tabular}
\end{table}

As a matter of notation we refer to \(p\) as the dimension of the
covariate space, \(p^{\mu}\), \(p^{e}\) and \(p^{\tau}\) as the
dimension of the signal covariates in the response function, the
propensity score function, and the CATE function, respectively. We set
the aforementioned dimensions as follows: \(p=100\), \(p^{\mu}=5\),
\(p^e=4\) and \(p^{\tau}\) varies with forthcoming simulation designs.
We note that such large-dimensional covariate set is quite unique as the
majority of simulation studies relies on low-dimensional covariate sets
\parencites(see e.g.)(){Kunzel2019}{Jacob2020}[or][]{Nie2021a}.\footnote{An exception is the simulation study of \textcite{Powers2018} \ul{who explicitly study
the estimation of heterogeneous treatment effects in high-dimensions.}} We
further define the sets of covariates such that
\(X^{p^{\tau}} \subset X^{p^{e}} \subset X^{p^{\mu}} \subset X^{p}\). By
doing so we make it difficult for the meta-learners to accurately fit
the functions and eliminate the spurious correlations between the
response and propensity score functions. Moreover, it also becomes
a non-trivial task to disentangle the confounding effects from the actual
treatment effect heterogeneity which the herein discussed meta-learners
are specifically designed for. Finally, a general overview of the
simulation study is provided in Table \ref{simtable}.

\subsubsection{Simulation 1: balanced treatment and constant zero CATE}\label{simulation-1-balanced-treatment-and-zero-constant-cate}

The first simulation design features our complicated sparse non-linear
nuisance functions as defined above in Equations \eqref{eq:friedman} and
\eqref{eq:ps} in contrast to a very simple CATE function. In fact, the
treatment effect here is defined as being constant and equal to zero:
\[\tau(x)=0\] with a balanced treatment assignment with the scaling
factor \(\alpha=\frac{1}{4}\) which results in approximately 50\%
treated and 50\% of control units. Such DGP with zero CATE serves as a
benchmark and should implicitly suit the S-learner as the treatment
indicator is not predictive for the outcome. Nevertheless the other
meta-learners with the exception of the T-learner should be also capable
of capturing the true zero effect as this is often a showcase example
when motivating the particular meta-learners as well as simulating their
performance
\parencites(see)(){Kunzel2019}{Kennedy2020}[and][for details]{Nie2021a}.

\subsubsection{Simulation 2: balanced treatment and complex nonlinear CATE}\label{simulation-2-balanced-treatment-and-complex-nonlinear-cate}

In the second simulation design we keep the balanced treatment
allocation but feature a highly complex and non-linear CATE function
resulting from a completely disjoint DGPs for the response function
under treatment and under control. As such the response function under
control is defined according to Equation \eqref{eq:friedman}, while the
response function under treatment is defined as a non-zero constant,
i.e. \(\mu_1(x) = 1\). The CATE is then defined as:
\[\tau(x)=\mu_1(x)-\mu_0(x).\] Such simulation setups have been
previously used also in \textcite{Kunzel2019} or in \textcite{Nie2021a}. In
this case the response functions, \(\mu_0(x)\) and \(\mu_1(x)\), are
uncorrelated and thus there is no advantage in pooling those two
together.
%\footnote{\ul{Notice, that in this special case, the CATE is
%indirectly a function of the covariates entering $\mu_0(x)$.}}
Rather, estimating these two functions separately is the best
strategy as there is nothing additional to learn from the other
treatment group. For this reason, the T-learner should perform best
here, however the meta-learners which also estimate the
response functions separately such as the X- and DR-learner are expected to
perform well too. Clearly, other meta-learners such as the S- and
R-learner which estimate the pooled response function have a
disadvantage as they first need to learn the disjoint structure.

\subsubsection{Simulation 3: highly unbalanced treatment and constant non-zero CATE}\label{simulation-3-highly-unbalanced-treatment-and-constant-non-zero-cate}

In our third simulation design we change the scaling factor in the
propensity score function to \(\alpha=\frac{1}{12}\) such that we
generate approximately 15\% treated units.\footnote{In contrast to
  \textcite{Kunzel2019} we do not specify the treatment imbalance as
  extreme as 1\% treated mostly for computational reasons. Due to our
  smallest sample size of \(N=500\) used in the simulations and the
  double sample-splitting procedure, it might occasionally happen that
  the estimated propensity scores would be exactly zero which would
  prevent estimation of the DR-learner as well as the R-learner due to
  the division by zero when constructing the pseudo-outcomes. In our
  specification, even with the share of the treated being \(16.77\%\) in
  expectation, the aforementioned issue with zero propensity scores
  still might occur. In such cases, we redraw the sample to ensure at
  least \(15\%\) of treated. However, this occurs only a handful of
  times out of 2000 draws in total and only for the smallest sample size
  considered. \textcite{Nie2021a} also use similar restrictions on the propensity
  scores in their simulations due to the very same issue.}
  We then model the treatment effect as a constant as for
example in \textcite{Kennedy2020} or \textcite{Nie2021a} and thus create a
scenario with highly complicated nuisance functions and very simple CATE
function given as:
\[\tau(x) = 1.\]
Accordingly, the X-learner should perform best in this scenario given
the high imbalance in the treatment assignment and the sparse CATE
function at the same time. In contrast, other meta-learners using the
propensity score for reweighting such as the DR- and R-learner might
perform worse due to potentially extreme propensity scores close to the
\(\{0,1\}\) bounds. Furthermore, the T-learner is clearly disadvantaged
in this scenario due to the high treatment imbalance as well as due to
the simple CATE function, whereas the S-learner is not expected to
work particularly well either due to the relatively bigger effect size
bounded away from zero.

\subsubsection{Simulation 4: unbalanced treatment and simple CATE}\label{simulation-4-unbalanced-treatment-and-simple-cate}

In our fourth simulation design we model the CATE function similarly to
the above design as a simple non-zero constant and combine it with an
indicator function as also used by \textcite{Kunzel2019} to add more
structure to the CATE. As such we define the treatment effect as:
\[\tau(x) = 1 + 1 \cdot\mathds{1}(x_1 > 0.5)\]
and otherwise keep the DGP same as in the third design while only
increasing the share of treated to roughly 25\% as is the case in the
simulations of \textcite{Nie2021a} by setting \(\alpha=\frac{1}{8}\). By
doing so we should theoretically shift the advantage from the X-learner
more onto the DR-learner as both meta-learners are motivated by complex
nuisance functions and a simple CATE function with the difference of the
X-learner being designed particularly for highly unbalanced treatment
allocation. Also the R-learner is expected to perform relatively well in
this scenario due to the less unbalanced treatment shares, whereas the
S- and T-learner are not expected to perform well for the same
reasons as in the above situations.

\subsubsection{Simulation 5: unbalanced treatment and linear CATE}\label{simulation-5-unbalanced-treatment-and-linear-cate}

The fifth simulation design features the same nuisance functions and
treatment share as the fourth design, however, here instead of the
indicator function we model the treatment effect as a low-dimensional
linear function as:
\[\tau(x)=1+\frac{1}{2}x_1+\frac{1}{2}x_2\]
as used in one of the simulation designs of \textcite{Nie2021a} where the
R-learner performed best and as such it is also expected to have an
advantage here. Yet again the X- and DR-learner should perform
comparatively well in this setting while the S- and T-learner not so
much for the very same reasons as stated above.

\subsubsection{Main Simulation: unbalanced treatment and nonlinear CATE}\label{simulation-6-unbalanced-treatment-and-nonlinear-cate}

In the last simulation design we create the most complex
scenario in which none of the meta-learners has an \textit{a priori}
advantage and thus presents our main simulation design of interest.
In this case not only the nuisances but also the CATE itself
is modelled as a smooth non-linear function of a slightly larger
dimension than in the previous settings, i.e. \(p^{\tau}=3\). Following
\textcite{Wager2018} we specify the CATE function as follows:
\[\tau(x)=1+\frac{4}{p^{\tau}}\sum^{p^{\tau}}_{j=1}\bigg(\frac{1}{1+e^{-12(x_j-0.5)}}-\frac{1}{2}\bigg).\]
We further keep the treatment share equal to about 25\% and the nuisance
functions as previously specified as well. Hence, the meta-learners need
to first account for the moderately unbalanced treatment shares, second
accurately estimate the complex nuisance functions and disentangle their
correlation, and third separate the treatment effect heterogeneity from
the selection effects by precisely estimating the non-linear CATE
function.

\subsection{Simulation Results}\label{c1sec:simresults}

For the analysis of the simulation results we focus on the Main Simulation design
as this is the most complex simulation design which does not
\textit{a priori} create conditions that would be advantageous for any
of the considered meta-learners. \ul{Additionally, we argue that intuitively
the performance in the most complex setting might be generalizable for simpler
settings, too.} We then summarize the results for the
rest of the simulation designs for which we provide the detailed results
in Appendix \ref{c1appendix:appendix-b-simulation-results}.
Supplementary results providing additional
measures, including the simulation error, bias, skewness, kurtosis, \ul{share of CATEs for which
the normality has been rejected,} as well as the correlation and variance ratio of the estimated and the true CATEs are presented in Appendix \ref{appendix-e-additional-results}.

\subsubsection{Results of Main Simulation: unbalanced treatment and nonlinear CATE}

\begin{figure}[ht]
\caption{\label{bigfigure:CATE6}CATE Results for Main Simulation}\label{fig:cate_results_app}
{\centering \includegraphics[scale=0.8]{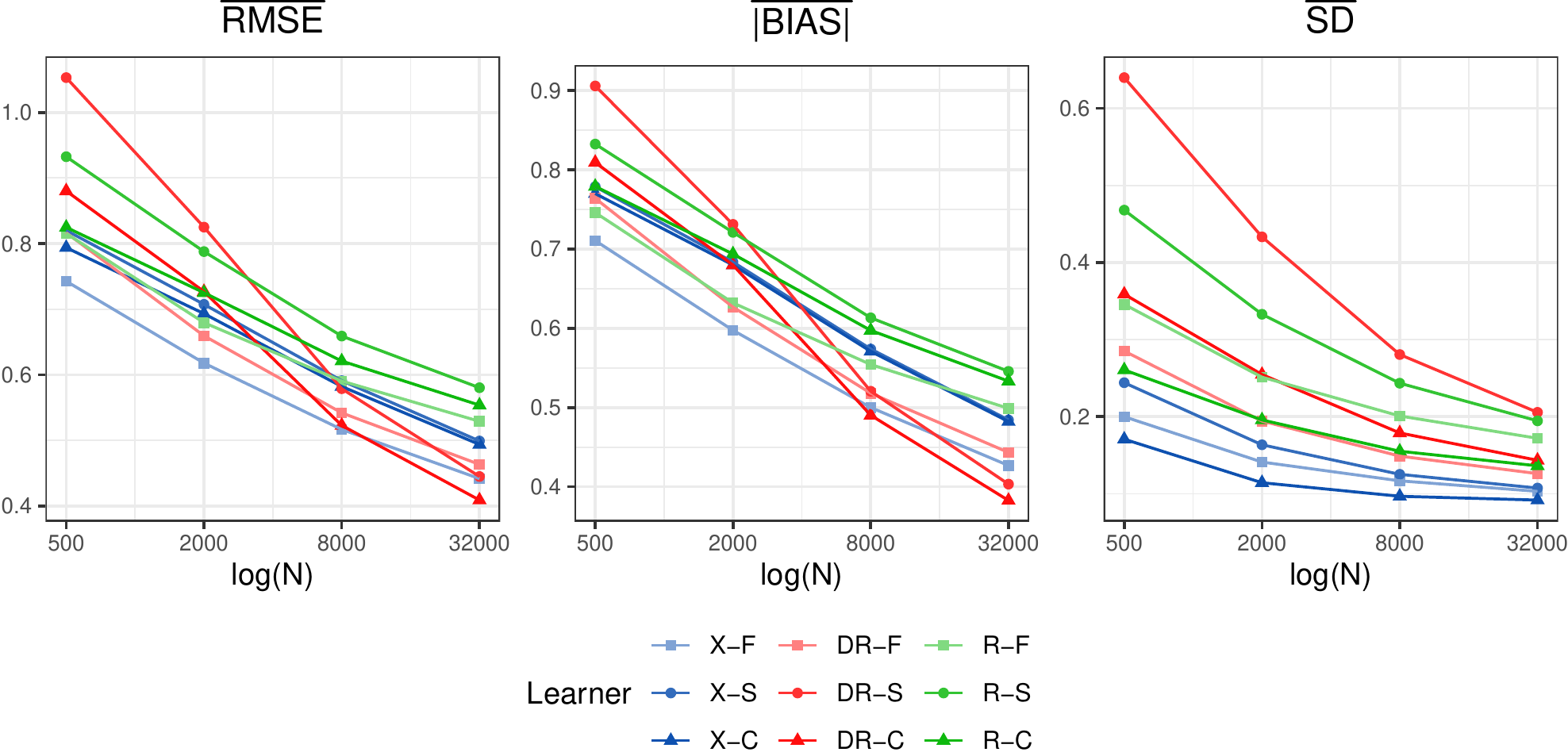} 

}
\vspace{0.25cm}
\caption*{\footnotesize{\textit{Note:} The results for $\overline{RMSE}$, $\overline{|BIAS|}$, and $\overline{SD}$ show the mean values of the root mean squared error, absolute bias, and standard deviation of all $10'000$ CATE estimates from the validation sample. The figure shows the results based on the increasing training samples of $\{500, 2'000, 8'000, 32'000\}$ observations displayed on the log scale. Additionally, X-F, DR-F, R-F denote the full-sample versions of the meta-learners, while X-S, DR-S, R-S and X-C, DR-C, R-C denote the sample-splitting and cross-fitting versions, respectively.}}
\end{figure}

The performance of the meta-learners in the Main Simulation is depicted in Table
\ref{bigtable:CATE6}. We report the results for the average values of
the RMSE, absolute bias, standard deviation and the Jarque-Bera test
statistic over the \(N^V=10'000\) predicted CATEs from the validation
sample. Figure \ref{fig:cate_results_app} details the performance of the
meta-learners \ul{implemented} in the full-sample, double sample-splitting and
double cross-fitting versions.

\begin{table}[ht]
\centering
\caption{\label{bigtable:CATE6}CATE Results for Main Simulation} 
\scalebox{0.83}{
%\resizebox{\textwidth}{!}{
\small
\setlength{\tabcolsep}{2.25pt}
\begin{tabular}{lrrrrrrrrrrrrrrrrrrr}
  \toprule
  & \multicolumn{4}{c}{$\overline{RMSE}$} & \phantom{..} & \multicolumn{4}{c}{$\overline{|BIAS|}$} & \phantom{..} & \multicolumn{4}{c}{$\overline{SD}$} & \phantom{..} & \multicolumn{4}{c}{$\overline{JB}$} \\
                      \cline{2-5} \cline{7-10} \cline{12-15} \cline{17-20}
                      & \textit{500} & \textit{2000} & \textit{8000} & \textit{32000} & & \textit{500} & \textit{2000} & \textit{8000} & \textit{32000} & & \textit{500} & \textit{2000} & \textit{8000} & \textit{32000} & & \textit{500} & \textit{2000} & \textit{8000} & \textit{32000} \\ \midrule
S & 0.878 & 0.749 & 0.651 & 0.570 &  & 0.867 & 0.739 & 0.641 & 0.560 &  & \textbf{0.108} & \textbf{0.096} & \textbf{0.091} & \textbf{0.088} &  & 7.140 & 2.888 & 2.173 & 1.936 \\ 
  S-W & 0.765 & 0.634 & 0.533 & 0.462 &  & 0.717 & 0.602 & 0.508 & 0.443 &  & 0.261 & 0.190 & 0.149 & 0.125 &  & \textbf{2.086} & 2.106 & 2.019 & 1.931 \\ 
  T & 0.766 & 0.634 & 0.533 & 0.462 &  & 0.719 & 0.602 & 0.509 & 0.442 &  & 0.260 & 0.190 & 0.149 & 0.125 &  & 2.603 & \textbf{2.085} & 2.016 & 1.924 \\ 
  X-F & \textbf{0.743} & \textbf{0.618} & \textbf{0.517} & 0.442 &  & \textbf{0.711} & \textbf{0.597} & 0.500 & 0.427 &  & 0.200 & 0.141 & 0.117 & 0.103 &  & 3.490 & 2.230 & 2.034 & 1.857 \\ 
  X-S & 0.820 & 0.707 & 0.591 & 0.499 &  & 0.779 & 0.684 & 0.574 & 0.484 &  & 0.244 & 0.164 & 0.125 & 0.107 &  & 5.146 & 2.680 & 2.157 & 1.929 \\ 
  X-C & 0.794 & 0.693 & 0.582 & 0.494 &  & 0.770 & 0.680 & 0.571 & 0.482 &  & 0.171 & 0.114 & 0.097 & 0.092 &  & 3.984 & 2.322 & \textbf{1.964} & \textbf{1.827} \\ 
  DR-F & 0.817 & 0.659 & 0.542 & 0.463 &  & 0.764 & 0.627 & 0.518 & 0.443 &  & 0.285 & 0.194 & 0.149 & 0.126 &  & 141.106 & 40.528 & 5.490 & 2.172 \\ 
  DR-S & 1.053 & 0.825 & 0.579 & 0.445 &  & 0.906 & 0.731 & 0.521 & 0.403 &  & 0.640 & 0.433 & 0.281 & 0.206 &  & 567.501 & 458.729 & 159.041 & 36.504 \\ 
  DR-C & 0.880 & 0.727 & 0.523 & \textbf{0.409} &  & 0.809 & 0.680 & \textbf{0.490} & \textbf{0.383} &  & 0.359 & 0.255 & 0.179 & 0.143 &  & 52.224 & 38.216 & 12.644 & 3.162 \\ 
  R-F & 0.815 & 0.679 & 0.590 & 0.529 &  & 0.746 & 0.632 & 0.554 & 0.499 &  & 0.346 & 0.251 & 0.201 & 0.172 &  & 4.583 & 3.499 & 2.225 & 1.983 \\ 
  R-S & 0.932 & 0.788 & 0.659 & 0.580 &  & 0.833 & 0.721 & 0.613 & 0.546 &  & 0.468 & 0.333 & 0.243 & 0.195 &  & 3.959 & 3.365 & 2.666 & 2.028 \\ 
  R-C & 0.825 & 0.725 & 0.621 & 0.554 &  & 0.779 & 0.694 & 0.597 & 0.533 &  & 0.261 & 0.196 & 0.155 & 0.136 &  & 2.416 & 2.184 & 2.036 & 1.959 \\ 
   \midrule \multicolumn{20}{c}{\parbox{18.65cm}{\setstretch{0.95}\footnotesize{\textit{Note:} The results for the $\overline{RMSE}$, $\overline{|BIAS|}$, $\overline{SD}$ and $\overline{JB}$ show the mean values of the root mean squared error, absolute bias, standard deviation and the Jarque-Bera test statistic of all $10'000$ CATE estimates from the validation sample. The critical values for the JB test statistic are 5.991 and 9.210 at the 5\% and 1\% level, respectively. Additionally, X-F, DR-F, R-F denote the full-sample versions of the meta-learners, while X-S, DR-S, R-S and X-C, DR-C, R-C denote the sample-splitting and cross-fitting versions, respectively. Bold numbers indicate the best performing meta-learner for given measure and sample size.}}}\\ \bottomrule
\end{tabular}
}
\end{table}
\vspace{0.55cm}

Starting with the most simple S-learner, we observe a competitive
performance in terms of the average RMSE for the smaller sample sizes
which, however, disappears for larger sample sizes. Taking a closer look
at the results reveals that the competitive performance of the S-learner
stems mainly from the very low standard deviation while being
substantially biased. Indeed, the variance of the S-learner is the
smallest among all meta-learners for all sample sizes. This is mainly
due to its tendency to predict effects close to zero if the treatment
indicator is not strongly predictive for the outcome as pointed out by
\textcite{Kunzel2019}. This explains also the high bias of this
estimator as the simulated CATEs vary between \(-1\) and \(3\) with only a small
proportions of these CATEs being equal to zero (see Figure
\ref{fig:simplot_6} in Appendix \ref{c1app:desc} for details).
\ul{Similar pattern with high bias of the S-learner with the Random Forest
as a base learner is documented also in the simulations of} \textcite{Lu2018}.
Nevertheless, the Jarque-Bera test does not indicate evidence against the
normality of the predicted CATE distribution, on average.

Considering the modified version of the S-learner with enforcement of
the treatment indicator into the forest splitting set, i.e. the
SW-learner, we notice almost identical performance to the one of the
T-learner. This result can be explained by an observation that once the
SW-learner finds the split based on the treatment indicator early within
the trees it mimics the disjoint structure of the T-learner. The rest of
the recursive partitioning is then very similar to the one of the
T-learner which has been also documented for the case of the S-learner
in the simulation experiments conducted by \textcite{Kunzel2019}. As a
result, it seems that enforcing the treatment indicator into the
splitting set helps to alleviate the high bias of the S-learner to some
degree, however, it increases the variance of the estimator at the same
time. Nevertheless, the bias-variance trade-off in this case results in
lower average RMSE in comparison to the S-learner and the SW-learner
might thus be preferred over the simple S-learner, when using the Random Forest
as a base learner. Overall, the
SW- and T-learner are very competitive in the smaller
sample sizes both in terms of the average RMSE as well as the average
absolute bias. However, with access to more training data these two
learners do not improve fast enough and are outperformed by the more
sophisticated learners in the largest sample size \ul{consisting of $32'000$
observations}. Concerning the
distribution of the predicted CATEs there seems to be on average no statistical
evidence against the normality, neither for the SW-learner nor for the T-learner.

In contrast to the above \ul{mentioned} meta-learners the X-learner makes use of the
additional estimation of nuisance functions. In its full-sample version
the X-learner performs best in terms of the average RMSE for all sample
sizes, except the largest one. The good RMSE performance stems partly
from the relatively low bias and partly from the relatively low variance
of the estimator as the X-learner exhibits the smallest average absolute
bias for the smaller sample sizes ($500$ and $2'000$), while having one of the lowest
average standard deviations throughout all sample sizes. Interestingly,
we only partly document the theoretical properties regarding the
sample-splitting and cross-fitting procedures. As such, the full-sample
version is the best performing one in terms of the average RMSE as well
as in terms of the average absolute bias across all sample sizes, which is a pattern
observed in the simulation experiments of \textcite{Jacob2020} as well.
Accordingly, the sample-splitting version exhibits not only higher
values of the average standard deviation but also higher values of the
average absolute bias. Nevertheless, we observe that the cross-fitting
version successfully regains the efficiency lost due to sample-splitting
as it exhibits steadily lower variance than the sample-splitting version
and even lower variance than the full-sample version, while having a
bias of roughly the same magnitude as the sample-splitting version. As
for the distribution of the predicted CATEs, on average, we do not observe evidence for
deviations from normality for any of the versions of the X-learner.
Additionally, we do not observe any major differences in the speed of
convergence between the different versions as can be seen in Figure
\ref{fig:cate_results_app}. Moreover, the absolute differences in the
performance measures among the different versions are small in
comparison to other meta-learners using nuisance functions. Albeit
rather surprising at the first sight, the explanation for this
phenomenon comes presumably from the different usage of the propensity
score by the X-learner in comparison to the R- and DR-learner. As such,
the R- and DR-learner use the propensity score \ul{together with the 
response functions} to construct a new
pseudo-outcome which is subsequently used to estimate the CATEs. In
contrast, the X-learner uses \ul{merely the response functions to create
the pseudo-outcome, while} the propensity score \ul{is used only} to reweight the
final CATE estimates and thus it does not enter into any additional
estimation step. Therefore, \ul{the X-learner might be less prone to overfitting bias}
which would \ul{partly} justify the full-sample estimation as described in
\textcite{Kunzel2019}.\footnote{Nonetheless, this insight might still substantiate
the need for sample-splitting, although only with two folds instead of three as
used here.}

Assessing the performance of the DR-learner reveals some interesting
insights. The first observation is that the cross-fitting version performs
best of all meta-learners in terms of the average RMSE for the
largest sample size of $32'000$ observations. This comes mainly from the low bias of this
estimator as the average absolute bias is the lowest among all learners
for the two largest sample sizes, while the average standard deviation
is relatively high. However, looking at \ul{the average value of} the Jarque-Bera statistic
suggests \ul{evidence against the} normality of the predicted CATEs
for all but the largest sample size. Inspecting the results
more closely reveals that the issue stems from heavy tails of the CATE
distributions. The extreme values of the predicted CATEs are mainly
caused by the propensity scores which are close to the \(\{0,1\}\)
bounds. Similar issues of the DR-learner due to extreme
propensity scores have also been documented in the simulation
experiments of \textcite{Knaus2021} as well as in the empirical application of
\textcite{Knaus2020b}. The second observation is that for the DR-learner we
clearly see how the theoretical arguments of sample-splitting and
cross-fitting translate into the finite sample properties of the
estimator. However, these can be documented only for large sample sizes.
As such, the bias of the sample-splitting version is smaller than the
one of the full-sample version in the largest sample size, while the
bias of the cross-fitted version is even slightly lower than the
sample-splitting version and is lower than the bias in the full-sample
version for both the largest \ul{($32'000$)} and the second largest
\ul{($8'000$)} sample considered.
For the smaller sample sizes \ul{($500$ and $2'000$)} we see that the reduction in the
overfitting bias is not large enough in comparison to the bias stemming
from the estimation of the CATE function. As such, for small sample
sizes the additional splitting of the sample does not leave enough
observations to learn the non-linear structure of the CATE. Considering
the variance of the estimator, we also observe the theoretically
expected pattern. The full-sample version of the DR-learner exhibits the
smallest average standard deviation throughout all sample sizes, while
the standard deviation for the sample-splitting version is roughly twice
as high. Nevertheless, the cross-fitting version successfully reduces
the standard deviation for all sample sizes and comes close to the
full-sample version, effectively regaining the lost efficiency of the
estimator due to sample-splitting. Overall, in terms of the average RMSE
this bias-variance trade-off results in favourable performance of the
sample-splitting version in the largest and of the cross-fitting version
in the two largest samples in comparison to the full-sample version.
Considering the distribution of the predicted CATEs we see that the
heavy tails problem due to extreme propensity scores is the worst for
the sample-splitting version, where even in the second largest sample size of $8'000$
observations, the normality is rejected for more than 50\% of the predicted CATEs from the
validation sample (compare the supplementary results in Appendix \ref{appendix-e-additional-results}). This stems from the smaller samples used
for estimation of the propensity scores which are more likely to yield
extreme values under unbalanced treatment assignment. We also observe
that this issue is less pronounced for the full-sample version. The third
and the last observation is yet the probably most noticeable pattern across
all performance measures, namely the fast convergence of the
sample-splitting and cross-fitting version of the DR-learner which is
substantially faster in comparison to all other meta-learners as can be
seen in Figure \ref{fig:cate_results_app}. As such the DR-learner
performs almost worst of all, both in terms of the average RMSE and
average absolute bias for the smallest sample size of $500$ observations,
but almost best of all for the largest sample size of $32'000$ observations.
This provides evidence that the
DR-learner is able to learn a \ul{highly} complex CATE function \ul{once enough
data becomes available} and \ul{additionally} highlights the need for sample-splitting
and cross-fitting in order to achieve the \ul{theoretically described optimal performance} \parencite{Kennedy2020}.

The performance of the R-learner is competitive with the other
meta-learners especially in smaller samples, particularly for the
full-sample version. In the smallest sample size of $500$ observations the R-learner
outperforms the DR-learner in terms of the average RMSE, irrespective of
the estimation procedure. However, with growing sample sizes the
performance evens out and eventually for the largest sample size of $32'000$ observations
the R-learner \ul{lags behind the majority of the estimators}. This is in contrast to previous
results from simulations of \textcite{Jacob2020} and \textcite{Knaus2021} where the R-learner exhibits rather good performance, albeit studied only in smaller samples.
The decomposition of the RMSE
shows that while the full-sample version of the R-learner exhibits
rather low bias, it suffers from a higher variance as can be seen in
Figure \ref{fig:cate_results_app}. Nonetheless, the distributions of the
predicted CATEs do not show \ul{on average} deviations from the normal distribution.
This is contrary to the DR-learner and illustrates the advantage of
the additional weighting step. As such, even though the R-learner uses
the propensity scores for reweighting to construct the modified outcome,
it successfully manages to downweight the modified outcomes based on
extreme propensity scores to alleviate the heavy tails issues observed
in the case of the DR-learner. In particular, even for the sample-splitting version
of the R-learner the share of predicted CATEs for which the normality is rejected is 
an order of magnitude lower in comparison to the DR-learner (see Appendix \ref{appendix-e-additional-results} for details). In terms of the estimation procedure, we
observe a similar pattern as for the X-learner in a sense that the
full-sample version performs better with respect to the average RMSE and
absolute bias, while the cross-fitting version helps to reduce the
variance of the estimator not only in comparison to the sample-splitting
version but even in comparison to the full-sample version. The
sample-splitting version exhibits higher values of the average absolute
bias and standard deviation for all sample sizes considered, while the
convergence rates are approximately same as for the full-sample and the
cross-fitting version. Hence, there is \ul{a lack of} indication that the overfitting
type of bias reduction could become relevant in bigger samples.
Similarly to the DR-learner, also for the R-learner the
differences between the different estimation procedures seem to be
higher than those for the X-learner which is again presumably due to the
different usage of the propensity scores.

Inspecting the results for the rest of the simulation designs reveals
further insights and helps to generalize the findings from the main and most
complex simulation design discussed \ul{above}. 

\subsubsection{Results of Simulation 1: balanced treatment and constant zero CATE}

Within the benchmark
Simulation 1 with zero constant CATE the S-learner, as expected, performs
best with respect to all performance measures across all sample sizes
(see Table \ref{bigtable:CATE1} in Appendix \ref{c1appendix:appendix-b-simulation-results}).
However, the results reveal poor
statistical properties of the S-learner as it appears to be
substantially biased and inconsistent as the absolute bias as well as
standard deviation increase with growing sample size.\footnote{A closer
  look on the estimation results reveals the reason for this phenomenon.
  With small sample sizes, the underlying trees of the S-learner's
  forest are quite shallow and barely split on the treatment indicator
  resulting in quite homogeneous CATE predictions which are very close
  to the actual zero effect. However, as the sample size increases, the
  chance of splits based on the treatment indicator increases which
  results in more heterogeneous effect predictions spread around zero.
  Accordingly, the bias as well as the standard deviation
  increase. Similar consistency issues of the forest-based S-learner seem
  to appear also in the simulations of \textcite{Kunzel2019} where the MSE
  rises with growing sample size for some designs and only stabilizes
  with very big sample sizes.} In general, the performance of the
S-learner is, in all simulation designs, plagued by the substantially
higher bias than all the other meta-learners, partially accompanied by
the consistency issues. The SW-learner is affected by the same issues as
the S-learner in Simulation 1 but manages to substantially reduce the bias
for the rest of the simulation designs and is generally
close to the performance of the T-learner as seen in the Main Simulation.

\subsubsection{Results of Simulation 2: balanced treatment and complex nonlinear CATE}

In Simulation 2 with balanced treatment and complex nonlinear CATE we
also observe, as expected, a very good RMSE performance of the T-learner
throughout all sample sizes (see Table \ref{bigtable:CATE2} in Appendix \ref{c1appendix:appendix-b-simulation-results}).
However, it exhibits quite high variance which is mostly due to the fact
\ul{that it estimates} two completely disjoint response functions for estimating
the CATE. Furthermore, in this design the R-learner in its full-sample
version performs particularly well, which comes rather as a surprise as
it pools the two disjoint response functions within the estimation
procedure. Nevertheless, the R-learner achieves even lower bias than the
T-learner for large samples, but with rather high variance which is a
pattern observed across all simulation designs.

\subsubsection{Results of Simulation 3: highly unbalanced treatment and constant non-zero CATE}

Simulation 3 features a highly unbalanced treatment assignment and a
constant CATE for which the X-learner performs best as expected,
throughout all sample sizes and irrespective of the estimation procedure
(see Table \ref{bigtable:CATE3} in Appendix \ref{c1appendix:appendix-b-simulation-results}). Indeed, the differences between
the particular versions, i.e.~full-sample, sample-splitting and
cross-fitting, are quite small which is in contrast to the R- and
DR-learner confirming the insights from the Main Simulation. Within
this highly unbalanced design the estimation of the propensity score
function plays a key role as in this case the estimated propensity
scores can get \ul{quite} often very close to zero. This, however, does not
affect the performance of the X-learner as it uses the propensity scores
in a fundamentally different way and even the most extreme \(\{0, 1\}\)
scores would be indeed admissible as pointed out by \textcite{Kunzel2019}.\footnote{This
would, however, violate the identification assumption of the common support.}
On the contrary, the results show that such extreme propensity scores
\ul{make now both the R-learner and the DR-learner unstable, with the instability
being particularly pronounced in the latter meta-learner.} In
the case of the DR-learner the heavy tail problem of the CATE
distribution \ul{is aggravated} by more unbalanced treatment
assignment as can be seen based on the Jarque-Bera statistic and also on the
higher variance of the estimator. While the R-learner manages to avoid
this issue by downweighting the observations with extreme propensity
scores in less unbalanced settings, \ul{it is not fully able to do so} when the
imbalance is very high and there is potentially a large proportion of
propensity scores close to 1. This translates into the higher values of
the Jarque-Bera statistic as well as to higher variance and higher bias,
too. These issues lead ultimately to bad performance in terms of the
average RMSE for both the R- and DR-learner.

\subsubsection{Results of Simulation 4: unbalanced treatment and simple CATE}

In Simulation 4 the imbalance in the treatment assignment is less
pronounced which should partly reduce the propensity score issues for
the R- and DR-learner. Within this simulation design we observe similar
patterns as for the Main Simulation. For the small and medium sized
samples the X-learner in the full-sample version performs best in terms
of the average RMSE,
while it gets outperformed by the DR-learner in its
cross-fitting version in the largest sample-size. While the R-learner's
performance is quite competitive in smaller samples, it lags behind in
larger samples as observed in other simulation designs. As a general
pattern, the X-learner remains quite stable with respect to the
estimation procedure whereas the DR-learner in its sample-splitting and
cross-fitting version exhibits substantially faster convergence than the
competing estimators. Nonetheless, based on the Jarque-Bera statistic,
the heavy tail issue is less pronounced but still present as can be seen
in Table \ref{bigtable:CATE4} in Appendix \ref{c1appendix:appendix-b-simulation-results}.

\subsubsection{Results of Simulation 5: unbalanced treatment and linear CATE}

Lastly, in Simulation 5 the CATE function gets more involved, while the
treatment assignment remains unchanged. The results once again resemble
the general pattern (for details see Table \ref{bigtable:CATE5} in Appendix \ref{c1appendix:appendix-b-simulation-results}).
As such the R-learner is competitive mainly in the smaller sample sizes,
in this case best performing in the cross-fitting version. The
DR-learner in the sample-splitting and cross-fitting version exhibits
faster convergence rates, however, in this case the considered sample
sizes are not large enough to outperform the X-learner. The X-learner
exhibits again little differences regarding the estimation procedure and
outperforms the other meta-learners in all performance measures across
all sample-sizes.

\subsection{Semi-synthetic Simulation}\label{c1sec:example}

In order to compare the performance of the meta-learners outside a
completely synthetic design of the above simulations we apply the estimators
in an arguably more realistic setting using an augmented real dataset. For
this purpose we use the dataset from the data challenge of the 2018
Atlantic Causal Inference Conference (2018 ACIC henceforth). This dataset
is particularly suitable for a comparison of the meta-learners for two
reasons. First, the dataset is based on a randomized control trial in
education, namely the National Study of Learning Mindsets (NSLM) by
\textcite{Yeager2019}, and thus provides us with a real data example.
Second, the dataset has been augmented to an observational setting with
measured confounding and known treatment effects \parencite{Carvalho2019}
which enables us to evaluate the performance of the meta-learners for
the estimation of CATEs.

The dataset includes a total of \(10'391\) observations with 10
covariates, a simulated continuous outcome and a binary treatment, while
the share of treated is approximately \(25\%\).\footnote{The dataset can be
  retrieved online from \href{https://github.com/grf-labs/grf/blob/master/experiments/acic18/synthetic_data.csv}{GitHub}. We neglect here the information about the additional school ID for
  simplicity and comparability reasons.} The variables are described in
Table \ref{tab:empdata} in Appendix \ref{appendix-c-empirical-example}. Additionally,
to create a sparse large-dimensional setting, similar to the synthetic
simulations, we augment the dataset further with \(p=90\) uniformly
distributed covariates, i.e.
\(X_{11,...,100} \sim \mathcal{U}\big([0, 1]^{n\times p}\big)\) with the
same correlation structure as used within the synthetic
simulations.\footnote{For more detailed descriptive statistics of the
  augmented real dataset including correlation heat map of the
  covariates see Appendix \ref{appendix-c-empirical-example}.} At a high level, we are interested
in estimating the treatment effects of an intervention to foster a belief
to develop intelligence in students on a measure of student achievement,
conditional on observed covariates. The CATEs were generated
according to the following specification:
\vspace{0.05cm}
\[\tau(x)=0.228 + 0.05 \cdot\mathds{1}(x_1 < 0.07) - 0.05 \cdot\mathds{1}(x_2 < - 0.69) - 0.08 \cdot\mathds{1}(c_1 \in {1, 13, 14})\]
and while the conditional independence assumption holds by
construction, the confounding has a complicated functional form. For a
detailed description of the data generating process used for the
augmentation see \textcite{Carvalho2019}.

Similarly as in the \ul{synthetic} simulations we estimate the CATEs
with all meta-learners and evaluate their performance \ul{with regard
to the point estimates}. For this purpose we perform an empirically grounded
semi-synthetic simulation study inspired, among others, by
\textcite{Lechner2019}, \textcite{Kunzel2019} and \textcite{Naghi2021}
where we first, set apart a
validation set of size \(N=1'000\) observations, and second, sample
\(R=\{2'000, 1'000, 500\}\) training sets each of sizes
\(N=\{500, 2'000, 8'000\}\) observations from the remaining data. We omit the biggest sample of \(N=32'000\) observations
due to the size restrictions of the dataset. We
report mean performance measures in a similar fashion as in the previous
simulation experiments.

\subsubsection{Results of Semi-synthetic Simulation}\label{c1sec:empirical-results}

\begin{figure}[ht]
\caption{\label{bigfigure:CATEEmpirical}CATE Results for Semi-synthetic Simulation}\label{fig:cate_emp}
{\centering \includegraphics[scale=0.8]{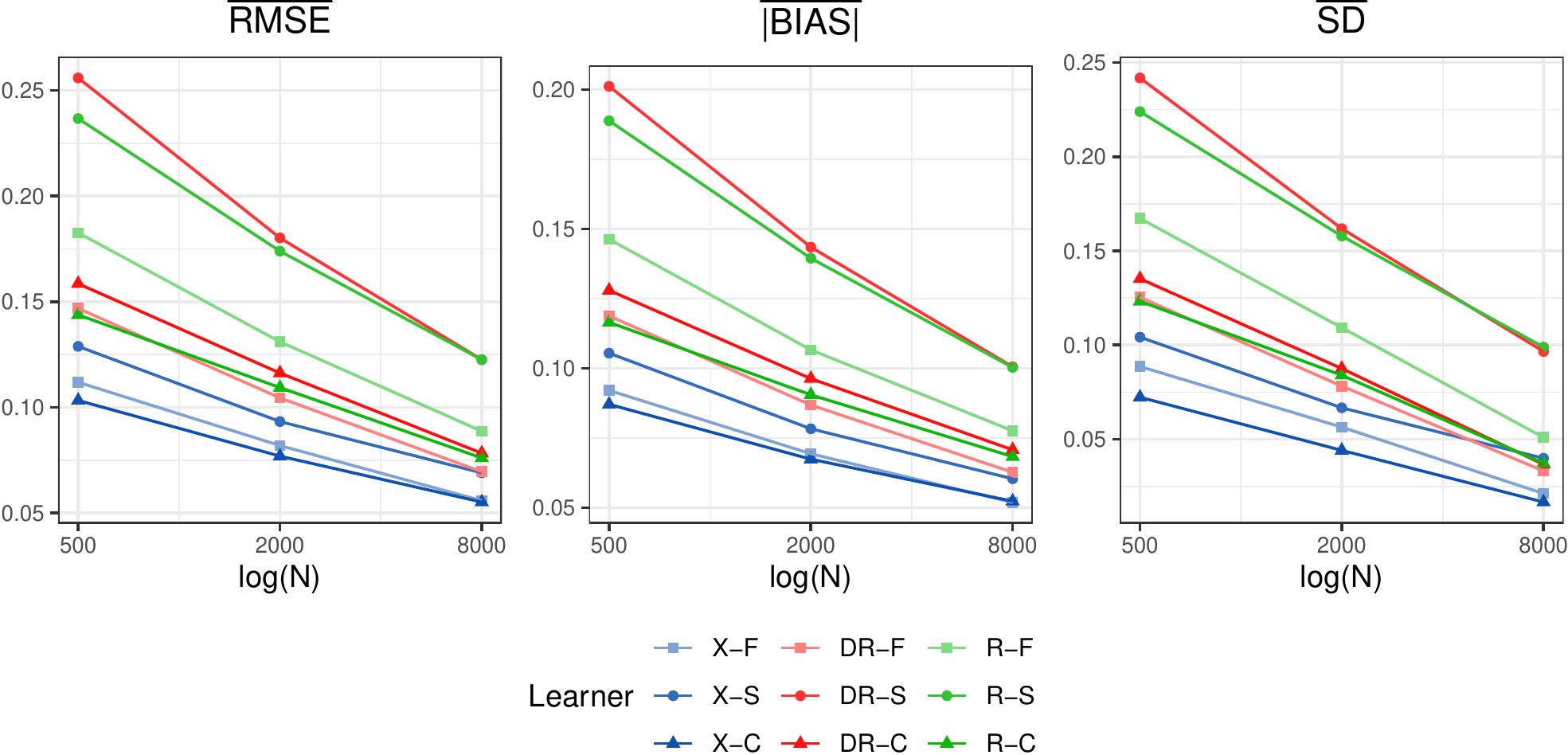} 

}
\vspace{0.25cm}
\caption*{\footnotesize{\textit{Note:} The results for $\overline{RMSE}$, $\overline{|BIAS|}$, and $\overline{SD}$ show the mean values of the root mean squared error, absolute bias, and standard deviation of all $1'000$ CATE estimates from the validation sample. The figure shows the results based on the increasing training samples of $\{500, 2'000, 8'000\}$ observations displayed on the log scale. Additionally, X-F, DR-F, R-F denote the full-sample versions of the meta-learners, while X-S, DR-S, R-S and X-C, DR-C, R-C denote the sample-splitting and cross-fitting versions, respectively.}}
\end{figure}

The CATE results of the Semi-synthetic Simulation for all meta-learners are
summarized in Table \ref{bigtable:CATEEmpirical}, while Figure
\ref{bigfigure:CATEEmpirical} provides details on the meta-learners in
the full-sample, double sample-splitting and double cross-fitting versions.

\begin{table}[ht]
\centering
\caption{\label{bigtable:CATEEmpirical}CATE Results for Semi-synthetic Simulation} 
\scalebox{0.83}{
%\resizebox{\textwidth}{!}{
\small
\setlength{\tabcolsep}{2.25pt}
\begin{tabular}{lrrrrrrrrrrrrrrr}
  \toprule
  & \multicolumn{3}{c}{$\overline{RMSE}$} & \phantom{..} & \multicolumn{3}{c}{$\overline{|BIAS|}$} & \phantom{..} & \multicolumn{3}{c}{$\overline{SD}$} & \phantom{..} & \multicolumn{3}{c}{$\overline{JB}$} \\
                      \cline{2-4} \cline{6-8} \cline{10-12} \cline{14-16}
                      & \textit{500} & \textit{2000} & \textit{8000} & & \textit{500} & \textit{2000} & \textit{8000} & & \textit{500} & \textit{2000} & \textit{8000} & & \textit{500} & \textit{2000} & \textit{8000} \\ \midrule
S & 0.175 & 0.127 & 0.093 &  & 0.171 & 0.121 & 0.090 &  & \textbf{0.035} & \textbf{0.035} & 0.023 &  & 165.803 & 7.372 & 2.054 \\ 
  S-W & 0.131 & 0.109 & 0.078 &  & 0.106 & 0.090 & 0.070 &  & 0.121 & 0.084 & 0.037 &  & 57.438 & 2.065 & \textbf{1.903} \\ 
  T & 0.150 & 0.111 & 0.079 &  & 0.122 & 0.092 & 0.071 &  & 0.127 & 0.084 & 0.037 &  & 2.050 & 2.047 & 2.082 \\ 
  X-F & 0.112 & 0.082 & 0.056 &  & 0.092 & 0.069 & \textbf{0.052} &  & 0.089 & 0.056 & 0.021 &  & 2.043 & 1.941 & 2.078 \\ 
  X-S & 0.129 & 0.093 & 0.069 &  & 0.105 & 0.078 & 0.060 &  & 0.104 & 0.067 & 0.040 &  & 2.129 & 2.594 & 2.004 \\ 
  X-C & \textbf{0.103} & \textbf{0.077} & \textbf{0.055} &  & \textbf{0.087} & \textbf{0.067} & \textbf{0.052} &  & 0.072 & 0.044 & \textbf{0.017} &  & \textbf{1.652} & \textbf{1.794} & 1.935 \\ 
  DR-F & 0.147 & 0.105 & 0.070 &  & 0.119 & 0.087 & 0.063 &  & 0.125 & 0.078 & 0.033 &  & 7.134 & 3.377 & 2.001 \\ 
  DR-S & 0.256 & 0.180 & 0.123 &  & 0.201 & 0.143 & 0.101 &  & 0.242 & 0.162 & 0.097 &  & 68.837 & 46.173 & 6.196 \\ 
  DR-C & 0.159 & 0.116 & 0.078 &  & 0.128 & 0.096 & 0.071 &  & 0.135 & 0.088 & 0.037 &  & 6.334 & 5.329 & 2.969 \\ 
  R-F & 0.183 & 0.131 & 0.089 &  & 0.146 & 0.107 & 0.078 &  & 0.167 & 0.109 & 0.051 &  & 3.819 & 3.862 & 2.027 \\ 
  R-S & 0.237 & 0.174 & 0.123 &  & 0.189 & 0.140 & 0.100 &  & 0.224 & 0.158 & 0.099 &  & 3.490 & 3.494 & 3.117 \\ 
  R-C & 0.144 & 0.109 & 0.076 &  & 0.117 & 0.091 & 0.068 &  & 0.123 & 0.084 & 0.037 &  & 2.060 & 2.222 & 2.225 \\ 
   \midrule \multicolumn{16}{c}{\parbox{14.25cm}{\setstretch{0.95}\footnotesize{\textit{Note:} The results for the $\overline{RMSE}$, $\overline{|BIAS|}$, $\overline{SD}$ and $\overline{JB}$ show the mean values of the root mean squared error, absolute bias, standard deviation and the Jarque-Bera test statistic of all $1'000$ CATE estimates from the validation sample. The critical values for the JB test statistic are 5.991 and 9.210 at the 5\% and 1\% level, respectively. Additionally, X-F, DR-F, R-F denote the full-sample versions of the meta-learners, while X-S, DR-S, R-S and X-C, DR-C, R-C denote the sample-splitting and cross-fitting versions, respectively. Bold numbers indicate the best performing meta-learner for given measure and sample size.}}}\\ \bottomrule
\end{tabular}
}
\end{table}
\vspace{0.55cm}

The results reveal a similar picture to the synthetic simulations in
general, with the largest similarities to Simulation 3 and 5 in
particular. Accordingly, the X-learner achieves the best performance in
terms of the average RMSE as well as average absolute bias in all
considered sample-sizes, regardless of the estimation procedure. This
emphasizes the good performance of the X-learner in settings
with unbalanced treatment assignment and sparse CATE function with
structural properties. In the largest sample size of $8'000$ observations, also the DR- and
R-learner come close to the performance of the X-learner in terms of the
average RMSE, while the simpler SW- and T-learner are competitive mainly
in the smaller sample sizes. We also observe a slightly faster
convergence of the sample-splitting and cross-fitting version of the
DR-learner as in the synthetic simulations, however, the limited sample
size in this case does not allow for a sufficiently large improvement to
outperform the X-learner. Given the smaller sample sizes in the
Semi-synthetic Simulation, we are not able to detect the bias-variance
trade-off and the sample-splitting versions always exhibit higher values
of the average RMSE, average absolute bias and average standard
deviation. This is particularly noticeable for the smallest sample
size of $500$ observations as there is \ul{essentially} not enough data left after splitting to
learn the correct CATE function. For all
meta-learners the cross-fitting versions then always perform
better in terms of the variance reduction and even lead to a
lower bias in comparison to the sample-splitting versions.
These results \ul{accentuate} the fact that the
benefits of sample-splitting in removing the overfitting bias become
apparent only for sufficiently large samples. Additionally, we
see larger discrepancies between the estimation versions of the
DR- and R-learner in comparison to very stable performance of the
X-learner, \ul{similarly as in the synthetic simulations.}
Lastly, the results on the distribution of the predicted
CATEs resemble those of the synthetic simulations with the heavy tail
problem of the DR-learner in its sample-splitting version as well as
deviations from normality of the S- and SW-learner.

\section{Discussion}\label{c1sec:disc}

Given the results of our synthetic and semi-synthetic simulations
there are several important findings for the estimation of heterogeneous causal effects by
the meta-learners and the associated usage of sample-splitting
and cross-fitting which are relevant for applied empirical work.

\subsection{Meta-Learners}\label{meta-learners-1}

In general, the results suggest that meta-learners that directly model
both the outcome equations and the selection process perform better,
especially in larger samples, which is in line with the insights from
the previous literature \parencite[see e.g.][]{Knaus2021}. Meta-learners
modelling only the outcome equations are competitive only in smaller
samples and tend to perform poorly in larger samples as they fail to
properly account for the selection into treatment.

In particular, we do not recommend the usage of the S-learner
for estimation of heterogeneous causal effects due to empirically documented
undesirable statistical properties such as high bias and consistency issues.
\ul{These results are in line with the simulation findings of} \textcite{Lu2018},
\ul{where the S-learner exhibits the highest bias across all simulation settings.}
The herein studied modification of the S-learner, the SW-learner, alleviates the
high bias of the S-learner, however, it does not solve the consistency
issues. Hence, enforcing the treatment variable into the splitting set
of the forest does not constitute an attractive option for estimation of
causal effects. In contrast, the T-learner does not suffer from high
bias or any consistency issues and has a stable performance as it uses
the full data sample without the need of sample-splitting due to
potential overfitting. Hence, the T-learner might be an interesting
option, if a large sample is not available for the empirical analysis.
\textcite{Alaa2018} \ul{similarly find the T-learner to uniformly outperform the
S-learner in estimating heterogeneous treatment effects.}
Related simulation studies \parencites{Jacob2020}{Knaus2021} find also a
relatively competitive performance of the T-learner, especially with the
Random Forest as a base learner.

Among the meta-learners based on the estimation of nuisance
functions, the X-learner performs very well not only in settings with highly
unbalanced but also in less unbalanced treatment shares with simple CATEs and
demonstrates the theoretically argued capability to learn such CATE
structures \parencite{Kunzel2019}. Moreover, the X-learner exhibits a
quite stable performance across all simulation designs with low bias and
very low variance, even in small samples. Additionally, due to its
particular usage of the propensity scores, the X-learner is not too
sensitive to the choice of the estimation procedure. As such, both the
full-sample version and the cross-fitting version of the
X-learner are viable options, regardless of the sample size. For these
reasons, we recommend to use the X-learner for CATE estimation if the
researcher is facing a situation with very low number of treated units
as well as in less unbalanced settings with potentially limited sample
size. In contrast to the X-learner, the DR-learner performs particularly well
in settings with nonlinear and complex CATEs if large enough
samples are available. However, it tends to be unstable in small samples
with unbalanced treatment assignment due to extreme propensity scores, which
relates to the results of \textcite{Jacob2020} and \textcite{Knaus2021}. 
Additionally, for the DR-learner the choice of the estimation procedure
is crucial as its sample-splitting and cross-fitting version exhibits
the fastest convergence rates of all meta-learners which highlights the
theoretical arguments provided in \textcite{Kennedy2020}. According to the
simulation evidence, we advice to employ the cross-fitting version of
the DR-learner for the CATE estimation in settings with rather balanced
treatment assignment and when large sample is available. Recently,
\textcite{Knaus2020b} proposed the \textit{normalized} DR-learner, that
\ul{addresses} the problem of unstable CATE predictions due to extreme
propensity scores which might be a viable option for smaller sample
sizes and settings with unbalanced treatment shares. Lastly, the simulation evidence
suggests that the R-learner is in comparison to the DR-learner less
prone to unstable performance due to extreme propensity scores. However,
its performance is competitive only in smaller samples, while the empirically
approximated speed of convergence is slower than the one of the DR-learner
and seems to depend on the CATE complexity as theoretically argued by \textcite{Nie2021a}.
With respect to the estimation procedure we do not find
a clear-cut evidence in favour of a particular version as both the
full-sample as well as the cross-fitting version exhibit comparably good
performance. Based on this evidence, the R-learner might be an
attractive option for estimation of CATEs if the treatment assignment is not too
unbalanced and if only a small sample is available. For comparable sample sizes,
\textcite{Knaus2021} also find the R-learner to have good performance in a variety of settings.

Overall, we point out that based on the simulation evidence,
for all meta-learners the approximate convergence
rates appear to be substantially slower than the parametric rate of $\sqrt{N}$.
This is expected given the insights from previous literature that the
estimation of more granular heterogeneous effects is a more difficult task
in comparison to the estimation of average effects \parencites(compare e.g.)(){Lechner2019}[or][]{Knaus2021}. However, we note that the approximate convergence rates differ considerably among the meta-learners and their specific implementations as documented in our simulation experiments.

\subsection{Estimation Procedures}\label{c1sec:estimation-procedures}

Our simulation evidence suggests that using the full sample for
estimation of both the nuisance functions as well as the CATE function
leads to a remarkably good performance in terms of both bias and variance
in finite samples. Recently, \textcite{Curth2021} also point out that the
full-sample estimation seems to work better in practice, especially for
small samples. In theory, we would expect lower variance yet higher
bias due to overfitting \parencite{Chernozhukov2018c}. The possible reason
for this phenomenon might in our case be due to the out-of-bag predictions
of the forest that we use throughout the simulation experiments. Even
though these predictions are not out-of-sample \textit{per se} they are
not directly based on the observations used for estimation and as such
might help to alleviate the overfitting problem when using full sample \parencite[compare][for a discussion of out-of-bag predictions in Random Forests]{Athey2019a}.
In the causal machine learning literature, such out-of-bag predictions
are for example also used in the case of the Generalized Random Forest
for the residualization \parencite{Athey2019}, similar to the one used in
the R-learner. In contrast to the full-sample estimation,
using the double sample-splitting for the
estimation of the nuisance functions, we effectively use only one third
of the available data. Theoretically, we should observe a smaller bias
but higher variance of the estimators. However, in almost all cases we
observe both higher bias as well as higher variance, particularly for
the small sample sizes. Nonetheless, we document the expected
bias-variance trade-off for the largest sample sizes. This stems mainly
from the fact that using only a third of the smaller samples does not
allow a sensible machine learning estimation of the highly non-linear nuisance functions
featured in our simulations. However, especially for the DR-learner we
do observe faster convergence rates for the sample-splitting version
which is compatible with the theoretical convergence arguments
\parencites{Newey2018}{Kennedy2020}. Hence, it seems to be the case that in
order to benefit from the double sample-splitting
the training sample must be of sufficient size, otherwise the full
sample estimation achieves a better performance. Lastly, the double
cross-fitting for estimation of the nuisance components effectively uses
all \ul{the available information from the} data and substantially reduces the variance of the
estimators, while keeping the bias low at the same time. \ul{This comes at the price of longer
computation time in comparison to the sample-splitting procedure as the estimation is repeated
multiple times. Nevertheless, the computation time of the cross-fitting procedure is on average comparable with the full-sample estimation} (see Appendix \ref{appendix-d-computation-time} for details).

Based on the above simulation evidence, it seems reasonable to always use the
full-sample estimation together with out-of-bag predictions (if
available) \ul{when} a relatively small sample is available to the applied empirical
researcher, whereas to use the double cross-fitting procedure \ul{when}
a relatively large data is accessible, regardless of the choice of
a meta-learner. On the contrary, the simulations do not provide any
evidence for an advantageous usage of the double sample-splitting over
the double cross-fitting, \ul{apart from the computational aspects}.

\section{Conclusion}\label{c1sec:end}

We investigate the finite sample performance of machine learning based meta-learners for the
estimation of heterogeneous causal effects with focus on the specific estimation
procedures related to data usage. In particular, we examine the
properties of double sample-splitting and double cross-fitting as
defined by \textcite{Newey2018} in contrast to using full sample for
estimation. For this purpose, we review several meta-learning algorithms
for estimation of causal effects and discuss their advantages and
disadvantages in particular empirical settings.
We conduct an extensive simulation study with data generating processes
involving highly non-linear functional forms and large-dimensional
feature space, while keeping the treatment effect specifications well-structured.
Furthermore, we perform a semi-synthetic simulation based on an augmented
real dataset to reflect an actual empirical setting. Moreover, we repeat
the simulation experiments for increasing sample sizes to empirically study the
convergence properties of the meta-learners. Based on our simulation
evidence, we provide a guideline for applied empirical researchers and
practitioners to better inform the decisions of applying certain method
and estimation procedure for their particular research objectives.

The results of our simulation study show that the choice of the
estimation procedure can indeed largely impact the performance of the
meta-learners in finite samples. On the one hand, we provide
simulation evidence for the theoretical arguments of the bias-variance
trade-off related to sample-splitting and cross-fitting which, however,
become apparent only if sufficiently large samples are used. On the
other hand, we document the adverse effects of these procedures in small
samples, \ul{when using machine learning}. Therefore, we argue that in empirical
studies based on small samples,
applied empirical researchers should use the full sample for \ul{machine learning}
estimation of both the nuisance functions as well as the treatment effect function
as the overfitting bias is in such cases of secondary importance.
However, for empirical analyses with access to large data samples, we
advocate for the usage of the double cross-fitting for the estimation of
treatment effects as the overfitting bias here becomes of primary
importance. The double cross-fitting procedure then effectively reduces
this overfitting bias and \ul{successfully preserves the full sample size
efficiency of the estimator.} Moreover, \ul{if computation time is not a constraint},
we discourage applied empirical researchers to use the double
sample-splitting procedure due to substantial \ul{increase in variance}, while
having no benefit over the double cross-fitting in terms of bias
reduction.

In contrast to the typical drawbacks of simulation studies, the particular
design of our simulation experiments with varying sample size and
varying treatment shares allows us to draw relevant conclusions that are
not solely dependent on the particular specification of the data
generating processes, but rely on the data characteristics that an
applied empirical researcher can observe without imposing arbitrary assumptions.
In particular, the simulation evidence implies a clear advantage for the
X-learner, when the researcher is confronted with highly unbalanced
treatment shares. This finding holds irrespective of the sample size at
hand and as such we recommend the usage of X-learner for estimation of
heterogeneous treatment effects whenever the share of treated or
controls is around \(15\%\) or less. With less unbalanced treatment
shares at around \(25\%\) of treated or controls, the size of the
available sample becomes decisive. For smaller samples with only few
hundred observations ($500$ and $2'000$), the simulation evidence again favours the usage
of the X-learner. However, for bigger samples with several thousand
observations ($8'000$ and $32'000$), our findings favour the DR-learner as it can successfully
learn \ul{highly} complex treatment effect function if enough data
is available. Finally, with perfectly balanced treatment shares, the
sample size matters less. In such cases, the DR-learner as well as the
R-learner are both the preferred estimators. However, we advise against
the usage of these two methods in highly unbalanced settings as their
performance becomes unstable due to extreme propensity scores. Finally,
concerning the simpler meta-learners, we explicitly argue against the
usage of the S-learner by applied empirical researchers for estimation of
heterogeneous treatment effects due to the herein documented
undesirable statistical properties, while the T-learner might be a reasonable
choice in small samples with balanced treatment shares.

Even though we shed light on certain finite sample issues of applying
different estimation procedures when using meta-learners for estimation
of heterogeneous causal effects, our findings raise new relevant questions. Most
importantly, the question of conducting statistical inference about the
estimated heterogeneous treatment effects is worth further
investigations. Based on the insights in this paper it would be of
interest to investigate the performance of the bootstrapping for
estimation of standard errors as studied by \textcite{Kunzel2019} for
meta-learners based on the double sample-splitting and double
cross-fitting procedures. \ul{Moreover, a comparison of such bootstrapping
inference procedure for meta-learners and the approaches used in the
Causal Forest literature such as the bootstrap of little bags in the
Generalized Random Forest}
\parencite{Athey2019} \ul{or the weight-based inference as in the
Modified Causal Forest} \parencite{Lechner2019} \ul{would be desirable.}
Furthermore, the performance difference in \ul{the point estimation}
using the out-of-bag vs.~in-sample predictions could provide
additional insights on the benefits of sample-splitting and
cross-fitting procedures and hence to assess the robustness of our
results to different types of base learners. Finally, a further
simulation comparison between the X-learner, the DR-learner and its
normalized version as proposed by \textcite{Knaus2020b} for highly
unbalanced settings would be of interest.

\pagebreak

\printbibliography

\pagebreak

\begin{appendix}

\section{Descriptive Statistics}\label{c1app:desc}

\subsection{Synthetic Simulations}
This appendix provides the descriptive statistics for the
data generated in the six synthetic simulation designs discussed in the main
text. For each simulation design we plot the distribution of the observed
realized outcomes, \(Y_i\), as well as the potential outcomes,
\(Y_i(0)\) and \(Y_i(1)\). Furthermore, we provide the distribution of
the treatment indicator, \(W_i\), together with the propensity score
distribution under treatment and under control to visualize the overlap
condition. Lastly, we plot the distribution of the true treatment
effects, \(\tau(X_i)\). Moreover, the plots include a correlation heat
map for the covariates \(X_i\). The respective figures for each
simulation design are listed below.

\pagebreak

\subsubsection{Simulation 1: balanced treatment and constant zero CATE}\label{simulation-1-balanced-treatment-and-constant-zero-cate}

\begin{figure}[h!]
\caption{Descriptive Statistics for the Validation Data in Simulation 1}\label{fig:simplot_1}
{\centering \includegraphics[scale=0.8]{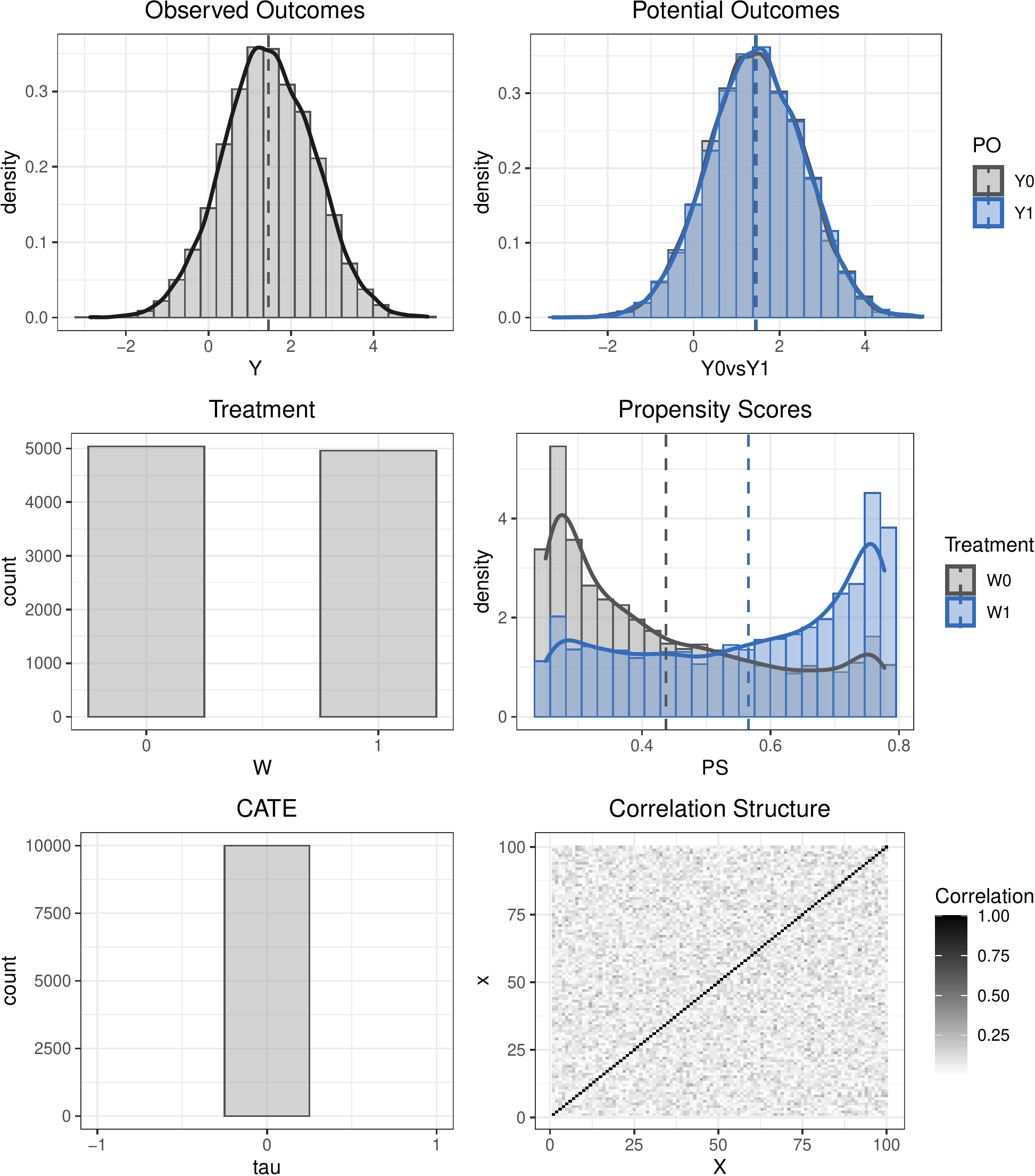} 

}
\end{figure}

\pagebreak

\subsubsection{Simulation 2: balanced treatment and complex nonlinear CATE}\label{simulation-2-balanced-treatment-and-complex-nonlinear-cate-1}

\begin{figure}[h!]
\caption{Descriptive Statistics for the Validation Data in Simulation 2}\label{fig:simplot_2}
{\centering \includegraphics[scale=0.8]{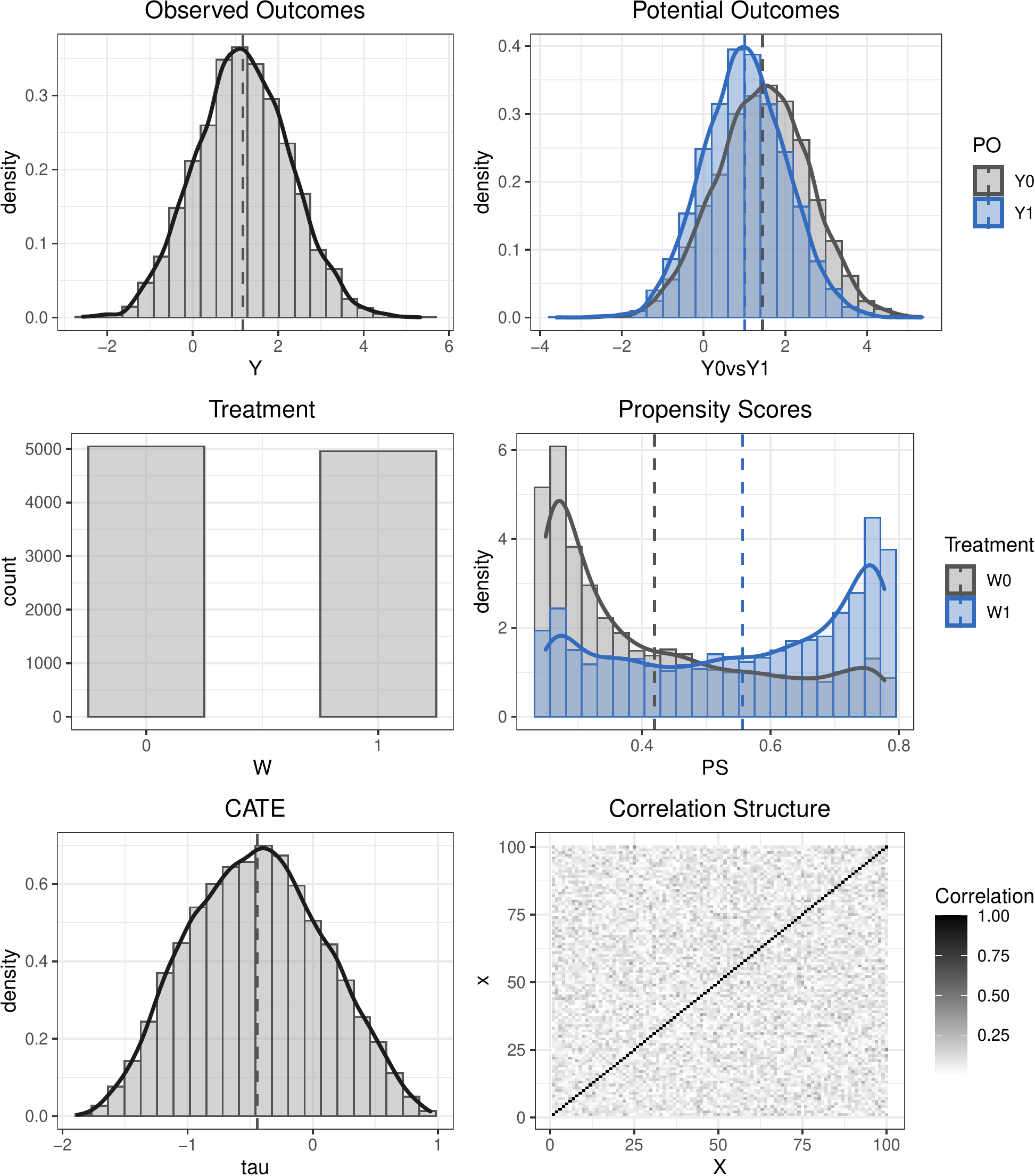} 

}
\end{figure}

\pagebreak

\subsubsection{Simulation 3: highly unbalanced treatment and constant non-zero CATE}\label{simulation-3-very-unbalanced-treatment-and-simple-cate}

\begin{figure}[h!]
\caption{Descriptive Statistics for the Validation Data in Simulation 3}\label{fig:simplot_3}
{\centering \includegraphics[scale=0.8]{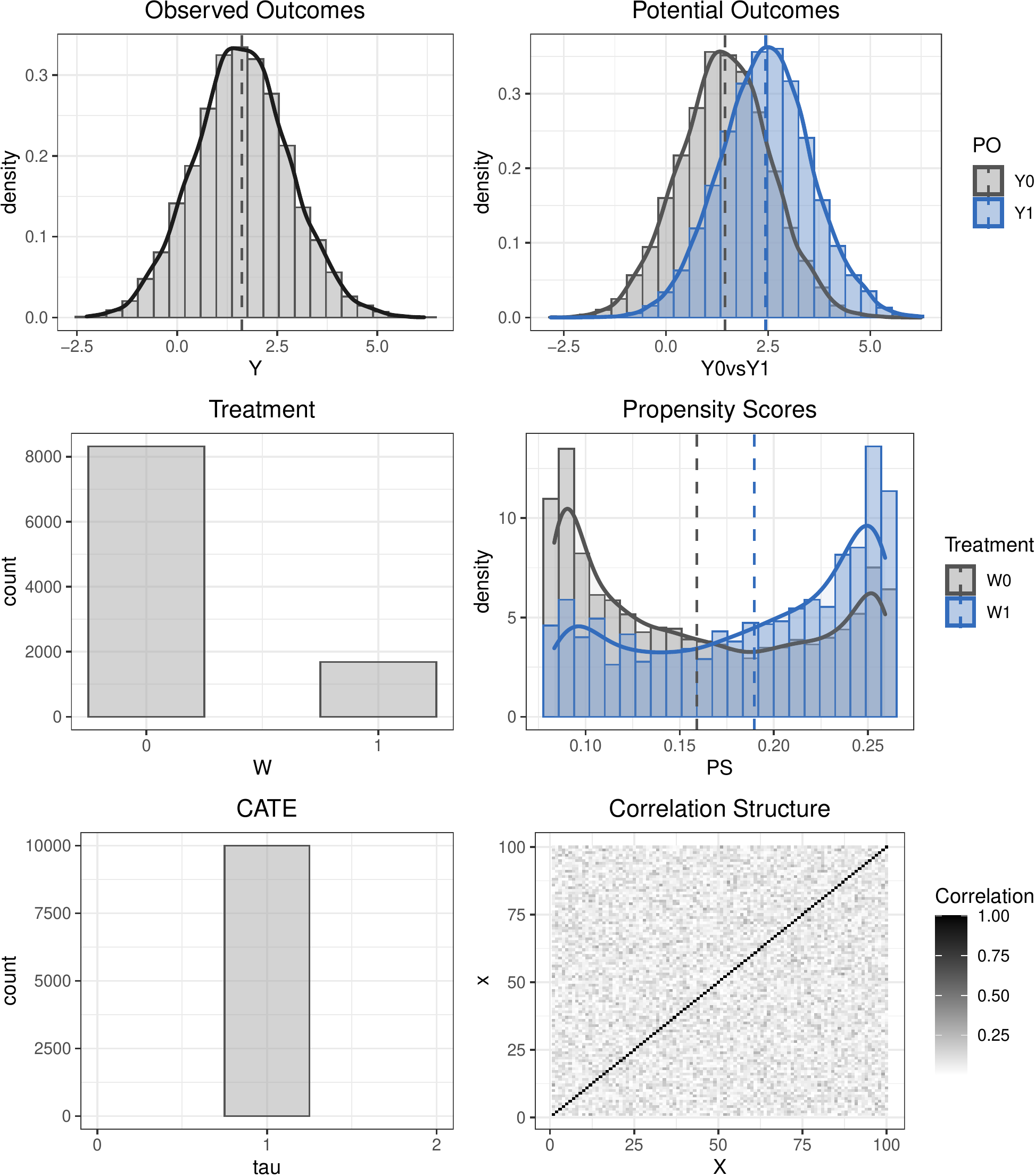} 

}
\end{figure}

\pagebreak

\subsubsection{Simulation 4: unbalanced treatment and simple CATE}\label{simulation-4-unbalanced-treatment-and-constant-nonzero-cate}

\begin{figure}[h!]
\caption{Descriptive Statistics for the Validation Data in Simulation 4}\label{fig:simplot_4}
{\centering \includegraphics[scale=0.8]{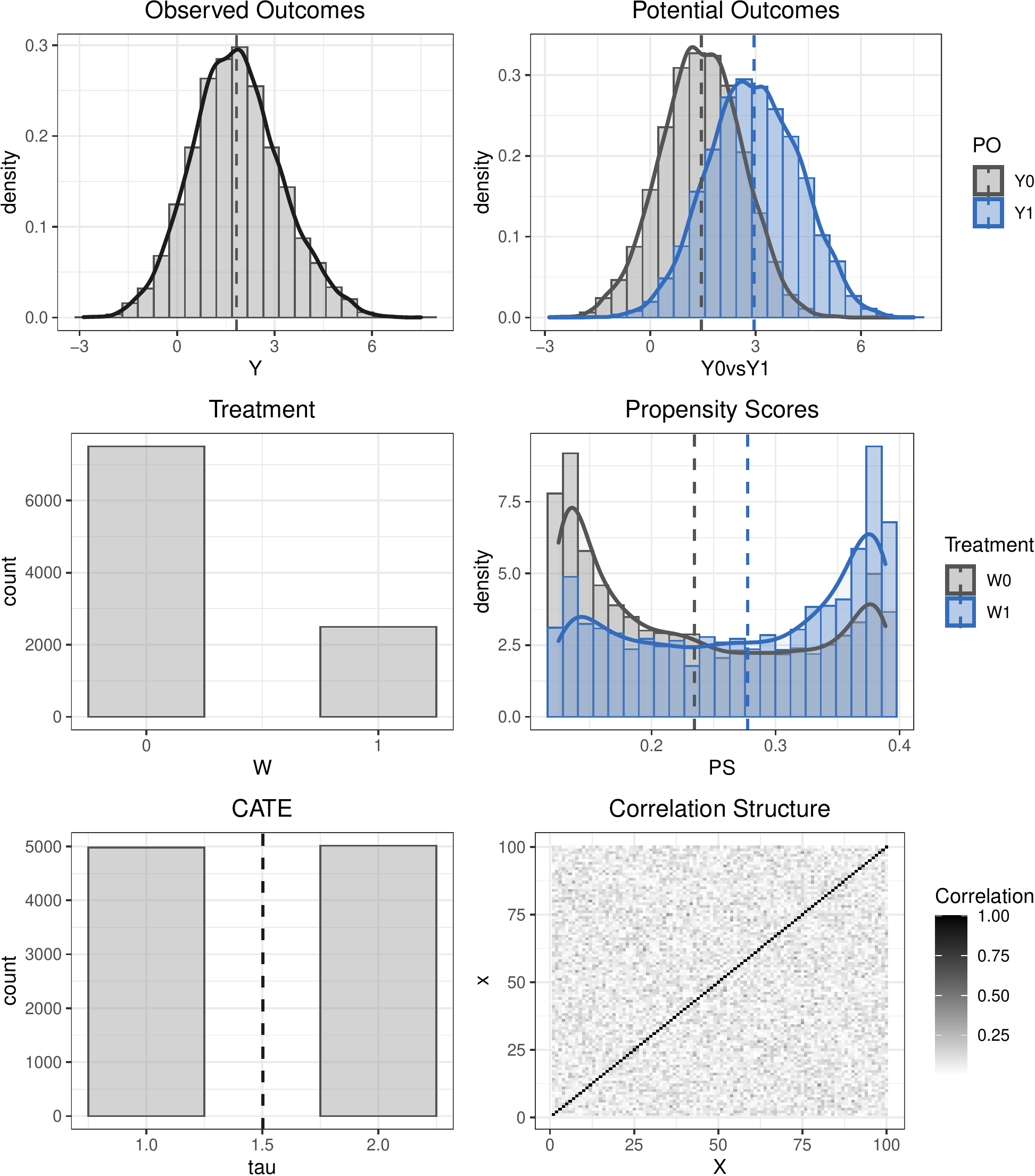} 

}
\end{figure}

\pagebreak

\subsubsection{Simulation 5: unbalanced treatment and linear CATE}\label{simulation-5-unbalanced-treatment-and-linear-cate-1}

\begin{figure}[h!]
\caption{Descriptive Statistics for the Validation Data in Simulation 5}\label{fig:simplot_5}
{\centering \includegraphics[scale=0.8]{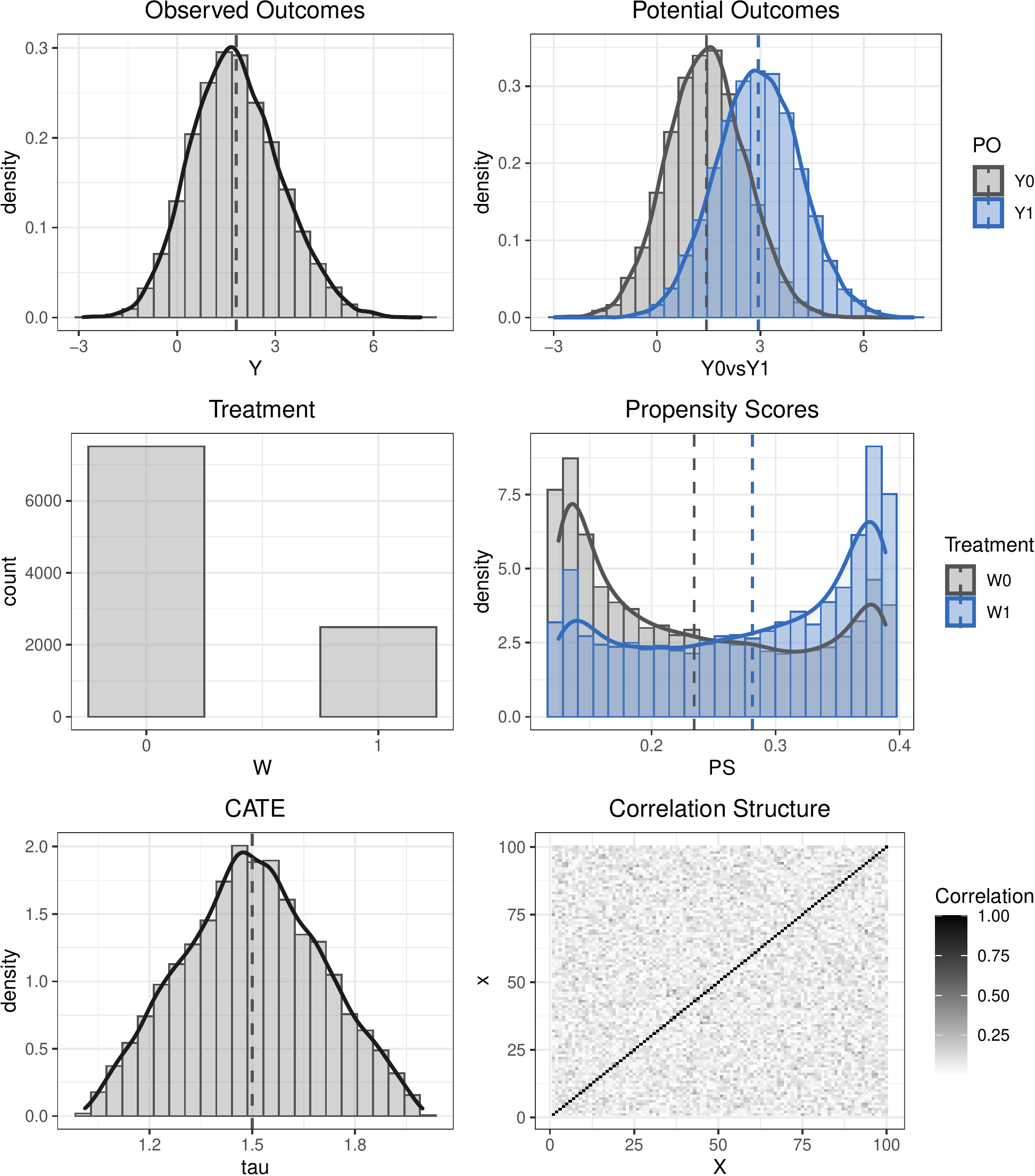} 

}
\end{figure}

\pagebreak

\subsubsection{Main Simulation: unbalanced treatment and nonlinear CATE}\label{simulation-6-unbalanced-treatment-and-nonlinear-cate-1}

\begin{figure}[h!]
\caption{Descriptive Statistics for the Validation Data in Main Simulation}\label{fig:simplot_6}
{\centering \includegraphics[scale=0.8]{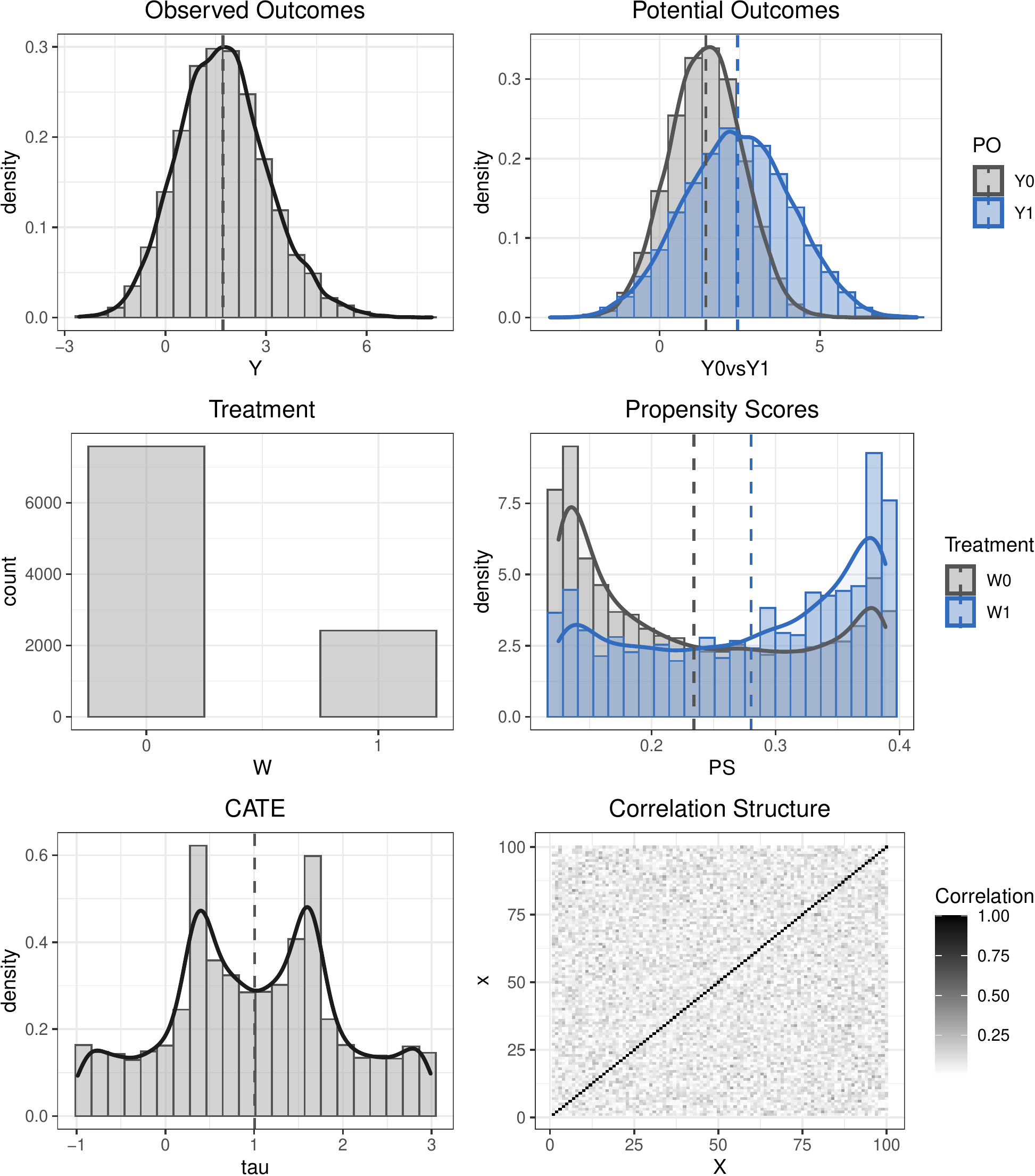} 

}
\end{figure}

\pagebreak

\subsection{Semi-synthetic Simulation}\label{appendix-c-empirical-example}

This appendix provides a comprehensive overview of the
variables in the augmented real dataset as well as descriptive
statistics thereof. Similarly to the results from the synthetic simulations,
we plot the distribution of the observed realized outcomes, \(Y_i\), as well
as the distribution of the treatment indicator, \(W_i\). Analogously, we
plot the distribution of the true treatment effects, \(\tau(X_i)\)
together with the correlation heat map for the covariates \(X_i\). The
respective figures for the distributions of the potential outcomes and
the propensity scores under treatment and under control are omitted due
to missing data availability for these quantities. The corresponding
figures and tables are listed below.\\

\begin{table}[ht]\centering
\caption{Variable description of the 2018 ACIC dataset. Source: \textcite{Carvalho2019}.}
    \label{tab:empdata}
    \begin{tabular}{|c|p{14cm}|}
    \toprule
     Variable & Description\\
    \midrule
     Y & outcome measure of achievement recorded post-treatment (continuous variable)\\
     \midrule
     W & treatment indicating receipt of the intervention (binary variable)\\
     \midrule
     S3 & student's self-reported expectations for success in the future, a proxy for prior achievement, measured prior to random assignment (ordered categorical variable)\\
     C1 & student's race/ethnicity (unordered categorical variable)\\
     C2 & student's identified gender (binary variable)\\
     C3 & student's first generation status, i.e. first in family to go to college (binary variable)\\
     \midrule
     XC & urbanicity of the school, i.e. rural, suburban, etc. (unordered categorical variable)\\
     X1 & school-level mean of students' fixed mindsets, reported prior to random assignment (continuous variable)\\
     X2 & school achievement level, measured by test scores and college preparation for the previous 4 cohorts of students (continuous variable)\\
     X3 & school racial/ethnic minority composition, i.e. percentage of student body that is Black, Latino, or Native American (continuous variable)\\
     X4 & school poverty concentration, i.e. percentage of students who are from families whose incomes fall below the federal poverty line (continuous variable)\\
     X5 & School size, i.e. total number of students in all four grade levels in the school (continuous variable)\\
     \bottomrule
    \end{tabular}
\end{table}

\pagebreak

\begin{figure}[h!]
\caption{Descriptive Statistics for the Validation Data in Semi-synthetic Simulation}\label{fig:emplot}
{\centering \includegraphics[scale=0.8]{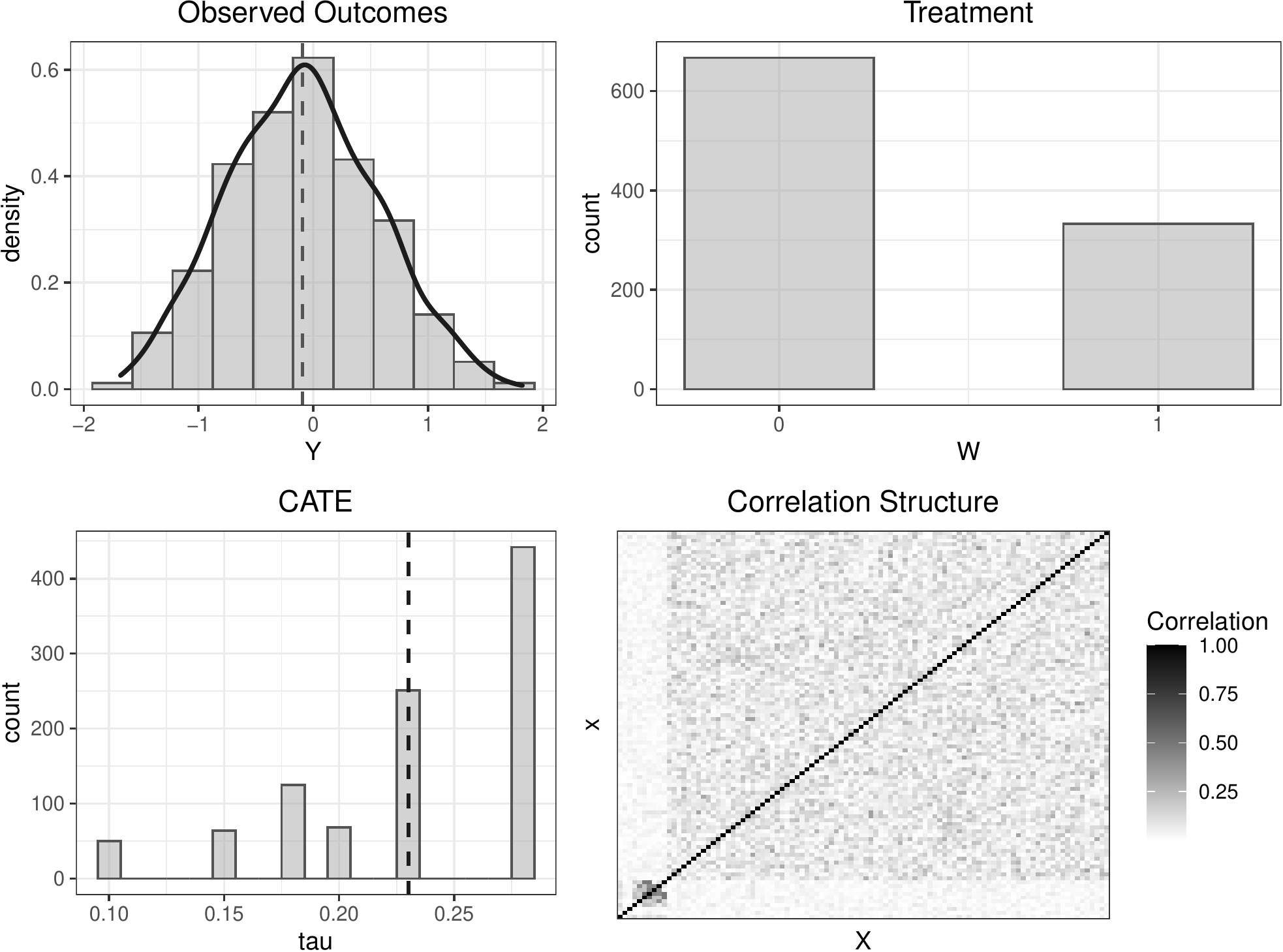} 

}
\end{figure}

\pagebreak
\clearpage

\section{Simulation Results}\label{appendix-b-simulation-results}

\subsection{Main Results}\label{c1appendix:appendix-b-simulation-results}

\subsubsection{Simulation 1: balanced treatment and constant zero CATE}\label{simulation-1-balanced-treatment-and-constant-zero-cate-1}

\begin{table}[ht]
\centering
\caption{\label{bigtable:CATE1}CATE Results for Simulation 1} 
\scalebox{0.83}{
%\resizebox{\textwidth}{!}{
\small
\setlength{\tabcolsep}{2.25pt}
\begin{tabular}{lrrrrrrrrrrrrrrrrrrr}
  \toprule
  & \multicolumn{4}{c}{$\overline{RMSE}$} & \phantom{..} & \multicolumn{4}{c}{$\overline{|BIAS|}$} & \phantom{..} & \multicolumn{4}{c}{$\overline{SD}$} & \phantom{..} & \multicolumn{4}{c}{$\overline{JB}$} \\
                      \cline{2-5} \cline{7-10} \cline{12-15} \cline{17-20}
                      & \textit{500} & \textit{2000} & \textit{8000} & \textit{32000} & & \textit{500} & \textit{2000} & \textit{8000} & \textit{32000} & & \textit{500} & \textit{2000} & \textit{8000} & \textit{32000} & & \textit{500} & \textit{2000} & \textit{8000} & \textit{32000} \\ \midrule
S & \textbf{0.008} & \textbf{0.009} & \textbf{0.013} & \textbf{0.018} &  & \textbf{0.005} & \textbf{0.006} & \textbf{0.010} & \textbf{0.014} &  & \textbf{0.007} & \textbf{0.008} & \textbf{0.012} & \textbf{0.016} &  & 21102.232 & 3656.421 & 285.128 & 12.451 \\ 
  S-W & 0.037 & 0.038 & 0.049 & 0.059 &  & 0.023 & 0.025 & 0.036 & 0.047 &  & 0.033 & 0.032 & 0.040 & 0.047 &  & 20410.604 & 4973.704 & 364.501 & 12.494 \\ 
  T & 0.225 & 0.168 & 0.128 & 0.101 &  & 0.180 & 0.135 & 0.103 & 0.082 &  & 0.206 & 0.149 & 0.109 & 0.083 &  & 2.071 & 2.280 & \textbf{2.000} & \textbf{1.912} \\ 
  X-F & 0.160 & 0.111 & 0.080 & 0.059 &  & 0.128 & 0.089 & 0.065 & 0.048 &  & 0.142 & 0.095 & 0.067 & 0.048 &  & 1.689 & 2.442 & 2.068 & 2.002 \\ 
  X-S & 0.186 & 0.127 & 0.091 & 0.067 &  & 0.149 & 0.103 & 0.074 & 0.055 &  & 0.162 & 0.106 & 0.073 & 0.053 &  & 1.916 & \textbf{2.152} & 2.180 & 2.125 \\ 
  X-C & 0.152 & 0.106 & 0.075 & 0.054 &  & 0.123 & 0.087 & 0.062 & 0.045 &  & 0.120 & 0.075 & 0.052 & 0.036 &  & \textbf{1.293} & 2.393 & 2.023 & 1.982 \\ 
  DR-F & 0.209 & 0.146 & 0.105 & 0.079 &  & 0.167 & 0.117 & 0.084 & 0.064 &  & 0.196 & 0.135 & 0.095 & 0.070 &  & 4.677 & 18.496 & 17.015 & 7.978 \\ 
  DR-S & 0.343 & 0.241 & 0.170 & 0.122 &  & 0.272 & 0.191 & 0.135 & 0.097 &  & 0.335 & 0.235 & 0.166 & 0.119 &  & 5.548 & 15.737 & 30.661 & 41.289 \\ 
  DR-C & 0.207 & 0.147 & 0.103 & 0.074 &  & 0.166 & 0.118 & 0.082 & 0.059 &  & 0.194 & 0.137 & 0.097 & 0.070 &  & 2.606 & 3.632 & 9.329 & 13.029 \\ 
  R-F & 0.261 & 0.192 & 0.145 & 0.114 &  & 0.208 & 0.153 & 0.116 & 0.092 &  & 0.249 & 0.180 & 0.132 & 0.102 &  & 5.911 & 23.800 & 26.134 & 14.008 \\ 
  R-S & 0.334 & 0.243 & 0.181 & 0.137 &  & 0.265 & 0.193 & 0.144 & 0.109 &  & 0.321 & 0.232 & 0.168 & 0.124 &  & 3.192 & 5.140 & 15.179 & 15.980 \\ 
  R-C & 0.208 & 0.156 & 0.118 & 0.092 &  & 0.166 & 0.125 & 0.096 & 0.075 &  & 0.186 & 0.135 & 0.098 & 0.072 &  & 2.117 & 2.586 & 3.209 & 3.779 \\ 
    \midrule \multicolumn{20}{c}{\parbox{19.25cm}{\setstretch{0.95}\footnotesize{\textit{Note:} The results for the $\overline{RMSE}$, $\overline{|BIAS|}$, $\overline{SD}$ and $\overline{JB}$ show the mean values of the root mean squared error, absolute bias, standard deviation and the Jarque-Bera test statistic of all $10'000$ CATE estimates from the validation sample. The critical values for the JB test statistic are 5.991 and 9.210 at the 5\% and 1\% level, respectively. Additionally, X-F, DR-F, R-F denote the full-sample versions of the meta-learners, while X-S, DR-S, R-S and X-C, DR-C, R-C denote the sample-splitting and cross-fitting versions, respectively. Bold numbers indicate the best performing meta-learner for given measure and sample size.}}}\\ \bottomrule
\end{tabular}
}
\end{table}
\vspace{0.25cm}
\begin{figure}[ht]
\caption{\label{bigfigure:CATE1}CATE Results for Simulation 1}\label{fig:unnamed-chunk-125}
{\centering \includegraphics[scale=0.8]{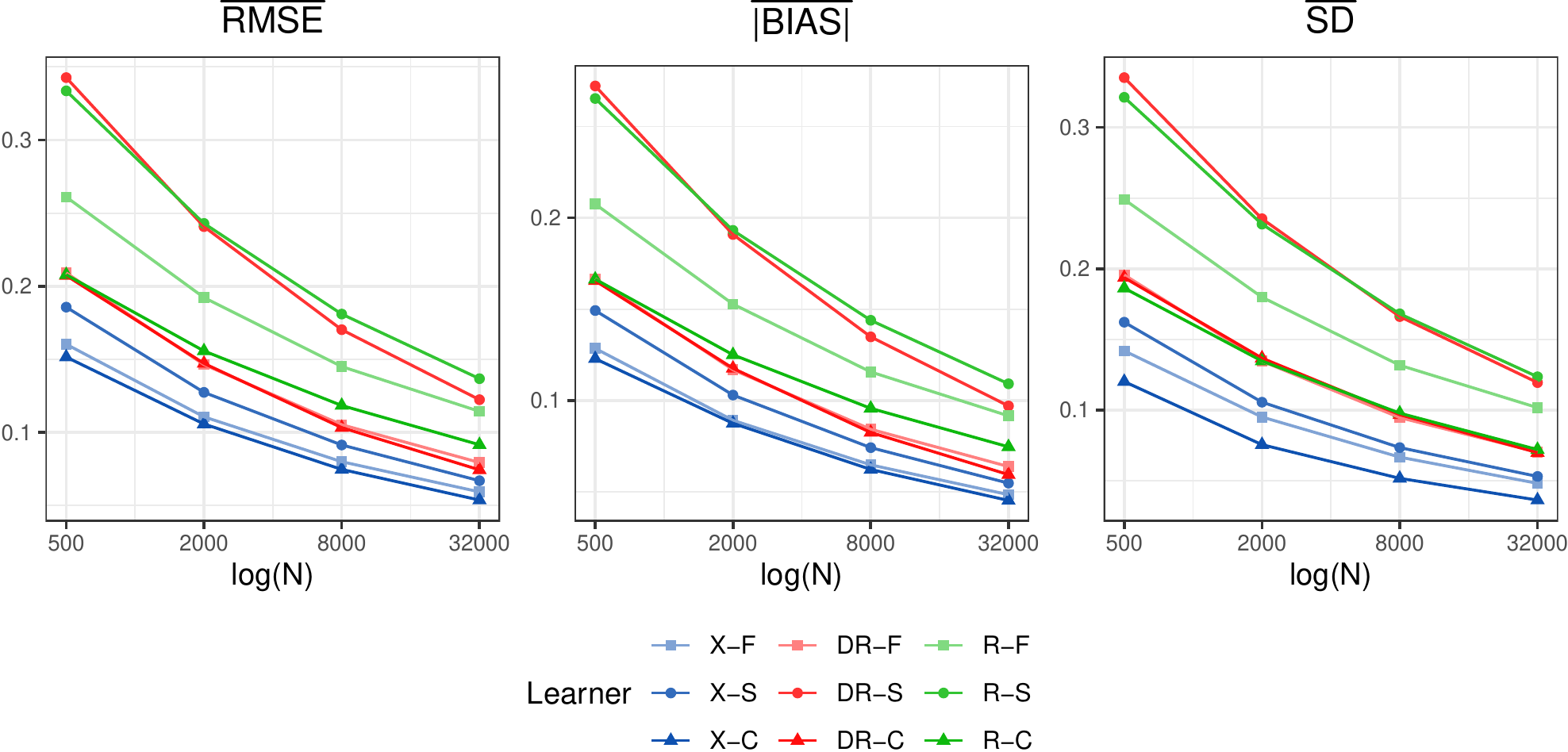} 

}
\vspace{0.25cm}
\caption*{\footnotesize{\textit{Note:} The results for $\overline{RMSE}$, $\overline{|BIAS|}$, and $\overline{SD}$ show the mean values of the root mean squared error, absolute bias, and standard deviation of all $10'000$ CATE estimates from the validation sample. The figure shows the results based on the increasing training samples of $\{500, 2'000, 8'000, 32'000\}$ observations displayed on the log scale. Additionally, X-F, DR-F, R-F denote the full-sample versions of the meta-learners, while X-S, DR-S, R-S and X-C, DR-C, R-C denote the sample-splitting and cross-fitting versions, respectively.}}
\end{figure}

\pagebreak

\subsubsection{Simulation 2: balanced treatment and complex nonlinear CATE}\label{simulation-2-balanced-treatment-and-complex-nonlinear-cate-2}

\begin{table}[ht]
\centering
\caption{\label{bigtable:CATE2}CATE Results for Simulation 2} 
\scalebox{0.83}{
%\resizebox{\textwidth}{!}{
\small
\setlength{\tabcolsep}{2.25pt}
\begin{tabular}{lrrrrrrrrrrrrrrrrrrr}
  \toprule
  & \multicolumn{4}{c}{$\overline{RMSE}$} & \phantom{..} & \multicolumn{4}{c}{$\overline{|BIAS|}$} & \phantom{..} & \multicolumn{4}{c}{$\overline{SD}$} & \phantom{..} & \multicolumn{4}{c}{$\overline{JB}$} \\
                      \cline{2-5} \cline{7-10} \cline{12-15} \cline{17-20}
                      & \textit{500} & \textit{2000} & \textit{8000} & \textit{32000} & & \textit{500} & \textit{2000} & \textit{8000} & \textit{32000} & & \textit{500} & \textit{2000} & \textit{8000} & \textit{32000} & & \textit{500} & \textit{2000} & \textit{8000} & \textit{32000} \\ \midrule
S & 0.527 & 0.442 & 0.374 & 0.326 &  & 0.522 & 0.434 & 0.366 & 0.317 &  & \textbf{0.055} & \textbf{0.068} & \textbf{0.066} & 0.064 &  & 711.571 & 17.960 & 4.186 & 2.537 \\ 
  S-W & 0.463 & \textbf{0.357} & \textbf{0.303} & \textbf{0.265} &  & 0.431 & \textbf{0.328} & 0.280 & 0.246 &  & 0.177 & 0.151 & 0.120 & 0.099 &  & 268.394 & 2.900 & 2.520 & 2.207 \\ 
  T & 0.434 & 0.358 & \textbf{0.303} & \textbf{0.265} &  & \textbf{0.392} & \textbf{0.328} & 0.280 & 0.246 &  & 0.204 & 0.154 & 0.120 & 0.099 &  & 2.206 & 2.464 & 2.466 & 2.250 \\ 
  X-F & \textbf{0.432} & 0.377 & 0.331 & 0.296 &  & 0.407 & 0.361 & 0.318 & 0.285 &  & 0.143 & 0.103 & 0.083 & 0.072 &  & 1.915 & 2.167 & \textbf{1.936} & 1.906 \\ 
  X-S & 0.460 & 0.411 & 0.362 & 0.321 &  & 0.432 & 0.393 & 0.349 & 0.310 &  & 0.160 & 0.110 & 0.085 & 0.073 &  & 2.048 & \textbf{2.046} & 2.156 & 1.957 \\ 
  X-C & 0.443 & 0.400 & 0.356 & 0.317 &  & 0.424 & 0.389 & 0.347 & 0.309 &  & 0.119 & 0.082 & 0.068 & \textbf{0.062} &  & \textbf{1.417} & 2.139 & 1.955 & \textbf{1.900} \\ 
  DR-F & 0.439 & 0.366 & 0.312 & 0.276 &  & 0.399 & 0.338 & 0.291 & 0.259 &  & 0.197 & 0.144 & 0.113 & 0.095 &  & 3.392 & 2.919 & 2.104 & 1.936 \\ 
  DR-S & 0.534 & 0.439 & 0.355 & 0.297 &  & 0.461 & 0.388 & 0.318 & 0.270 &  & 0.328 & 0.236 & 0.173 & 0.136 &  & 5.134 & 8.959 & 4.011 & 2.371 \\ 
  DR-C & 0.451 & 0.388 & 0.322 & 0.276 &  & 0.413 & 0.361 & 0.302 & 0.259 &  & 0.193 & 0.146 & 0.114 & 0.095 &  & 2.498 & 2.525 & 2.158 & 1.980 \\ 
  R-F & 0.458 & 0.373 & 0.307 & 0.266 &  & 0.404 & 0.333 & \textbf{0.277} & \textbf{0.241} &  & 0.251 & 0.188 & 0.146 & 0.122 &  & 4.201 & 4.206 & 2.437 & 2.021 \\ 
  R-S & 0.529 & 0.439 & 0.356 & 0.298 &  & 0.458 & 0.389 & 0.319 & 0.269 &  & 0.318 & 0.236 & 0.178 & 0.140 &  & 2.989 & 3.550 & 4.630 & 3.400 \\ 
  R-C & 0.449 & 0.388 & 0.322 & 0.274 &  & 0.413 & 0.361 & 0.301 & 0.256 &  & 0.187 & 0.145 & 0.116 & 0.097 &  & 2.195 & 2.280 & 2.107 & 1.940 \\ 
    \midrule \multicolumn{20}{c}{\parbox{18.3cm}{\setstretch{0.95}\footnotesize{\textit{Note:} The results for the $\overline{RMSE}$, $\overline{|BIAS|}$, $\overline{SD}$ and $\overline{JB}$ show the mean values of the root mean squared error, absolute bias, standard deviation and the Jarque-Bera test statistic of all $10'000$ CATE estimates from the validation sample. The critical values for the JB test statistic are 5.991 and 9.210 at the 5\% and 1\% level, respectively. Additionally, X-F, DR-F, R-F denote the full-sample versions of the meta-learners, while X-S, DR-S, R-S and X-C, DR-C, R-C denote the sample-splitting and cross-fitting versions, respectively. Bold numbers indicate the best performing meta-learner for given measure and sample size.}}}\\ \bottomrule
\end{tabular}
}
\end{table}
\vspace{0.25cm}
\begin{figure}[ht]
\caption{\label{bigfigure:CATE2}CATE Results for Simulation 2}\label{fig:unnamed-chunk-127}
{\centering \includegraphics[scale=0.8]{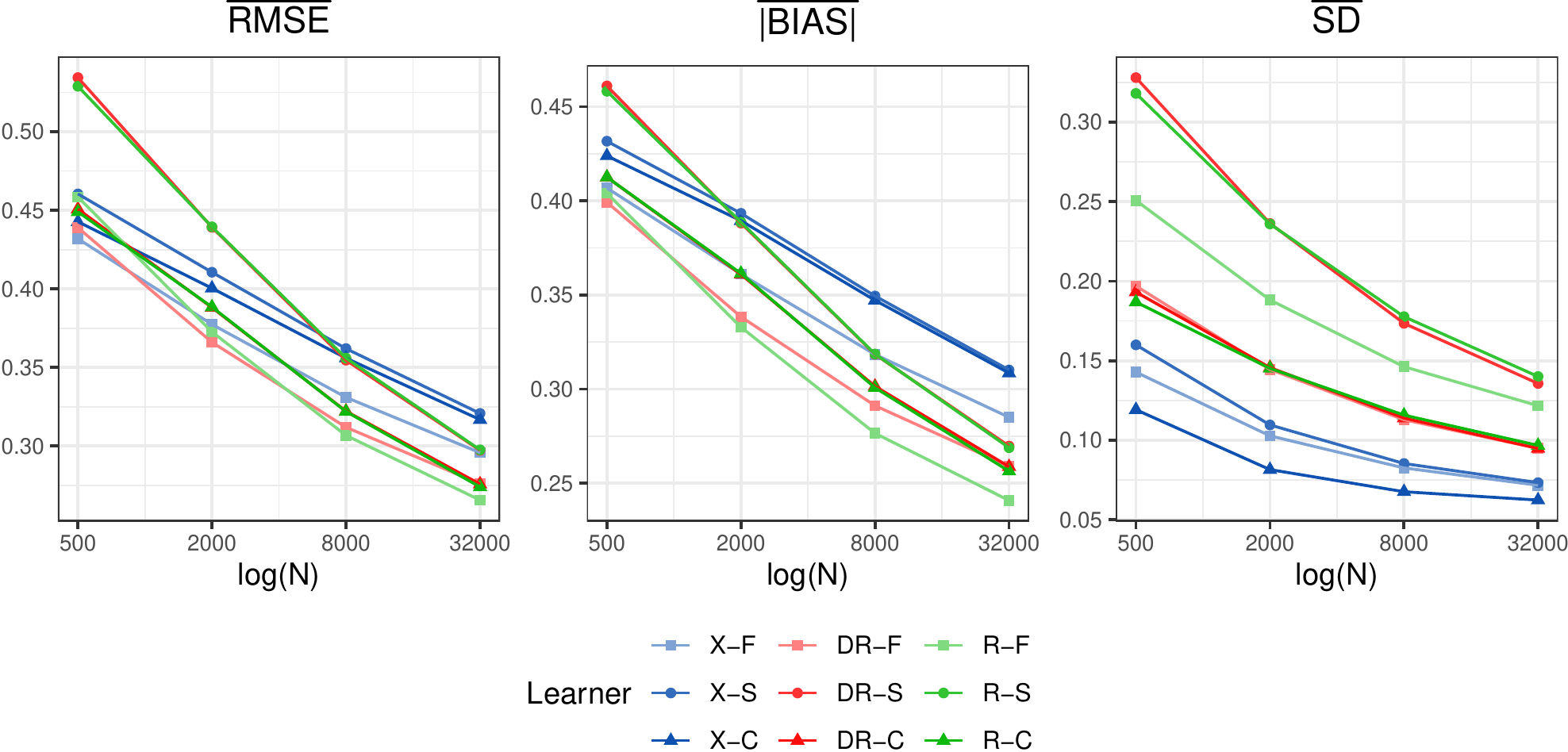} 

}
\vspace{0.25cm}
\caption*{\footnotesize{\textit{Note:} The results for $\overline{RMSE}$, $\overline{|BIAS|}$, and $\overline{SD}$ show the mean values of the root mean squared error, absolute bias, and standard deviation of all $10'000$ CATE estimates from the validation sample. The figure shows the results based on the increasing training samples of $\{500, 2'000, 8'000, 32'000\}$ observations displayed on the log scale. Additionally, X-F, DR-F, R-F denote the full-sample versions of the meta-learners, while X-S, DR-S, R-S and X-C, DR-C, R-C denote the sample-splitting and cross-fitting versions, respectively.}}
\end{figure}

\pagebreak

\subsubsection{Simulation 3: highly unbalanced treatment and constant non-zero CATE}\label{simulation-3-very-unbalanced-treatment-and-simple-cate-1}

\begin{table}[ht]
\centering
\caption{\label{bigtable:CATE3}CATE Results for Simulation 3} 
\scalebox{0.83}{
%\resizebox{\textwidth}{!}{
\small
\setlength{\tabcolsep}{2.25pt}
\begin{tabular}{lrrrrrrrrrrrrrrrrrrr}
  \toprule
  & \multicolumn{4}{c}{$\overline{RMSE}$} & \phantom{..} & \multicolumn{4}{c}{$\overline{|BIAS|}$} & \phantom{..} & \multicolumn{4}{c}{$\overline{SD}$} & \phantom{..} & \multicolumn{4}{c}{$\overline{JB}$} \\
                      \cline{2-5} \cline{7-10} \cline{12-15} \cline{17-20}
                      & \textit{500} & \textit{2000} & \textit{8000} & \textit{32000} & & \textit{500} & \textit{2000} & \textit{8000} & \textit{32000} & & \textit{500} & \textit{2000} & \textit{8000} & \textit{32000} & & \textit{500} & \textit{2000} & \textit{8000} & \textit{32000} \\ \midrule
S & 0.645 & 0.475 & 0.359 & 0.279 &  & 0.638 & 0.468 & 0.352 & 0.272 &  & \textbf{0.099} & \textbf{0.084} & 0.072 & 0.062 &  & 2.888 & 2.667 & \textbf{2.111} & 1.981 \\ 
  S-W & 0.246 & 0.191 & 0.146 & 0.111 &  & 0.197 & 0.154 & 0.119 & 0.091 &  & 0.233 & 0.163 & 0.121 & 0.090 &  & 4.611 & 2.281 & 2.385 & 1.993 \\ 
  T & 0.244 & 0.191 & 0.146 & 0.111 &  & 0.195 & 0.154 & 0.119 & 0.091 &  & 0.227 & 0.164 & 0.121 & 0.090 &  & 2.806 & 2.271 & 2.243 & \textbf{1.964} \\ 
  X-F & 0.180 & 0.123 & 0.090 & 0.068 &  & 0.144 & 0.098 & 0.072 & 0.054 &  & 0.175 & 0.118 & 0.085 & 0.061 &  & 3.663 & 2.552 & 4.441 & 2.820 \\ 
  X-S & 0.226 & 0.149 & 0.110 & 0.078 &  & 0.180 & 0.119 & 0.087 & 0.062 &  & 0.219 & 0.143 & 0.104 & 0.072 &  & \textbf{2.367} & 2.678 & 3.058 & 3.192 \\ 
  X-C & \textbf{0.159} & \textbf{0.102} & \textbf{0.073} & \textbf{0.054} &  & \textbf{0.127} & \textbf{0.081} & \textbf{0.059} & \textbf{0.043} &  & 0.150 & 0.092 & \textbf{0.064} & \textbf{0.046} &  & 6.541 & \textbf{1.969} & 2.263 & 1.994 \\ 
  DR-F & 0.287 & 0.202 & 0.146 & 0.110 &  & 0.222 & 0.158 & 0.116 & 0.089 &  & 0.279 & 0.188 & 0.129 & 0.093 &  & 3060.536 & 812.294 & 244.016 & 38.545 \\ 
  DR-S & 0.649 & 0.502 & 0.334 & 0.218 &  & 0.475 & 0.365 & 0.250 & 0.168 &  & 0.645 & 0.498 & 0.329 & 0.213 &  & 1276.433 & 1496.545 & 795.711 & 258.858 \\ 
  DR-C & 0.364 & 0.290 & 0.197 & 0.131 &  & 0.282 & 0.223 & 0.153 & 0.104 &  & 0.359 & 0.283 & 0.189 & 0.124 &  & 112.249 & 149.500 & 126.268 & 43.274 \\ 
  R-F & 0.441 & 0.366 & 0.293 & 0.243 &  & 0.348 & 0.287 & 0.232 & 0.195 &  & 0.435 & 0.354 & 0.273 & 0.215 &  & 14.590 & 27.616 & 19.045 & 8.171 \\ 
  R-S & 0.573 & 0.461 & 0.366 & 0.285 &  & 0.453 & 0.363 & 0.288 & 0.227 &  & 0.570 & 0.454 & 0.353 & 0.264 &  & 7.887 & 12.474 & 18.638 & 11.149 \\ 
  R-C & 0.295 & 0.262 & 0.220 & 0.184 &  & 0.235 & 0.208 & 0.176 & 0.150 &  & 0.289 & 0.249 & 0.199 & 0.152 &  & 2.822 & 3.570 & 4.324 & 3.162 \\ 
   \midrule \multicolumn{20}{c}{\parbox{19.2cm}{\setstretch{0.95}\footnotesize{\textit{Note:} The results for the $\overline{RMSE}$, $\overline{|BIAS|}$, $\overline{SD}$ and $\overline{JB}$ show the mean values of the root mean squared error, absolute bias, standard deviation and the Jarque-Bera test statistic of all $10'000$ CATE estimates from the validation sample. The critical values for the JB test statistic are 5.991 and 9.210 at the 5\% and 1\% level, respectively. Additionally, X-F, DR-F, R-F denote the full-sample versions of the meta-learners, while X-S, DR-S, R-S and X-C, DR-C, R-C denote the sample-splitting and cross-fitting versions, respectively. Bold numbers indicate the best performing meta-learner for given measure and sample size.}}}\\ \bottomrule
\end{tabular}
}
\end{table}
\vspace{0.25cm}
\begin{figure}[ht]
\caption{\label{bigfigure:CATE3}CATE Results for Simulation 3}\label{fig:unnamed-chunk-129}
{\centering \includegraphics[scale=0.8]{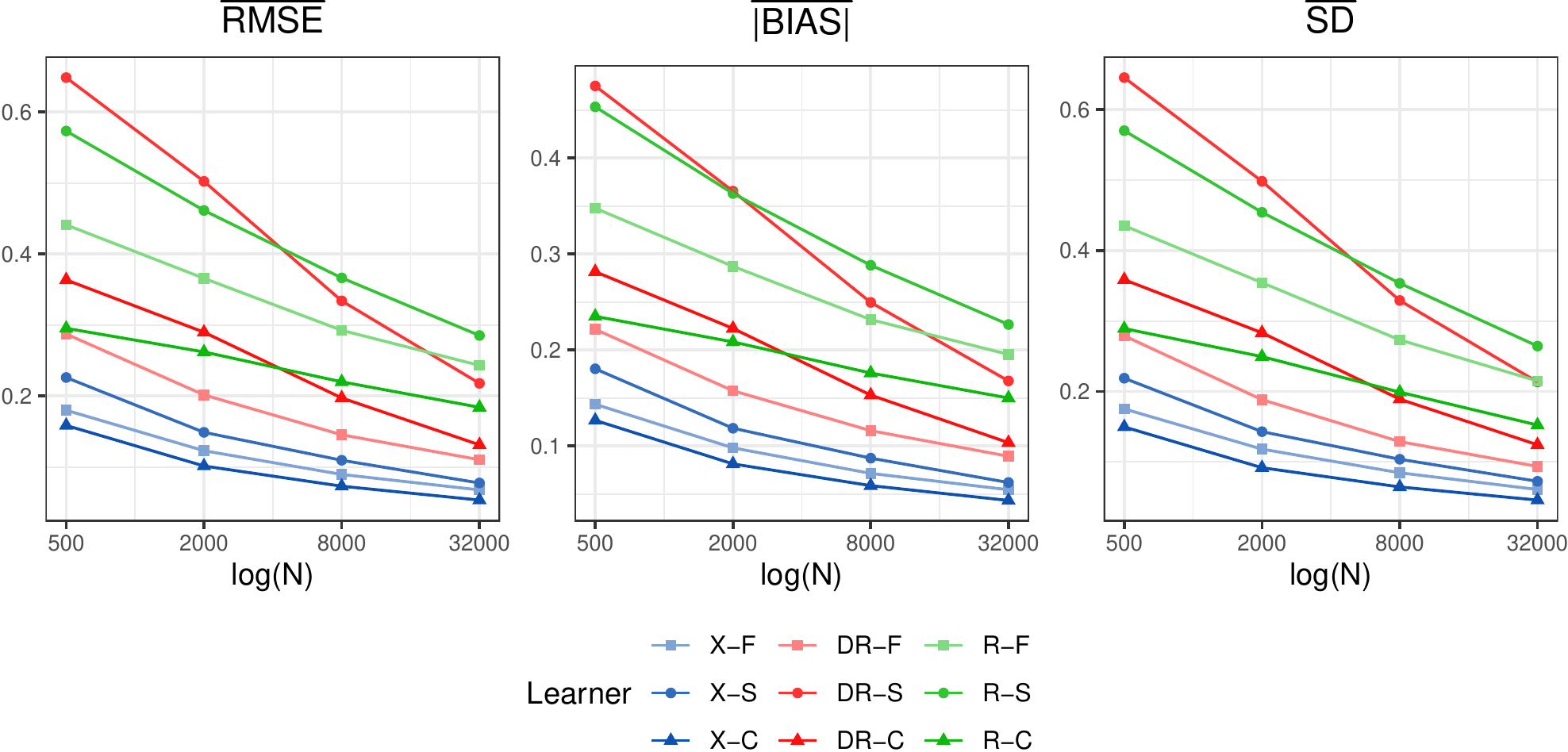} 

}
\vspace{0.25cm}
\caption*{\footnotesize{\textit{Note:} The results for $\overline{RMSE}$, $\overline{|BIAS|}$, and $\overline{SD}$ show the mean values of the root mean squared error, absolute bias, and standard deviation of all $10'000$ CATE estimates from the validation sample. The figure shows the results based on the increasing training samples of $\{500, 2'000, 8'000, 32'000\}$ observations displayed on the log scale. Additionally, X-F, DR-F, R-F denote the full-sample versions of the meta-learners, while X-S, DR-S, R-S and X-C, DR-C, R-C denote the sample-splitting and cross-fitting versions, respectively.}}
\end{figure}

\pagebreak

\subsubsection{Simulation 4: unbalanced treatment and simple CATE}\label{simulation-4-unbalanced-treatment-and-constant-nonzero-cate-1}

\begin{table}[ht]
\centering
\caption{\label{bigtable:CATE4}CATE Results for Simulation 4} 
\scalebox{0.83}{
%\resizebox{\textwidth}{!}{
\small
\setlength{\tabcolsep}{2.25pt}
\begin{tabular}{lrrrrrrrrrrrrrrrrrrr}
  \toprule
  & \multicolumn{4}{c}{$\overline{RMSE}$} & \phantom{..} & \multicolumn{4}{c}{$\overline{|BIAS|}$} & \phantom{..} & \multicolumn{4}{c}{$\overline{SD}$} & \phantom{..} & \multicolumn{4}{c}{$\overline{JB}$} \\
                      \cline{2-5} \cline{7-10} \cline{12-15} \cline{17-20}
                      & \textit{500} & \textit{2000} & \textit{8000} & \textit{32000} & & \textit{500} & \textit{2000} & \textit{8000} & \textit{32000} & & \textit{500} & \textit{2000} & \textit{8000} & \textit{32000} & & \textit{500} & \textit{2000} & \textit{8000} & \textit{32000} \\ \midrule
S & 0.834 & 0.616 & 0.472 & 0.370 &  & 0.825 & 0.606 & 0.462 & 0.361 &  & \textbf{0.105} & \textbf{0.090} & 0.078 & 0.069 &  & \textbf{1.935} & 2.120 & 2.075 & 1.951 \\ 
  S-W & 0.443 & 0.336 & 0.258 & 0.206 &  & \textbf{0.390} & \textbf{0.300} & \textbf{0.233} & 0.187 &  & 0.229 & 0.162 & 0.120 & 0.093 &  & 2.575 & 2.330 & 2.124 & 1.957 \\ 
  T & 0.443 & 0.335 & 0.258 & 0.206 &  & \textbf{0.390} & \textbf{0.300} & \textbf{0.233} & 0.187 &  & 0.229 & 0.163 & 0.120 & 0.093 &  & 2.552 & 2.278 & 2.130 & 1.938 \\ 
  X-F & \textbf{0.428} & \textbf{0.329} & \textbf{0.247} & 0.191 &  & 0.394 & 0.308 & \textbf{0.233} & 0.180 &  & 0.171 & 0.114 & 0.083 & 0.064 &  & 3.731 & 2.312 & 2.210 & 1.978 \\ 
  X-S & 0.501 & 0.399 & 0.307 & 0.232 &  & 0.456 & 0.375 & 0.291 & 0.220 &  & 0.213 & 0.136 & 0.097 & 0.073 &  & 6.602 & 2.762 & 2.298 & 2.113 \\ 
  X-C & 0.477 & 0.385 & 0.300 & 0.227 &  & 0.453 & 0.374 & 0.292 & 0.220 &  & 0.148 & 0.091 & \textbf{0.068} & \textbf{0.055} &  & 5.617 & \textbf{2.026} & \textbf{2.023} & \textbf{1.898} \\ 
  DR-F & 0.510 & 0.369 & 0.275 & 0.214 &  & 0.454 & 0.334 & 0.251 & 0.196 &  & 0.249 & 0.165 & 0.117 & 0.088 &  & 116.949 & 156.252 & 41.933 & 5.158 \\ 
  DR-S & 0.728 & 0.537 & 0.339 & 0.230 &  & 0.591 & 0.445 & 0.279 & 0.190 &  & 0.549 & 0.377 & 0.247 & 0.171 &  & 497.136 & 530.045 & 407.510 & 97.233 \\ 
  DR-C & 0.565 & 0.435 & 0.269 & \textbf{0.182} &  & 0.493 & 0.388 & 0.236 & \textbf{0.159} &  & 0.308 & 0.215 & 0.145 & 0.104 &  & 51.595 & 50.726 & 42.424 & 15.770 \\ 
  R-F & 0.537 & 0.426 & 0.349 & 0.293 &  & 0.454 & 0.364 & 0.303 & 0.258 &  & 0.337 & 0.250 & 0.192 & 0.151 &  & 8.764 & 13.754 & 7.349 & 2.839 \\ 
  R-S & 0.653 & 0.521 & 0.412 & 0.338 &  & 0.539 & 0.438 & 0.352 & 0.294 &  & 0.460 & 0.332 & 0.245 & 0.184 &  & 7.025 & 6.592 & 7.512 & 5.029 \\ 
  R-C & 0.524 & 0.440 & 0.361 & 0.303 &  & 0.469 & 0.402 & 0.333 & 0.282 &  & 0.246 & 0.185 & 0.142 & 0.113 &  & 2.700 & 2.900 & 2.732 & 2.318 \\ 
   \midrule \multicolumn{20}{c}{\parbox{18.75cm}{\setstretch{0.95}\footnotesize{\textit{Note:} The results for the $\overline{RMSE}$, $\overline{|BIAS|}$, $\overline{SD}$ and $\overline{JB}$ show the mean values of the root mean squared error, absolute bias, standard deviation and the Jarque-Bera test statistic of all $10'000$ CATE estimates from the validation sample. The critical values for the JB test statistic are 5.991 and 9.210 at the 5\% and 1\% level, respectively. Additionally, X-F, DR-F, R-F denote the full-sample versions of the meta-learners, while X-S, DR-S, R-S and X-C, DR-C, R-C denote the sample-splitting and cross-fitting versions, respectively. Bold numbers indicate the best performing meta-learner for given measure and sample size.}}}\\ \bottomrule
\end{tabular}
}
\end{table}
\vspace{0.25cm}
\begin{figure}[ht]
\caption{\label{bigfigure:CATE4}CATE Results for Simulation 4}\label{fig:unnamed-chunk-131}
{\centering \includegraphics[scale=0.8]{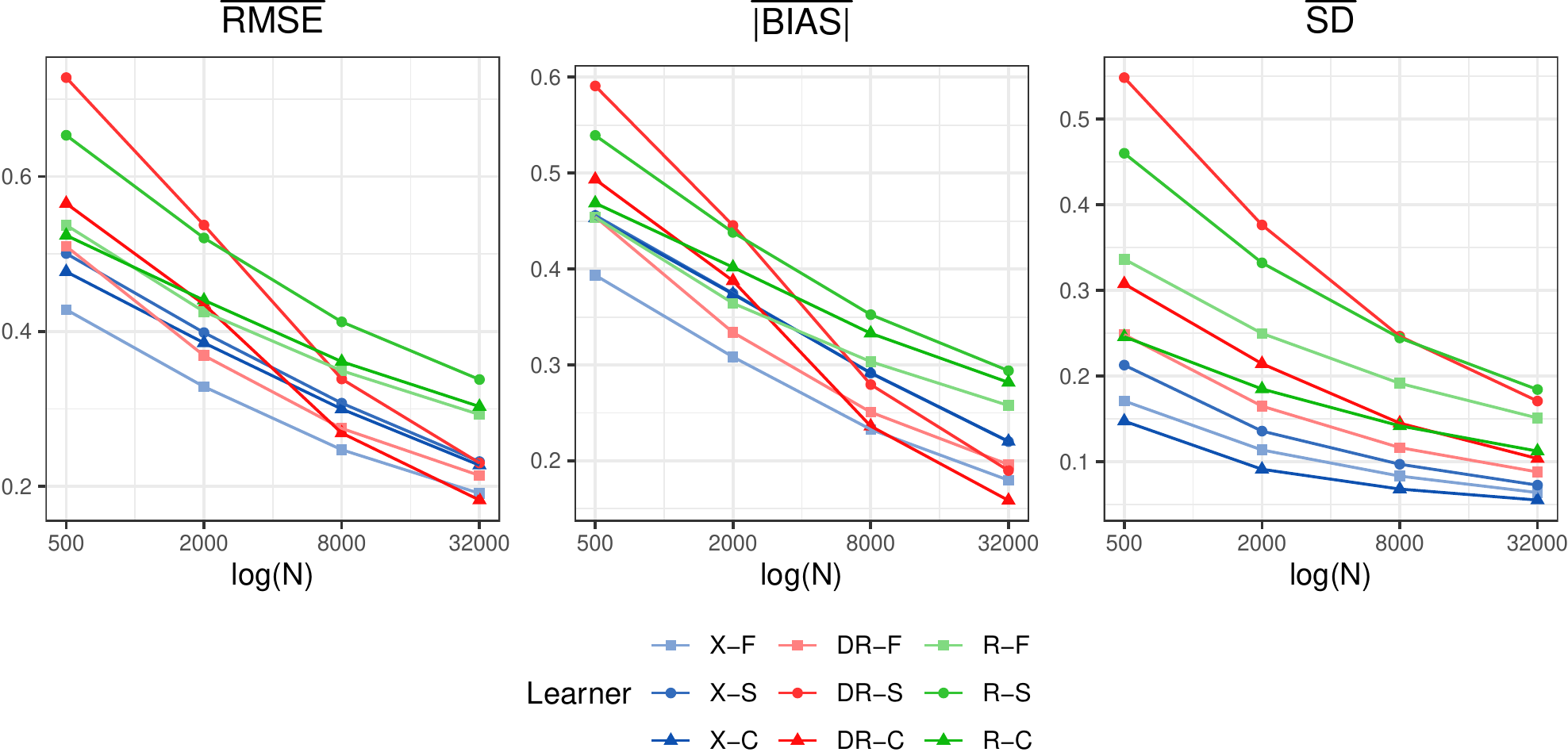} 

}
\vspace{0.25cm}
\caption*{\footnotesize{\textit{Note:} The results for $\overline{RMSE}$, $\overline{|BIAS|}$, and $\overline{SD}$ show the mean values of the root mean squared error, absolute bias, and standard deviation of all $10'000$ CATE estimates from the validation sample. The figure shows the results based on the increasing training samples of $\{500, 2'000, 8'000, 32'000\}$ observations displayed on the log scale. Additionally, X-F, DR-F, R-F denote the full-sample versions of the meta-learners, while X-S, DR-S, R-S and X-C, DR-C, R-C denote the sample-splitting and cross-fitting versions, respectively.}}
\end{figure}

\pagebreak

\subsubsection{Simulation 5: unbalanced treatment and linear CATE}\label{simulation-5-unbalanced-treatment-and-linear-cate-2}

\begin{table}[ht]
\centering
\caption{\label{bigtable:CATE5}CATE Results for Simulation 5} 
\scalebox{0.83}{
%\resizebox{\textwidth}{!}{
\small
\setlength{\tabcolsep}{2.25pt}
\begin{tabular}{lrrrrrrrrrrrrrrrrrrr}
  \toprule
  & \multicolumn{4}{c}{$\overline{RMSE}$} & \phantom{..} & \multicolumn{4}{c}{$\overline{|BIAS|}$} & \phantom{..} & \multicolumn{4}{c}{$\overline{SD}$} & \phantom{..} & \multicolumn{4}{c}{$\overline{JB}$} \\
                      \cline{2-5} \cline{7-10} \cline{12-15} \cline{17-20}
                      & \textit{500} & \textit{2000} & \textit{8000} & \textit{32000} & & \textit{500} & \textit{2000} & \textit{8000} & \textit{32000} & & \textit{500} & \textit{2000} & \textit{8000} & \textit{32000} & & \textit{500} & \textit{2000} & \textit{8000} & \textit{32000} \\ \midrule
S & 0.823 & 0.606 & 0.461 & 0.358 &  & 0.817 & 0.599 & 0.454 & 0.351 &  & \textbf{0.101} & 0.087 & 0.075 & 0.066 &  & \textbf{1.796} & 2.150 & 2.057 & 1.986 \\ 
  S-W & 0.305 & 0.244 & 0.196 & 0.164 &  & 0.255 & 0.209 & 0.170 & 0.145 &  & 0.222 & 0.159 & 0.117 & 0.089 &  & 2.457 & 2.189 & 2.054 & 1.957 \\ 
  T & 0.305 & 0.244 & 0.196 & 0.164 &  & 0.255 & 0.209 & 0.171 & 0.145 &  & 0.222 & 0.159 & 0.117 & 0.089 &  & 2.497 & 2.173 & 2.026 & \textbf{1.936} \\ 
  X-F & 0.237 & \textbf{0.178} & \textbf{0.137} & \textbf{0.109} &  & \textbf{0.200} & \textbf{0.154} & \textbf{0.120} & \textbf{0.097} &  & 0.164 & 0.110 & 0.078 & 0.058 &  & 3.329 & 2.228 & 2.102 & 2.022 \\ 
  X-S & 0.276 & 0.210 & 0.163 & 0.126 &  & 0.230 & 0.181 & 0.143 & 0.112 &  & 0.202 & 0.130 & 0.092 & 0.067 &  & 6.639 & 2.811 & 2.296 & 2.421 \\ 
  X-C & \textbf{0.231} & 0.182 & 0.144 & 0.114 &  & \textbf{0.200} & 0.165 & 0.132 & 0.105 &  & 0.139 & \textbf{0.086} & \textbf{0.061} & \textbf{0.046} &  & 4.474 & \textbf{2.014} & \textbf{2.004} & 2.037 \\ 
  DR-F & 0.314 & 0.248 & 0.203 & 0.166 &  & 0.258 & 0.212 & 0.179 & 0.148 &  & 0.237 & 0.159 & 0.112 & 0.084 &  & 123.780 & 364.063 & 249.515 & 26.116 \\ 
  DR-S & 0.556 & 0.413 & 0.298 & 0.215 &  & 0.428 & 0.322 & 0.239 & 0.176 &  & 0.515 & 0.362 & 0.242 & 0.167 &  & 453.484 & 685.910 & 651.725 & 174.087 \\ 
  DR-C & 0.354 & 0.280 & 0.217 & 0.162 &  & 0.287 & 0.232 & 0.185 & 0.140 &  & 0.289 & 0.205 & 0.140 & 0.099 &  & 50.509 & 61.849 & 72.770 & 22.771 \\ 
  R-F & 0.392 & 0.312 & 0.254 & 0.223 &  & 0.314 & 0.253 & 0.211 & 0.190 &  & 0.339 & 0.253 & 0.187 & 0.146 &  & 12.888 & 28.931 & 17.700 & 4.568 \\ 
  R-S & 0.500 & 0.388 & 0.305 & 0.248 &  & 0.398 & 0.311 & 0.248 & 0.206 &  & 0.457 & 0.336 & 0.246 & 0.179 &  & 9.107 & 10.173 & 16.684 & 12.581 \\ 
  R-C & 0.311 & 0.262 & 0.222 & 0.194 &  & 0.256 & 0.219 & 0.191 & 0.172 &  & 0.241 & 0.185 & 0.139 & 0.104 &  & 2.925 & 3.617 & 3.874 & 3.072 \\ 
   \midrule \multicolumn{20}{c}{\parbox{18.8cm}{\setstretch{0.95}\footnotesize{\textit{Note:} The results for the $\overline{RMSE}$, $\overline{|BIAS|}$, $\overline{SD}$ and $\overline{JB}$ show the mean values of the root mean squared error, absolute bias, standard deviation and the Jarque-Bera test statistic of all $10'000$ CATE estimates from the validation sample. The critical values for the JB test statistic are 5.991 and 9.210 at the 5\% and 1\% level, respectively. Additionally, X-F, DR-F, R-F denote the full-sample versions of the meta-learners, while X-S, DR-S, R-S and X-C, DR-C, R-C denote the sample-splitting and cross-fitting versions, respectively. Bold numbers indicate the best performing meta-learner for given measure and sample size.}}}\\ \bottomrule
\end{tabular}
}
\end{table}
\vspace{0.25cm}
\begin{figure}[ht]
\caption{\label{bigfigure:CATE5}CATE Results for Simulation 5}\label{fig:unnamed-chunk-133}
{\centering \includegraphics[scale=0.8]{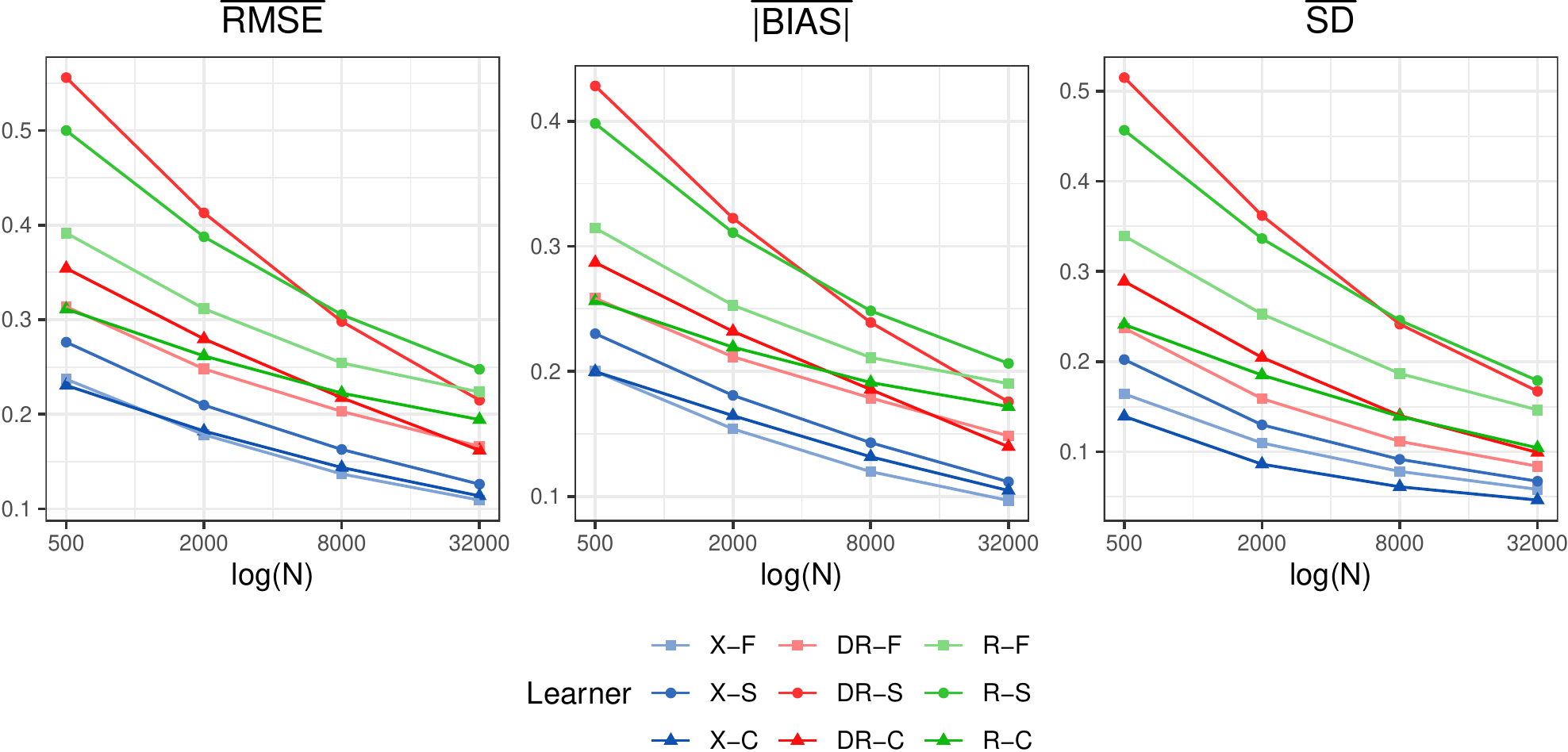} 

}
\vspace{0.25cm}
\caption*{\footnotesize{\textit{Note:} The results for $\overline{RMSE}$, $\overline{|BIAS|}$, and $\overline{SD}$ show the mean values of the root mean squared error, absolute bias, and standard deviation of all $10'000$ CATE estimates from the validation sample. The figure shows the results based on the increasing training samples of $\{500, 2'000, 8'000, 32'000\}$ observations displayed on the log scale. Additionally, X-F, DR-F, R-F denote the full-sample versions of the meta-learners, while X-S, DR-S, R-S and X-C, DR-C, R-C denote the sample-splitting and cross-fitting versions, respectively.}}
\end{figure}

\pagebreak

\begin{landscape}

\pagestyle{empty}%

\pagebreak

\subsection{Supplementary Results}\label{appendix-e-additional-results}

\ul{This appendix provides supplementary results based on additional performance measures, complementing those from Section} \ref{c1sec:performance-measures}. \ul{To
understand the simulation noise and thus the precision the average RMSE is measured with, we compute the standard error of the average RMSE following} \textcite{Knaus2021} as:
\[SE(\overline{RMSE}) = \sqrt{\frac{1}{R} \sum^R_{r=1} \bigg(\frac{1}{N^V}\sum^{N^V}_{i=1} \big( \tau(X_i) - \hat{\tau}^r(X_i) \big)^2 - \overline{RMSE} \bigg)^2}.\]
Additionally, besides the absolute bias, we evaluate also the bias without the absolute value given by:
\[BIAS \big(\hat{\tau}(X_i)\big)= \frac{1}{R} \sum^R_{r=1} \bigg( \tau(X_i) - \hat{\tau}^r(X_i) \bigg)\]
We further evaluate also the components of the Jarque-Bera statistic separately, namely the skewness, i.e. \(S\big(\hat{\tau}(X_i)\big)\) and the kurtosis, i.e. \(K\big(\hat{\tau}(X_i)\big)\) defined by:
\[S \big(\hat{\tau}(X_i)\big)= \frac{\frac{1}{R} \sum^R_{r=1} \big( \hat{\tau}^r(X_i) - \frac{1}{R} \sum^R_{r=1} \hat{\tau}^r(X_i) \big)^3}{\bigg(\frac{1}{R} \sum^R_{r=1} \big( \hat{\tau}^r(X_i) - \frac{1}{R} \sum^R_{r=1} \hat{\tau}^r(X_i) \big)^2\bigg)^{3/2}}
\quad \text{ and } \quad
K \big(\hat{\tau}(X_i)\big)= \frac{\frac{1}{R} \sum^R_{r=1} \big( \hat{\tau}^r(X_i) - \frac{1}{R} \sum^R_{r=1} \hat{\tau}^r(X_i) \big)^4}{\bigg(\frac{1}{R} \sum^R_{r=1} \big( \hat{\tau}^r(X_i) - \frac{1}{R} \sum^R_{r=1} \hat{\tau}^r(X_i) \big)^2\bigg)^{2}}.\] 
As in the main simulation results, we report the averages of the above measures over the validation sample \(N^V\). \ul{Complementary to the average values of the Jarque-Bera statistic presented in the main text, herein we report the share of CATEs for which the normality gets rejected at the 5\% level}. In order to further evaluate the performance on the replication level we compute the correlation between the true and the estimated treatment effects given by:
\[CORR= \frac{1}{R} \sum^R_{r=1} \bigg(\rho \big( \boldsymbol{\tau}, \hat{\boldsymbol{\tau}}^r \big) \bigg)\]
where \(\boldsymbol{\tau}\) is a vector of size \(N^V\) containing the true treatment effects from the validation sample and \(\hat{\boldsymbol{\tau}}^r\) is a vector of size \(N^V\) containing the estimated treatment effects for the validation sample at the replication \(r\), while \(\rho(\cdot)\) denotes the correlation function. Similarly, we compute also the variance ratio of the true and the estimated treatment effects as follows:
\[VARR= \frac{1}{R} \sum^R_{r=1} \bigg(\frac{Var(\hat{\boldsymbol{\tau}}^r)}{Var(\boldsymbol{\tau})} \bigg)\]
where \(Var(\cdot)\) denotes the variance. The full results including the main and the supplementary performance measures are listed in Tables \ref{bigtableplus:CATE1} - \ref{bigtableplus:CATEEmpirical} below.

\pagebreak

\subsubsection{Simulation 1: balanced treatment and constant zero CATE}\label{simulation-1-balanced-treatment-and-constant-zero-cate-2}

\begin{table}[ht]
\centering
\caption{\label{bigtableplus:CATE1}CATE Results for Simulation 1} 
\scalebox{0.8}{
\begin{tabular}{lrrrrrrrrrrrrrrrrrrrrrrrr}
  \toprule
  & \multicolumn{4}{c}{$\overline{RMSE}$} & & \multicolumn{4}{c}{$SE(\overline{RMSE})$} & & \multicolumn{4}{c}{$\overline{|BIAS|}$} & & \multicolumn{4}{c}{$\overline{BIAS}$} & & \multicolumn{4}{c}{$\overline{SD}$} \\
                      \cline{2-5} \cline{7-10} \cline{12-15} \cline{17-20} \cline{22-25}
                      & \textit{500} & \textit{2000} & \textit{8000} & \textit{32000} & & \textit{500} & \textit{2000} & \textit{8000} & \textit{32000} & & \textit{500} & \textit{2000} & \textit{8000} & \textit{32000} & & \textit{500} & \textit{2000} & \textit{8000} & \textit{32000} & & \textit{500} & \textit{2000} & \textit{8000} & \textit{32000} \\ \midrule
S & 0.008 & 0.009 & 0.013 & 0.018 &  & 0.005 & 0.005 & 0.006 & 0.004 &  & 0.005 & 0.006 & 0.010 & 0.014 &  & -0.003 & -0.004 & -0.006 & -0.008 &  & 0.007 & 0.008 & 0.012 & 0.016 \\ 
  S-W & 0.037 & 0.038 & 0.049 & 0.059 &  & 0.028 & 0.024 & 0.024 & 0.015 &  & 0.023 & 0.025 & 0.036 & 0.047 &  & -0.017 & -0.020 & -0.028 & -0.033 &  & 0.033 & 0.032 & 0.040 & 0.047 \\ 
  T & 0.225 & 0.168 & 0.128 & 0.101 &  & 0.046 & 0.024 & 0.013 & 0.007 &  & 0.180 & 0.135 & 0.103 & 0.082 &  & -0.087 & -0.072 & -0.060 & -0.048 &  & 0.206 & 0.149 & 0.109 & 0.083 \\ 
  X-F & 0.160 & 0.111 & 0.080 & 0.059 &  & 0.054 & 0.026 & 0.014 & 0.006 &  & 0.128 & 0.089 & 0.065 & 0.048 &  & -0.074 & -0.055 & -0.041 & -0.029 &  & 0.142 & 0.095 & 0.067 & 0.048 \\ 
  X-S & 0.186 & 0.127 & 0.091 & 0.067 &  & 0.071 & 0.034 & 0.018 & 0.008 &  & 0.149 & 0.103 & 0.074 & 0.055 &  & -0.090 & -0.071 & -0.053 & -0.037 &  & 0.162 & 0.106 & 0.073 & 0.053 \\ 
  X-C & 0.152 & 0.106 & 0.075 & 0.054 &  & 0.068 & 0.034 & 0.018 & 0.008 &  & 0.123 & 0.087 & 0.062 & 0.045 &  & -0.092 & -0.074 & -0.052 & -0.036 &  & 0.120 & 0.075 & 0.052 & 0.036 \\ 
  DR-F & 0.209 & 0.146 & 0.105 & 0.079 &  & 0.044 & 0.021 & 0.011 & 0.006 &  & 0.167 & 0.117 & 0.084 & 0.064 &  & -0.072 & -0.055 & -0.043 & -0.032 &  & 0.196 & 0.135 & 0.095 & 0.070 \\ 
  DR-S & 0.343 & 0.241 & 0.170 & 0.122 &  & 0.071 & 0.029 & 0.013 & 0.006 &  & 0.272 & 0.191 & 0.135 & 0.097 &  & -0.069 & -0.045 & -0.030 & -0.017 &  & 0.335 & 0.235 & 0.166 & 0.119 \\ 
  DR-C & 0.207 & 0.147 & 0.103 & 0.074 &  & 0.046 & 0.020 & 0.009 & 0.004 &  & 0.166 & 0.118 & 0.082 & 0.059 &  & -0.072 & -0.050 & -0.029 & -0.015 &  & 0.194 & 0.137 & 0.097 & 0.070 \\ 
  R-F & 0.261 & 0.192 & 0.145 & 0.114 &  & 0.039 & 0.020 & 0.012 & 0.006 &  & 0.208 & 0.153 & 0.116 & 0.092 &  & -0.077 & -0.066 & -0.058 & -0.048 &  & 0.249 & 0.180 & 0.132 & 0.102 \\ 
  R-S & 0.334 & 0.243 & 0.181 & 0.137 &  & 0.073 & 0.032 & 0.018 & 0.010 &  & 0.265 & 0.193 & 0.144 & 0.109 &  & -0.089 & -0.072 & -0.064 & -0.055 &  & 0.321 & 0.232 & 0.168 & 0.124 \\ 
  R-C & 0.208 & 0.156 & 0.118 & 0.092 &  & 0.050 & 0.026 & 0.015 & 0.008 &  & 0.166 & 0.125 & 0.096 & 0.075 &  & -0.092 & -0.077 & -0.065 & -0.054 &  & 0.186 & 0.135 & 0.098 & 0.072 \\ 
    \midrule & \multicolumn{4}{c}{$\overline{SKEW}$} & & \multicolumn{4}{c}{$\overline{KURT}$} & & \multicolumn{4}{c}{$JB\%$} & & \multicolumn{4}{c}{$CORR$} & & \multicolumn{4}{c}{$VARR$} \\
                      \cline{2-5} \cline{7-10} \cline{12-15} \cline{17-20} \cline{22-25}
                      & \textit{500} & \textit{2000} & \textit{8000} & \textit{32000} & & \textit{500} & \textit{2000} & \textit{8000} & \textit{32000} & & \textit{500} & \textit{2000} & \textit{8000} & \textit{32000} & & \textit{500} & \textit{2000} & \textit{8000} & \textit{32000} & & \textit{500} & \textit{2000} & \textit{8000} & \textit{32000} \\ \midrule S  & 2.638 & 1.809 & 0.929 & 0.307 &  & 17.334 & 10.674 & 5.654 & 3.539 &  & 1.000 & 1.000 & 1.000 & 0.539 &  &  &  &  &  &  &  &  &  &  \\ 
  S-W  & 2.744 & 2.171 & 1.150 & 0.339 &  & 17.212 & 12.179 & 6.003 & 3.493 &  & 1.000 & 1.000 & 1.000 & 0.586 &  &  &  &  &  &  &  &  &  &  \\ 
  T  & 0.005 & -0.009 & -0.004 & -0.008 &  & 3.016 & 3.038 & 3.006 & 2.983 &  & 0.058 & 0.075 & 0.052 & 0.044 &  &  &  &  &  &  &  &  &  &  \\ 
  X-F  & 0.010 & -0.030 & -0.001 & -0.022 &  & 2.990 & 3.058 & 3.002 & 2.985 &  & 0.028 & 0.091 & 0.055 & 0.051 &  &  &  &  &  &  &  &  &  &  \\ 
  X-S  & 0.017 & -0.023 & 0.003 & -0.008 &  & 2.951 & 3.023 & 3.029 & 3.003 &  & 0.036 & 0.066 & 0.063 & 0.060 &  &  &  &  &  &  &  &  &  &  \\ 
  X-C  & 0.002 & -0.040 & 0.003 & -0.018 &  & 2.985 & 3.061 & 3.006 & 2.986 &  & 0.010 & 0.086 & 0.052 & 0.049 &  &  &  &  &  &  &  &  &  &  \\ 
  DR-F  & -0.008 & -0.031 & -0.020 & -0.010 &  & 3.112 & 3.235 & 3.196 & 3.100 &  & 0.211 & 0.253 & 0.154 & 0.086 &  &  &  &  &  &  &  &  &  &  \\ 
  DR-S  & -0.009 & -0.040 & -0.060 & -0.040 &  & 3.150 & 3.308 & 3.410 & 3.359 &  & 0.271 & 0.359 & 0.274 & 0.164 &  &  &  &  &  &  &  &  &  &  \\ 
  DR-C  & -0.013 & -0.031 & -0.039 & -0.019 &  & 3.054 & 3.099 & 3.158 & 3.128 &  & 0.102 & 0.158 & 0.152 & 0.099 &  &  &  &  &  &  &  &  &  &  \\ 
  R-F  & 0.017 & 0.015 & 0.013 & 0.007 &  & 3.145 & 3.296 & 3.256 & 3.147 &  & 0.275 & 0.296 & 0.181 & 0.104 &  &  &  &  &  &  &  &  &  &  \\ 
  R-S  & 0.027 & 0.009 & 0.014 & 0.012 &  & 3.077 & 3.161 & 3.251 & 3.222 &  & 0.150 & 0.216 & 0.211 & 0.132 &  &  &  &  &  &  &  &  &  &  \\ 
  R-C  & 0.003 & 0.003 & 0.001 & 0.008 &  & 3.024 & 3.065 & 3.081 & 3.067 &  & 0.065 & 0.092 & 0.104 & 0.083 &  &  &  &  &  &  &  &  &  &  \\ 
   \midrule \multicolumn{25}{c}{\parbox{28.75cm}{\footnotesize{\textit{Note:} The results for the $\overline{RMSE}$, $\overline{|BIAS|}$, $\overline{BIAS}$, $\overline{SD}$, $\overline{SKEW}$, and $\overline{KURT}$ show the mean values of the root mean squared error, absolute bias, bias, standard deviation, skewness and kurtosis of all $10'000$ CATE estimates from the validation sample. $SE(\overline{RMSE})$ depicts the standard error of the average RMSE and $JB\%$ presents the share of CATEs for which the Jarque-Bera test has been rejected at the 5\% level. The results for $CORR$ and $VARR$ show the values of the correlation and variance ratio between the true and the estimated CATEs over all replications. Additionally, X-F, DR-F, R-F denote the full-sample versions of the meta-learners, while X-S, DR-S, R-S and X-C, DR-C, R-C denote the sample-splitting and cross-fitting versions, respectively.}}}\\ \bottomrule
\end{tabular}
}
\end{table}

\pagebreak

\subsubsection{Simulation 2: balanced treatment and complex nonlinear CATE}\label{simulation-2-balanced-treatment-and-complex-nonlinear-cate-3}

\begin{table}[ht]
\centering
\caption{\label{bigtableplus:CATE2}CATE Results for Simulation 2} 
\scalebox{0.8}{
\begin{tabular}{lrrrrrrrrrrrrrrrrrrrrrrrr}
  \toprule
  & \multicolumn{4}{c}{$\overline{RMSE}$} & & \multicolumn{4}{c}{$SE(\overline{RMSE})$} & & \multicolumn{4}{c}{$\overline{|BIAS|}$} & & \multicolumn{4}{c}{$\overline{BIAS}$} & & \multicolumn{4}{c}{$\overline{SD}$} \\
                      \cline{2-5} \cline{7-10} \cline{12-15} \cline{17-20} \cline{22-25}
                      & \textit{500} & \textit{2000} & \textit{8000} & \textit{32000} & & \textit{500} & \textit{2000} & \textit{8000} & \textit{32000} & & \textit{500} & \textit{2000} & \textit{8000} & \textit{32000} & & \textit{500} & \textit{2000} & \textit{8000} & \textit{32000} & & \textit{500} & \textit{2000} & \textit{8000} & \textit{32000} \\ \midrule
S & 0.527 & 0.442 & 0.374 & 0.326 &  & 0.114 & 0.092 & 0.075 & 0.064 &  & 0.522 & 0.434 & 0.366 & 0.317 &  & -0.369 & -0.258 & -0.190 & -0.143 &  & 0.055 & 0.068 & 0.066 & 0.064 \\ 
  S-W & 0.463 & 0.357 & 0.303 & 0.265 &  & 0.084 & 0.050 & 0.045 & 0.041 &  & 0.431 & 0.328 & 0.280 & 0.246 &  & -0.159 & -0.036 & -0.029 & -0.023 &  & 0.177 & 0.151 & 0.120 & 0.099 \\ 
  T & 0.434 & 0.358 & 0.303 & 0.265 &  & 0.056 & 0.049 & 0.045 & 0.041 &  & 0.392 & 0.328 & 0.280 & 0.246 &  & -0.040 & -0.034 & -0.029 & -0.023 &  & 0.204 & 0.154 & 0.120 & 0.099 \\ 
  X-F & 0.432 & 0.377 & 0.331 & 0.296 &  & 0.069 & 0.066 & 0.061 & 0.056 &  & 0.407 & 0.361 & 0.318 & 0.285 &  & -0.031 & -0.022 & -0.017 & -0.012 &  & 0.143 & 0.103 & 0.083 & 0.072 \\ 
  X-S & 0.460 & 0.411 & 0.362 & 0.321 &  & 0.071 & 0.071 & 0.067 & 0.061 &  & 0.432 & 0.393 & 0.349 & 0.310 &  & -0.041 & -0.029 & -0.021 & -0.015 &  & 0.160 & 0.110 & 0.085 & 0.073 \\ 
  X-C & 0.443 & 0.400 & 0.356 & 0.317 &  & 0.076 & 0.075 & 0.070 & 0.063 &  & 0.424 & 0.389 & 0.347 & 0.309 &  & -0.042 & -0.032 & -0.022 & -0.014 &  & 0.119 & 0.082 & 0.068 & 0.062 \\ 
  DR-F & 0.439 & 0.366 & 0.312 & 0.276 &  & 0.058 & 0.053 & 0.048 & 0.044 &  & 0.399 & 0.338 & 0.291 & 0.259 &  & -0.032 & -0.026 & -0.021 & -0.016 &  & 0.197 & 0.144 & 0.113 & 0.095 \\ 
  DR-S & 0.534 & 0.439 & 0.355 & 0.297 &  & 0.059 & 0.050 & 0.044 & 0.038 &  & 0.461 & 0.388 & 0.318 & 0.270 &  & -0.027 & -0.017 & -0.013 & -0.008 &  & 0.328 & 0.236 & 0.173 & 0.136 \\ 
  DR-C & 0.451 & 0.388 & 0.322 & 0.276 &  & 0.060 & 0.057 & 0.050 & 0.044 &  & 0.413 & 0.361 & 0.302 & 0.259 &  & -0.030 & -0.021 & -0.013 & -0.006 &  & 0.193 & 0.146 & 0.114 & 0.095 \\ 
  R-F & 0.458 & 0.373 & 0.307 & 0.266 &  & 0.052 & 0.045 & 0.039 & 0.034 &  & 0.404 & 0.333 & 0.277 & 0.241 &  & -0.039 & -0.034 & -0.030 & -0.027 &  & 0.251 & 0.188 & 0.146 & 0.122 \\ 
  R-S & 0.529 & 0.439 & 0.356 & 0.298 &  & 0.060 & 0.050 & 0.043 & 0.038 &  & 0.458 & 0.389 & 0.319 & 0.269 &  & -0.042 & -0.036 & -0.033 & -0.030 &  & 0.318 & 0.236 & 0.178 & 0.140 \\ 
  R-C & 0.449 & 0.388 & 0.322 & 0.274 &  & 0.061 & 0.058 & 0.050 & 0.044 &  & 0.413 & 0.361 & 0.301 & 0.256 &  & -0.045 & -0.040 & -0.034 & -0.028 &  & 0.187 & 0.145 & 0.116 & 0.097 \\ 
    \midrule & \multicolumn{4}{c}{$\overline{SKEW}$} & & \multicolumn{4}{c}{$\overline{KURT}$} & & \multicolumn{4}{c}{$JB\%$} & & \multicolumn{4}{c}{$CORR$} & & \multicolumn{4}{c}{$VARR$} \\
                      \cline{2-5} \cline{7-10} \cline{12-15} \cline{17-20} \cline{22-25}
                      & \textit{500} & \textit{2000} & \textit{8000} & \textit{32000} & & \textit{500} & \textit{2000} & \textit{8000} & \textit{32000} & & \textit{500} & \textit{2000} & \textit{8000} & \textit{32000} & & \textit{500} & \textit{2000} & \textit{8000} & \textit{32000} & & \textit{500} & \textit{2000} & \textit{8000} & \textit{32000} \\ \midrule S  & -1.172 & -0.282 & -0.138 & -0.087 &  & 4.608 & 3.106 & 3.021 & 3.003 &  & 1.000 & 0.832 & 0.206 & 0.087 &  & 0.430 & 0.737 & 0.850 & 0.890 &  & 772.103 & 31.143 & 11.674 & 6.871 \\ 
  S-W  & -0.820 & -0.054 & -0.003 & 0.003 &  & 3.494 & 3.015 & 3.027 & 2.999 &  & 1.000 & 0.116 & 0.075 & 0.058 &  & 0.490 & 0.728 & 0.846 & 0.896 &  & 51.036 & 5.794 & 4.379 & 3.494 \\ 
  T  & 0.000 & -0.007 & 0.002 & 0.004 &  & 3.030 & 3.043 & 3.025 & 3.002 &  & 0.070 & 0.085 & 0.072 & 0.057 &  & 0.465 & 0.722 & 0.846 & 0.896 &  & 6.865 & 5.692 & 4.382 & 3.492 \\ 
  X-F  & 0.001 & -0.012 & 0.007 & 0.001 &  & 3.010 & 3.026 & 2.984 & 2.971 &  & 0.045 & 0.067 & 0.045 & 0.043 &  & 0.458 & 0.723 & 0.847 & 0.891 &  & 20.826 & 14.657 & 8.854 & 5.962 \\ 
  X-S  & 0.008 & -0.014 & 0.011 & -0.007 &  & 2.931 & 3.022 & 3.006 & 2.981 &  & 0.043 & 0.058 & 0.062 & 0.050 &  & 0.266 & 0.553 & 0.785 & 0.868 &  & 25.303 & 22.204 & 13.868 & 8.249 \\ 
  X-C  & 0.000 & -0.022 & 0.007 & -0.006 &  & 2.993 & 3.029 & 2.983 & 2.975 &  & 0.015 & 0.065 & 0.045 & 0.045 &  & 0.391 & 0.677 & 0.835 & 0.886 &  & 55.123 & 33.303 & 15.678 & 8.554 \\ 
  DR-F  & -0.003 & -0.013 & -0.000 & 0.000 &  & 3.082 & 3.054 & 3.002 & 2.976 &  & 0.151 & 0.100 & 0.052 & 0.046 &  & 0.429 & 0.707 & 0.842 & 0.891 &  & 7.946 & 6.991 & 5.202 & 4.024 \\ 
  DR-S  & -0.002 & -0.021 & -0.009 & -0.010 &  & 3.135 & 3.199 & 3.090 & 3.016 &  & 0.247 & 0.249 & 0.113 & 0.063 &  & 0.224 & 0.455 & 0.718 & 0.841 &  & 3.595 & 4.565 & 4.468 & 3.649 \\ 
  DR-C  & -0.006 & -0.017 & -0.005 & 0.001 &  & 3.057 & 3.047 & 3.005 & 2.978 &  & 0.099 & 0.087 & 0.056 & 0.051 &  & 0.359 & 0.628 & 0.828 & 0.895 &  & 9.232 & 8.713 & 5.869 & 4.112 \\ 
  R-F  & 0.012 & 0.006 & 0.015 & 0.010 &  & 3.107 & 3.086 & 3.018 & 2.986 &  & 0.200 & 0.132 & 0.064 & 0.050 &  & 0.398 & 0.668 & 0.824 & 0.884 &  & 4.473 & 4.449 & 3.642 & 3.005 \\ 
  R-S  & 0.014 & 0.003 & 0.017 & 0.011 &  & 3.072 & 3.108 & 3.073 & 3.008 &  & 0.130 & 0.151 & 0.097 & 0.059 &  & 0.227 & 0.456 & 0.716 & 0.840 &  & 3.805 & 4.558 & 4.304 & 3.512 \\ 
  R-C  & 0.001 & 0.002 & 0.013 & 0.017 &  & 3.033 & 3.033 & 3.000 & 2.983 &  & 0.069 & 0.074 & 0.053 & 0.048 &  & 0.363 & 0.630 & 0.828 & 0.897 &  & 9.803 & 8.713 & 5.702 & 3.970 \\ 
   \midrule \multicolumn{25}{c}{\parbox{29.0cm}{\footnotesize{\textit{Note:} The results for the $\overline{RMSE}$, $\overline{|BIAS|}$, $\overline{BIAS}$, $\overline{SD}$, $\overline{SKEW}$, and $\overline{KURT}$ show the mean values of the root mean squared error, absolute bias, bias, standard deviation, skewness and kurtosis of all $10'000$ CATE estimates from the validation sample. $SE(\overline{RMSE})$ depicts the standard error of the average RMSE and $JB\%$ presents the share of CATEs for which the Jarque-Bera test has been rejected at the 5\% level. The results for $CORR$ and $VARR$ show the values of the correlation and variance ratio between the true and the estimated CATEs over all replications. Additionally, X-F, DR-F, R-F denote the full-sample versions of the meta-learners, while X-S, DR-S, R-S and X-C, DR-C, R-C denote the sample-splitting and cross-fitting versions, respectively.}}}\\ \bottomrule
\end{tabular}
}
\end{table}

\pagebreak

\subsubsection{Simulation 3: highly unbalanced treatment and constant non-zero CATE}\label{simulation-3-very-unbalanced-treatment-and-simple-cate-2}

\begin{table}[ht]
\centering
\caption{\label{bigtableplus:CATE3}CATE Results for Simulation 3} 
\scalebox{0.8}{
\begin{tabular}{lrrrrrrrrrrrrrrrrrrrrrrrr}
  \toprule
  & \multicolumn{4}{c}{$\overline{RMSE}$} & & \multicolumn{4}{c}{$SE(\overline{RMSE})$} & & \multicolumn{4}{c}{$\overline{|BIAS|}$} & & \multicolumn{4}{c}{$\overline{BIAS}$} & & \multicolumn{4}{c}{$\overline{SD}$} \\
                      \cline{2-5} \cline{7-10} \cline{12-15} \cline{17-20} \cline{22-25}
                      & \textit{500} & \textit{2000} & \textit{8000} & \textit{32000} & & \textit{500} & \textit{2000} & \textit{8000} & \textit{32000} & & \textit{500} & \textit{2000} & \textit{8000} & \textit{32000} & & \textit{500} & \textit{2000} & \textit{8000} & \textit{32000} & & \textit{500} & \textit{2000} & \textit{8000} & \textit{32000} \\ \midrule
S & 0.645 & 0.475 & 0.359 & 0.279 &  & 0.077 & 0.043 & 0.025 & 0.014 &  & 0.638 & 0.468 & 0.352 & 0.272 &  & 0.638 & 0.468 & 0.352 & 0.272 &  & 0.099 & 0.084 & 0.072 & 0.062 \\ 
  S-W & 0.246 & 0.191 & 0.146 & 0.111 &  & 0.059 & 0.025 & 0.015 & 0.009 &  & 0.197 & 0.154 & 0.119 & 0.091 &  & -0.045 & -0.050 & -0.042 & -0.032 &  & 0.233 & 0.163 & 0.121 & 0.090 \\ 
  T & 0.244 & 0.191 & 0.146 & 0.111 &  & 0.053 & 0.025 & 0.015 & 0.008 &  & 0.195 & 0.154 & 0.119 & 0.091 &  & -0.057 & -0.050 & -0.041 & -0.032 &  & 0.227 & 0.164 & 0.121 & 0.090 \\ 
  X-F & 0.180 & 0.123 & 0.090 & 0.068 &  & 0.060 & 0.024 & 0.012 & 0.006 &  & 0.144 & 0.098 & 0.072 & 0.054 &  & -0.041 & -0.033 & -0.025 & -0.016 &  & 0.175 & 0.118 & 0.085 & 0.061 \\ 
  X-S & 0.226 & 0.149 & 0.110 & 0.078 &  & 0.095 & 0.040 & 0.019 & 0.007 &  & 0.180 & 0.119 & 0.087 & 0.062 &  & -0.058 & -0.042 & -0.034 & -0.021 &  & 0.219 & 0.143 & 0.104 & 0.072 \\ 
  X-C & 0.159 & 0.102 & 0.073 & 0.054 &  & 0.074 & 0.031 & 0.015 & 0.007 &  & 0.127 & 0.081 & 0.059 & 0.043 &  & -0.053 & -0.043 & -0.033 & -0.021 &  & 0.150 & 0.092 & 0.064 & 0.046 \\ 
  DR-F & 0.287 & 0.202 & 0.146 & 0.110 &  & 0.058 & 0.022 & 0.012 & 0.007 &  & 0.222 & 0.158 & 0.116 & 0.089 &  & -0.049 & -0.040 & -0.029 & -0.017 &  & 0.279 & 0.188 & 0.129 & 0.093 \\ 
  DR-S & 0.649 & 0.502 & 0.334 & 0.218 &  & 0.204 & 0.093 & 0.039 & 0.018 &  & 0.475 & 0.365 & 0.250 & 0.168 &  & -0.052 & -0.038 & -0.025 & -0.007 &  & 0.645 & 0.498 & 0.329 & 0.213 \\ 
  DR-C & 0.364 & 0.290 & 0.197 & 0.131 &  & 0.075 & 0.032 & 0.016 & 0.008 &  & 0.282 & 0.223 & 0.153 & 0.104 &  & -0.048 & -0.036 & -0.026 & -0.005 &  & 0.359 & 0.283 & 0.189 & 0.124 \\ 
  R-F & 0.441 & 0.366 & 0.293 & 0.243 &  & 0.048 & 0.028 & 0.022 & 0.020 &  & 0.348 & 0.287 & 0.232 & 0.195 &  & 0.032 & 0.040 & 0.043 & 0.048 &  & 0.435 & 0.354 & 0.273 & 0.215 \\ 
  R-S & 0.573 & 0.461 & 0.366 & 0.285 &  & 0.141 & 0.057 & 0.034 & 0.025 &  & 0.453 & 0.363 & 0.288 & 0.227 &  & 0.032 & 0.038 & 0.042 & 0.044 &  & 0.570 & 0.454 & 0.353 & 0.264 \\ 
  R-C & 0.295 & 0.262 & 0.220 & 0.184 &  & 0.044 & 0.022 & 0.017 & 0.019 &  & 0.235 & 0.208 & 0.176 & 0.150 &  & 0.030 & 0.041 & 0.043 & 0.046 &  & 0.289 & 0.249 & 0.199 & 0.152 \\ 
    \midrule & \multicolumn{4}{c}{$\overline{SKEW}$} & & \multicolumn{4}{c}{$\overline{KURT}$} & & \multicolumn{4}{c}{$JB\%$} & & \multicolumn{4}{c}{$CORR$} & & \multicolumn{4}{c}{$VARR$} \\
                      \cline{2-5} \cline{7-10} \cline{12-15} \cline{17-20} \cline{22-25}
                      & \textit{500} & \textit{2000} & \textit{8000} & \textit{32000} & & \textit{500} & \textit{2000} & \textit{8000} & \textit{32000} & & \textit{500} & \textit{2000} & \textit{8000} & \textit{32000} & & \textit{500} & \textit{2000} & \textit{8000} & \textit{32000} & & \textit{500} & \textit{2000} & \textit{8000} & \textit{32000} \\ \midrule S  & 0.060 & 0.062 & 0.037 & 0.027 &  & 2.991 & 3.007 & 2.999 & 2.983 &  & 0.113 & 0.101 & 0.059 & 0.052 &  &  &  &  &  &  &  &  &  &  \\ 
  S-W  & -0.070 & -0.005 & -0.011 & -0.003 &  & 3.102 & 3.039 & 3.028 & 2.988 &  & 0.277 & 0.075 & 0.067 & 0.049 &  &  &  &  &  &  &  &  &  &  \\ 
  T  & -0.024 & -0.007 & -0.009 & -0.003 &  & 3.062 & 3.042 & 3.024 & 2.987 &  & 0.121 & 0.076 & 0.067 & 0.046 &  &  &  &  &  &  &  &  &  &  \\ 
  X-F  & -0.055 & -0.015 & -0.023 & -0.004 &  & 3.080 & 3.069 & 3.111 & 3.035 &  & 0.186 & 0.095 & 0.123 & 0.068 &  &  &  &  &  &  &  &  &  &  \\ 
  X-S  & -0.017 & 0.030 & -0.012 & -0.010 &  & 3.085 & 3.074 & 3.095 & 3.083 &  & 0.084 & 0.107 & 0.110 & 0.091 &  &  &  &  &  &  &  &  &  &  \\ 
  X-C  & -0.101 & -0.014 & -0.025 & -0.001 &  & 3.127 & 3.020 & 3.021 & 2.998 &  & 0.430 & 0.053 & 0.067 & 0.049 &  &  &  &  &  &  &  &  &  &  \\ 
  DR-F  & -0.006 & -0.035 & -0.041 & -0.020 &  & 5.349 & 4.944 & 3.984 & 3.340 &  & 0.949 & 0.674 & 0.337 & 0.147 &  &  &  &  &  &  &  &  &  &  \\ 
  DR-S  & -0.011 & -0.070 & -0.096 & -0.095 &  & 6.502 & 7.416 & 6.268 & 4.523 &  & 1.000 & 0.995 & 0.817 & 0.397 &  &  &  &  &  &  &  &  &  &  \\ 
  DR-C  & -0.033 & -0.047 & -0.056 & -0.055 &  & 4.033 & 4.468 & 4.258 & 3.538 &  & 0.996 & 0.928 & 0.557 & 0.229 &  &  &  &  &  &  &  &  &  &  \\ 
  R-F  & -0.026 & -0.025 & -0.012 & 0.006 &  & 3.325 & 3.511 & 3.375 & 3.159 &  & 0.658 & 0.542 & 0.282 & 0.121 &  &  &  &  &  &  &  &  &  &  \\ 
  R-S  & 0.035 & -0.003 & -0.018 & -0.005 &  & 3.222 & 3.382 & 3.464 & 3.292 &  & 0.468 & 0.520 & 0.378 & 0.183 &  &  &  &  &  &  &  &  &  &  \\ 
  R-C  & -0.015 & -0.011 & -0.006 & 0.009 &  & 3.067 & 3.129 & 3.159 & 3.086 &  & 0.117 & 0.173 & 0.178 & 0.105 &  &  &  &  &  &  &  &  &  &  \\ 
   \midrule \multicolumn{25}{c}{\parbox{28.5cm}{\footnotesize{\textit{Note:} The results for the $\overline{RMSE}$, $\overline{|BIAS|}$, $\overline{BIAS}$, $\overline{SD}$, $\overline{SKEW}$, and $\overline{KURT}$ show the mean values of the root mean squared error, absolute bias, bias, standard deviation, skewness and kurtosis of all $10'000$ CATE estimates from the validation sample. $SE(\overline{RMSE})$ depicts the standard error of the average RMSE and $JB\%$ presents the share of CATEs for which the Jarque-Bera test has been rejected at the 5\% level. The results for $CORR$ and $VARR$ show the values of the correlation and variance ratio between the true and the estimated CATEs over all replications. Additionally, X-F, DR-F, R-F denote the full-sample versions of the meta-learners, while X-S, DR-S, R-S and X-C, DR-C, R-C denote the sample-splitting and cross-fitting versions, respectively.}}}\\ \bottomrule
\end{tabular}
}
\end{table}

\pagebreak

\subsubsection{Simulation 4: unbalanced treatment and simple CATE}\label{simulation-4-unbalanced-treatment-and-constant-nonzero-cate-2}

\begin{table}[ht]
\centering
\caption{\label{bigtableplus:CATE4}CATE Results for Simulation 4} 
\scalebox{0.8}{
\begin{tabular}{lrrrrrrrrrrrrrrrrrrrrrrrr}
  \toprule
  & \multicolumn{4}{c}{$\overline{RMSE}$} & & \multicolumn{4}{c}{$SE(\overline{RMSE})$} & & \multicolumn{4}{c}{$\overline{|BIAS|}$} & & \multicolumn{4}{c}{$\overline{BIAS}$} & & \multicolumn{4}{c}{$\overline{SD}$} \\
                      \cline{2-5} \cline{7-10} \cline{12-15} \cline{17-20} \cline{22-25}
                      & \textit{500} & \textit{2000} & \textit{8000} & \textit{32000} & & \textit{500} & \textit{2000} & \textit{8000} & \textit{32000} & & \textit{500} & \textit{2000} & \textit{8000} & \textit{32000} & & \textit{500} & \textit{2000} & \textit{8000} & \textit{32000} & & \textit{500} & \textit{2000} & \textit{8000} & \textit{32000} \\ \midrule
S & 0.834 & 0.616 & 0.472 & 0.370 &  & 0.121 & 0.106 & 0.091 & 0.076 &  & 0.825 & 0.606 & 0.462 & 0.361 &  & 0.825 & 0.605 & 0.458 & 0.354 &  & 0.105 & 0.090 & 0.078 & 0.069 \\ 
  S-W & 0.443 & 0.336 & 0.258 & 0.206 &  & 0.049 & 0.032 & 0.024 & 0.020 &  & 0.390 & 0.300 & 0.233 & 0.187 &  & -0.077 & -0.071 & -0.059 & -0.048 &  & 0.229 & 0.162 & 0.120 & 0.093 \\ 
  T & 0.443 & 0.335 & 0.258 & 0.206 &  & 0.049 & 0.033 & 0.024 & 0.020 &  & 0.390 & 0.300 & 0.233 & 0.187 &  & -0.076 & -0.071 & -0.059 & -0.048 &  & 0.229 & 0.163 & 0.120 & 0.093 \\ 
  X-F & 0.428 & 0.329 & 0.247 & 0.191 &  & 0.040 & 0.026 & 0.017 & 0.011 &  & 0.394 & 0.308 & 0.233 & 0.180 &  & -0.061 & -0.052 & -0.038 & -0.027 &  & 0.171 & 0.114 & 0.083 & 0.064 \\ 
  X-S & 0.501 & 0.399 & 0.307 & 0.232 &  & 0.052 & 0.031 & 0.020 & 0.014 &  & 0.456 & 0.375 & 0.291 & 0.220 &  & -0.077 & -0.064 & -0.050 & -0.034 &  & 0.213 & 0.136 & 0.097 & 0.073 \\ 
  X-C & 0.477 & 0.385 & 0.300 & 0.227 &  & 0.033 & 0.021 & 0.015 & 0.010 &  & 0.453 & 0.374 & 0.292 & 0.220 &  & -0.079 & -0.067 & -0.048 & -0.034 &  & 0.148 & 0.091 & 0.068 & 0.055 \\ 
  DR-F & 0.510 & 0.369 & 0.275 & 0.214 &  & 0.046 & 0.033 & 0.020 & 0.013 &  & 0.454 & 0.334 & 0.251 & 0.196 &  & -0.064 & -0.055 & -0.038 & -0.027 &  & 0.249 & 0.165 & 0.117 & 0.088 \\ 
  DR-S & 0.728 & 0.537 & 0.339 & 0.230 &  & 0.122 & 0.059 & 0.030 & 0.016 &  & 0.591 & 0.445 & 0.279 & 0.190 &  & -0.055 & -0.050 & -0.034 & -0.014 &  & 0.549 & 0.377 & 0.247 & 0.171 \\ 
  DR-C & 0.565 & 0.435 & 0.269 & 0.182 &  & 0.043 & 0.035 & 0.021 & 0.012 &  & 0.493 & 0.388 & 0.236 & 0.159 &  & -0.069 & -0.054 & -0.030 & -0.010 &  & 0.308 & 0.215 & 0.145 & 0.104 \\ 
  R-F & 0.537 & 0.426 & 0.349 & 0.293 &  & 0.046 & 0.030 & 0.021 & 0.015 &  & 0.454 & 0.364 & 0.303 & 0.258 &  & -0.029 & -0.022 & -0.012 & -0.005 &  & 0.337 & 0.250 & 0.192 & 0.151 \\ 
  R-S & 0.653 & 0.521 & 0.412 & 0.338 &  & 0.092 & 0.049 & 0.032 & 0.023 &  & 0.539 & 0.438 & 0.352 & 0.294 &  & -0.035 & -0.032 & -0.019 & -0.014 &  & 0.460 & 0.332 & 0.245 & 0.184 \\ 
  R-C & 0.524 & 0.440 & 0.361 & 0.303 &  & 0.032 & 0.026 & 0.018 & 0.015 &  & 0.469 & 0.402 & 0.333 & 0.282 &  & -0.039 & -0.030 & -0.018 & -0.011 &  & 0.246 & 0.185 & 0.142 & 0.113 \\ 
    \midrule & \multicolumn{4}{c}{$\overline{SKEW}$} & & \multicolumn{4}{c}{$\overline{KURT}$} & & \multicolumn{4}{c}{$JB\%$} & & \multicolumn{4}{c}{$CORR$} & & \multicolumn{4}{c}{$VARR$} \\
                      \cline{2-5} \cline{7-10} \cline{12-15} \cline{17-20} \cline{22-25}
                      & \textit{500} & \textit{2000} & \textit{8000} & \textit{32000} & & \textit{500} & \textit{2000} & \textit{8000} & \textit{32000} & & \textit{500} & \textit{2000} & \textit{8000} & \textit{32000} & & \textit{500} & \textit{2000} & \textit{8000} & \textit{32000} & & \textit{500} & \textit{2000} & \textit{8000} & \textit{32000} \\ \midrule S  & 0.015 & 0.028 & 0.030 & 0.019 &  & 3.010 & 3.000 & 2.993 & 2.976 &  & 0.050 & 0.056 & 0.056 & 0.047 &  & 0.598 & 0.816 & 0.904 & 0.942 &  & 24.819 & 10.142 & 5.669 & 3.726 \\ 
  S-W  & -0.027 & -0.003 & -0.001 & -0.005 &  & 3.047 & 3.026 & 3.002 & 2.982 &  & 0.101 & 0.076 & 0.058 & 0.044 &  & 0.507 & 0.763 & 0.878 & 0.926 &  & 4.538 & 3.298 & 2.466 & 2.002 \\ 
  T  & -0.027 & -0.005 & -0.002 & -0.006 &  & 3.044 & 3.025 & 3.003 & 2.977 &  & 0.094 & 0.076 & 0.060 & 0.044 &  & 0.509 & 0.764 & 0.878 & 0.927 &  & 4.528 & 3.295 & 2.467 & 1.994 \\ 
  X-F  & -0.064 & -0.017 & -0.010 & -0.013 &  & 3.065 & 3.020 & 3.006 & 2.966 &  & 0.192 & 0.076 & 0.068 & 0.048 &  & 0.644 & 0.877 & 0.952 & 0.977 &  & 9.886 & 5.412 & 3.200 & 2.330 \\ 
  X-S  & -0.071 & -0.027 & -0.009 & -0.014 &  & 3.185 & 3.066 & 3.038 & 2.992 &  & 0.426 & 0.112 & 0.073 & 0.057 &  & 0.337 & 0.734 & 0.909 & 0.963 &  & 14.543 & 9.199 & 4.845 & 2.973 \\ 
  X-C  & -0.103 & -0.029 & -0.017 & -0.010 &  & 3.089 & 2.962 & 2.996 & 2.959 &  & 0.374 & 0.049 & 0.052 & 0.042 &  & 0.496 & 0.850 & 0.945 & 0.975 &  & 31.636 & 11.894 & 5.217 & 3.047 \\ 
  DR-F  & -0.058 & -0.059 & -0.034 & -0.011 &  & 3.848 & 3.745 & 3.319 & 3.054 &  & 0.798 & 0.443 & 0.200 & 0.073 &  & 0.242 & 0.717 & 0.894 & 0.949 &  & 4.993 & 4.567 & 3.159 & 2.400 \\ 
  DR-S  & -0.067 & -0.115 & -0.101 & -0.076 &  & 5.197 & 5.478 & 4.775 & 3.768 &  & 0.999 & 0.933 & 0.587 & 0.267 &  & 0.061 & 0.324 & 0.742 & 0.892 &  & 1.351 & 1.832 & 1.821 & 1.578 \\ 
  DR-C  & -0.026 & -0.054 & -0.059 & -0.047 &  & 3.683 & 3.762 & 3.578 & 3.234 &  & 0.958 & 0.710 & 0.371 & 0.146 &  & 0.107 & 0.498 & 0.870 & 0.948 &  & 3.482 & 4.144 & 2.451 & 1.768 \\ 
  R-F  & -0.018 & -0.007 & -0.002 & 0.001 &  & 3.215 & 3.280 & 3.152 & 3.031 &  & 0.443 & 0.342 & 0.165 & 0.067 &  & 0.251 & 0.524 & 0.712 & 0.825 &  & 2.355 & 2.844 & 2.901 & 2.686 \\ 
  R-S  & 0.049 & 0.024 & 0.010 & 0.003 &  & 3.182 & 3.217 & 3.212 & 3.106 &  & 0.412 & 0.314 & 0.211 & 0.110 &  & 0.105 & 0.308 & 0.564 & 0.739 &  & 1.873 & 2.403 & 2.835 & 2.879 \\ 
  R-C  & -0.028 & 0.000 & 0.002 & -0.004 &  & 3.053 & 3.076 & 3.060 & 3.016 &  & 0.110 & 0.118 & 0.098 & 0.066 &  & 0.174 & 0.476 & 0.731 & 0.846 &  & 5.354 & 5.714 & 4.761 & 3.722 \\ 
   \midrule \multicolumn{25}{c}{\parbox{28.75cm}{\footnotesize{\textit{Note:} The results for the $\overline{RMSE}$, $\overline{|BIAS|}$, $\overline{BIAS}$, $\overline{SD}$, $\overline{SKEW}$, and $\overline{KURT}$ show the mean values of the root mean squared error, absolute bias, bias, standard deviation, skewness and kurtosis of all $10'000$ CATE estimates from the validation sample. $SE(\overline{RMSE})$ depicts the standard error of the average RMSE and $JB\%$ presents the share of CATEs for which the Jarque-Bera test has been rejected at the 5\% level. The results for $CORR$ and $VARR$ show the values of the correlation and variance ratio between the true and the estimated CATEs over all replications. Additionally, X-F, DR-F, R-F denote the full-sample versions of the meta-learners, while X-S, DR-S, R-S and X-C, DR-C, R-C denote the sample-splitting and cross-fitting versions, respectively.}}}\\ \bottomrule
\end{tabular}
}
\end{table}

\pagebreak

\subsubsection{Simulation 5: unbalanced treatment and linear CATE}\label{simulation-5-unbalanced-treatment-and-linear-cate-3}

\begin{table}[ht]
\centering
\caption{\label{bigtableplus:CATE5}CATE Results for Simulation 5} 
\scalebox{0.8}{
\begin{tabular}{lrrrrrrrrrrrrrrrrrrrrrrrr}
  \toprule
  & \multicolumn{4}{c}{$\overline{RMSE}$} & & \multicolumn{4}{c}{$SE(\overline{RMSE})$} & & \multicolumn{4}{c}{$\overline{|BIAS|}$} & & \multicolumn{4}{c}{$\overline{BIAS}$} & & \multicolumn{4}{c}{$\overline{SD}$} \\
                      \cline{2-5} \cline{7-10} \cline{12-15} \cline{17-20} \cline{22-25}
                      & \textit{500} & \textit{2000} & \textit{8000} & \textit{32000} & & \textit{500} & \textit{2000} & \textit{8000} & \textit{32000} & & \textit{500} & \textit{2000} & \textit{8000} & \textit{32000} & & \textit{500} & \textit{2000} & \textit{8000} & \textit{32000} & & \textit{500} & \textit{2000} & \textit{8000} & \textit{32000} \\ \midrule
S & 0.823 & 0.606 & 0.461 & 0.358 &  & 0.069 & 0.043 & 0.031 & 0.028 &  & 0.817 & 0.599 & 0.454 & 0.351 &  & 0.817 & 0.599 & 0.454 & 0.351 &  & 0.101 & 0.087 & 0.075 & 0.066 \\ 
  S-W & 0.305 & 0.244 & 0.196 & 0.164 &  & 0.046 & 0.029 & 0.021 & 0.017 &  & 0.255 & 0.209 & 0.170 & 0.145 &  & -0.076 & -0.067 & -0.054 & -0.044 &  & 0.222 & 0.159 & 0.117 & 0.089 \\ 
  T & 0.305 & 0.244 & 0.196 & 0.164 &  & 0.046 & 0.029 & 0.021 & 0.017 &  & 0.255 & 0.209 & 0.171 & 0.145 &  & -0.076 & -0.068 & -0.055 & -0.044 &  & 0.222 & 0.159 & 0.117 & 0.089 \\ 
  X-F & 0.237 & 0.178 & 0.137 & 0.109 &  & 0.044 & 0.026 & 0.017 & 0.014 &  & 0.200 & 0.154 & 0.120 & 0.097 &  & -0.062 & -0.052 & -0.038 & -0.028 &  & 0.164 & 0.110 & 0.078 & 0.058 \\ 
  X-S & 0.276 & 0.210 & 0.163 & 0.126 &  & 0.068 & 0.032 & 0.021 & 0.015 &  & 0.230 & 0.181 & 0.143 & 0.112 &  & -0.074 & -0.065 & -0.050 & -0.034 &  & 0.202 & 0.130 & 0.092 & 0.067 \\ 
  X-C & 0.231 & 0.182 & 0.144 & 0.114 &  & 0.049 & 0.030 & 0.022 & 0.017 &  & 0.200 & 0.165 & 0.132 & 0.105 &  & -0.078 & -0.067 & -0.048 & -0.034 &  & 0.139 & 0.086 & 0.061 & 0.046 \\ 
  DR-F & 0.314 & 0.248 & 0.203 & 0.166 &  & 0.041 & 0.025 & 0.023 & 0.020 &  & 0.258 & 0.212 & 0.179 & 0.148 &  & -0.064 & -0.054 & -0.038 & -0.027 &  & 0.237 & 0.159 & 0.112 & 0.084 \\ 
  DR-S & 0.556 & 0.413 & 0.298 & 0.215 &  & 0.136 & 0.053 & 0.024 & 0.016 &  & 0.428 & 0.322 & 0.239 & 0.176 &  & -0.053 & -0.051 & -0.034 & -0.014 &  & 0.515 & 0.362 & 0.242 & 0.167 \\ 
  DR-C & 0.354 & 0.280 & 0.217 & 0.162 &  & 0.054 & 0.024 & 0.019 & 0.016 &  & 0.287 & 0.232 & 0.185 & 0.140 &  & -0.068 & -0.053 & -0.030 & -0.010 &  & 0.289 & 0.205 & 0.140 & 0.099 \\ 
  R-F & 0.392 & 0.312 & 0.254 & 0.223 &  & 0.038 & 0.023 & 0.020 & 0.019 &  & 0.314 & 0.253 & 0.211 & 0.190 &  & -0.021 & -0.016 & -0.009 & -0.003 &  & 0.339 & 0.253 & 0.187 & 0.146 \\ 
  R-S & 0.500 & 0.388 & 0.305 & 0.248 &  & 0.105 & 0.042 & 0.024 & 0.020 &  & 0.398 & 0.311 & 0.248 & 0.206 &  & -0.023 & -0.024 & -0.014 & -0.010 &  & 0.457 & 0.336 & 0.246 & 0.179 \\ 
  R-C & 0.311 & 0.262 & 0.222 & 0.194 &  & 0.037 & 0.022 & 0.021 & 0.023 &  & 0.256 & 0.219 & 0.191 & 0.172 &  & -0.028 & -0.021 & -0.013 & -0.007 &  & 0.241 & 0.185 & 0.139 & 0.104 \\ 
    \midrule & \multicolumn{4}{c}{$\overline{SKEW}$} & & \multicolumn{4}{c}{$\overline{KURT}$} & & \multicolumn{4}{c}{$JB\%$} & & \multicolumn{4}{c}{$CORR$} & & \multicolumn{4}{c}{$VARR$} \\
                      \cline{2-5} \cline{7-10} \cline{12-15} \cline{17-20} \cline{22-25}
                      & \textit{500} & \textit{2000} & \textit{8000} & \textit{32000} & & \textit{500} & \textit{2000} & \textit{8000} & \textit{32000} & & \textit{500} & \textit{2000} & \textit{8000} & \textit{32000} & & \textit{500} & \textit{2000} & \textit{8000} & \textit{32000} & & \textit{500} & \textit{2000} & \textit{8000} & \textit{32000} \\ \midrule S  & 0.004 & 0.030 & 0.027 & 0.026 &  & 2.981 & 2.998 & 2.991 & 2.984 &  & 0.034 & 0.062 & 0.054 & 0.050 &  & 0.211 & 0.415 & 0.561 & 0.672 &  & 6.435 & 4.582 & 4.002 & 3.502 \\ 
  S-W  & -0.024 & -0.005 & -0.007 & -0.012 &  & 3.045 & 3.023 & 3.000 & 2.988 &  & 0.087 & 0.067 & 0.054 & 0.048 &  & -0.035 & 0.115 & 0.332 & 0.524 &  & 1.094 & 1.571 & 2.232 & 2.654 \\ 
  T  & -0.024 & -0.005 & -0.006 & -0.008 &  & 3.045 & 3.025 & 2.996 & 2.981 &  & 0.091 & 0.065 & 0.053 & 0.047 &  & -0.034 & 0.115 & 0.331 & 0.524 &  & 1.093 & 1.573 & 2.230 & 2.652 \\ 
  X-F  & -0.056 & -0.015 & -0.008 & -0.012 &  & 3.061 & 3.020 & 3.015 & 2.987 &  & 0.163 & 0.067 & 0.059 & 0.051 &  & 0.202 & 0.481 & 0.728 & 0.849 &  & 2.966 & 3.413 & 3.225 & 2.767 \\ 
  X-S  & -0.063 & -0.025 & -0.013 & -0.016 &  & 3.202 & 3.087 & 3.032 & 3.009 &  & 0.442 & 0.120 & 0.075 & 0.063 &  & 0.093 & 0.270 & 0.563 & 0.780 &  & 3.071 & 3.496 & 3.862 & 3.255 \\ 
  X-C  & -0.081 & -0.026 & -0.011 & -0.021 &  & 3.101 & 2.960 & 2.999 & 2.983 &  & 0.260 & 0.047 & 0.051 & 0.054 &  & 0.154 & 0.400 & 0.713 & 0.861 &  & 7.940 & 7.788 & 6.045 & 3.922 \\ 
  DR-F  & -0.064 & -0.072 & -0.047 & -0.026 &  & 3.874 & 4.167 & 3.685 & 3.167 &  & 0.803 & 0.500 & 0.250 & 0.101 &  & -0.038 & 0.013 & 0.159 & 0.442 &  & 0.965 & 1.702 & 2.879 & 3.695 \\ 
  DR-S  & -0.073 & -0.121 & -0.132 & -0.113 &  & 5.095 & 5.798 & 5.324 & 4.050 &  & 0.998 & 0.933 & 0.601 & 0.285 &  & -0.012 & 0.016 & 0.101 & 0.345 &  & 0.251 & 0.361 & 0.691 & 1.205 \\ 
  DR-C  & -0.029 & -0.062 & -0.077 & -0.061 &  & 3.678 & 3.871 & 3.805 & 3.335 &  & 0.957 & 0.750 & 0.407 & 0.167 &  & -0.012 & 0.030 & 0.167 & 0.515 &  & 0.664 & 1.013 & 1.885 & 2.627 \\ 
  R-F  & -0.028 & -0.017 & -0.002 & -0.007 &  & 3.282 & 3.454 & 3.273 & 3.069 &  & 0.563 & 0.445 & 0.203 & 0.082 &  & 0.004 & 0.055 & 0.114 & 0.150 &  & 0.400 & 0.631 & 1.037 & 1.510 \\ 
  R-S  & 0.043 & 0.020 & 0.004 & -0.010 &  & 3.234 & 3.309 & 3.373 & 3.229 &  & 0.525 & 0.438 & 0.302 & 0.154 &  & -0.014 & 0.016 & 0.063 & 0.124 &  & 0.310 & 0.419 & 0.680 & 1.134 \\ 
  R-C  & -0.029 & -0.002 & -0.002 & -0.001 &  & 3.064 & 3.117 & 3.124 & 3.055 &  & 0.129 & 0.165 & 0.146 & 0.081 &  & -0.027 & 0.024 & 0.102 & 0.193 &  & 0.923 & 1.223 & 1.840 & 2.659 \\ 
   \midrule \multicolumn{25}{c}{\parbox{28.35cm}{\footnotesize{\textit{Note:} The results for the $\overline{RMSE}$, $\overline{|BIAS|}$, $\overline{BIAS}$, $\overline{SD}$, $\overline{SKEW}$, and $\overline{KURT}$ show the mean values of the root mean squared error, absolute bias, bias, standard deviation, skewness and kurtosis of all $10'000$ CATE estimates from the validation sample. $SE(\overline{RMSE})$ depicts the standard error of the average RMSE and $JB\%$ presents the share of CATEs for which the Jarque-Bera test has been rejected at the 5\% level. The results for $CORR$ and $VARR$ show the values of the correlation and variance ratio between the true and the estimated CATEs over all replications. Additionally, X-F, DR-F, R-F denote the full-sample versions of the meta-learners, while X-S, DR-S, R-S and X-C, DR-C, R-C denote the sample-splitting and cross-fitting versions, respectively.}}}\\ \bottomrule
\end{tabular}
}
\end{table}

\pagebreak

\subsubsection{Main Simulation: unbalanced treatment and nonlinear CATE}\label{simulation-6-unbalanced-treatment-and-nonlinear-cate-3}

\begin{table}[ht]
\centering
\caption{\label{bigtableplus:CATE6}CATE Results for Main Simulation} 
\scalebox{0.8}{
\begin{tabular}{lrrrrrrrrrrrrrrrrrrrrrrrr}
  \toprule
  & \multicolumn{4}{c}{$\overline{RMSE}$} & & \multicolumn{4}{c}{$SE(\overline{RMSE})$} & & \multicolumn{4}{c}{$\overline{|BIAS|}$} & & \multicolumn{4}{c}{$\overline{BIAS}$} & & \multicolumn{4}{c}{$\overline{SD}$} \\
                      \cline{2-5} \cline{7-10} \cline{12-15} \cline{17-20} \cline{22-25}
                      & \textit{500} & \textit{2000} & \textit{8000} & \textit{32000} & & \textit{500} & \textit{2000} & \textit{8000} & \textit{32000} & & \textit{500} & \textit{2000} & \textit{8000} & \textit{32000} & & \textit{500} & \textit{2000} & \textit{8000} & \textit{32000} & & \textit{500} & \textit{2000} & \textit{8000} & \textit{32000} \\ \midrule
S & 0.878 & 0.749 & 0.651 & 0.570 &  & 0.203 & 0.169 & 0.142 & 0.121 &  & 0.867 & 0.739 & 0.641 & 0.560 &  & 0.578 & 0.413 & 0.305 & 0.229 &  & 0.108 & 0.096 & 0.091 & 0.088 \\ 
  S-W & 0.765 & 0.634 & 0.533 & 0.462 &  & 0.123 & 0.107 & 0.093 & 0.082 &  & 0.717 & 0.602 & 0.508 & 0.443 &  & -0.135 & -0.121 & -0.099 & -0.081 &  & 0.261 & 0.190 & 0.149 & 0.125 \\ 
  T & 0.766 & 0.634 & 0.533 & 0.462 &  & 0.123 & 0.107 & 0.093 & 0.081 &  & 0.719 & 0.602 & 0.509 & 0.442 &  & -0.139 & -0.121 & -0.099 & -0.081 &  & 0.260 & 0.190 & 0.149 & 0.125 \\ 
  X-F & 0.743 & 0.618 & 0.517 & 0.442 &  & 0.128 & 0.111 & 0.095 & 0.082 &  & 0.711 & 0.597 & 0.500 & 0.427 &  & -0.124 & -0.102 & -0.077 & -0.060 &  & 0.200 & 0.141 & 0.117 & 0.103 \\ 
  X-S & 0.820 & 0.707 & 0.591 & 0.499 &  & 0.137 & 0.127 & 0.109 & 0.093 &  & 0.779 & 0.684 & 0.574 & 0.484 &  & -0.147 & -0.123 & -0.096 & -0.073 &  & 0.244 & 0.164 & 0.125 & 0.107 \\ 
  X-C & 0.794 & 0.693 & 0.582 & 0.494 &  & 0.144 & 0.132 & 0.112 & 0.095 &  & 0.770 & 0.680 & 0.571 & 0.482 &  & -0.151 & -0.126 & -0.095 & -0.072 &  & 0.171 & 0.114 & 0.097 & 0.092 \\ 
  DR-F & 0.817 & 0.659 & 0.542 & 0.463 &  & 0.126 & 0.112 & 0.097 & 0.085 &  & 0.764 & 0.627 & 0.518 & 0.443 &  & -0.116 & -0.095 & -0.067 & -0.049 &  & 0.285 & 0.194 & 0.149 & 0.126 \\ 
  DR-S & 1.053 & 0.825 & 0.579 & 0.445 &  & 0.133 & 0.097 & 0.076 & 0.064 &  & 0.906 & 0.731 & 0.521 & 0.403 &  & -0.102 & -0.085 & -0.053 & -0.021 &  & 0.640 & 0.433 & 0.281 & 0.206 \\ 
  DR-C & 0.880 & 0.727 & 0.523 & 0.409 &  & 0.118 & 0.112 & 0.088 & 0.072 &  & 0.809 & 0.680 & 0.490 & 0.383 &  & -0.118 & -0.088 & -0.049 & -0.017 &  & 0.359 & 0.255 & 0.179 & 0.143 \\ 
  R-F & 0.815 & 0.679 & 0.590 & 0.529 &  & 0.112 & 0.101 & 0.095 & 0.090 &  & 0.746 & 0.632 & 0.554 & 0.499 &  & -0.115 & -0.100 & -0.081 & -0.066 &  & 0.346 & 0.251 & 0.201 & 0.172 \\ 
  R-S & 0.932 & 0.788 & 0.659 & 0.580 &  & 0.120 & 0.110 & 0.100 & 0.095 &  & 0.833 & 0.721 & 0.613 & 0.546 &  & -0.126 & -0.117 & -0.095 & -0.081 &  & 0.468 & 0.333 & 0.243 & 0.195 \\ 
  R-C & 0.825 & 0.725 & 0.621 & 0.554 &  & 0.130 & 0.123 & 0.110 & 0.102 &  & 0.779 & 0.694 & 0.597 & 0.533 &  & -0.131 & -0.115 & -0.094 & -0.077 &  & 0.261 & 0.196 & 0.155 & 0.136 \\ 
    \midrule & \multicolumn{4}{c}{$\overline{SKEW}$} & & \multicolumn{4}{c}{$\overline{KURT}$} & & \multicolumn{4}{c}{$JB\%$} & & \multicolumn{4}{c}{$CORR$} & & \multicolumn{4}{c}{$VARR$} \\
                      \cline{2-5} \cline{7-10} \cline{12-15} \cline{17-20} \cline{22-25}
                      & \textit{500} & \textit{2000} & \textit{8000} & \textit{32000} & & \textit{500} & \textit{2000} & \textit{8000} & \textit{32000} & & \textit{500} & \textit{2000} & \textit{8000} & \textit{32000} & & \textit{500} & \textit{2000} & \textit{8000} & \textit{32000} & & \textit{500} & \textit{2000} & \textit{8000} & \textit{32000} \\ \midrule S  & 0.115 & 0.071 & 0.047 & 0.024 &  & 3.006 & 2.966 & 2.990 & 2.968 &  & 0.466 & 0.115 & 0.062 & 0.045 &  & 0.624 & 0.831 & 0.904 & 0.934 &  & 92.890 & 27.182 & 12.760 & 7.598 \\ 
  S-W  & -0.024 & -0.016 & -0.011 & -0.026 &  & 3.004 & 2.999 & 2.988 & 2.967 &  & 0.055 & 0.056 & 0.054 & 0.044 &  & 0.524 & 0.798 & 0.891 & 0.922 &  & 13.575 & 8.180 & 5.033 & 3.633 \\ 
  T  & -0.035 & -0.017 & -0.014 & -0.027 &  & 3.035 & 2.996 & 2.986 & 2.967 &  & 0.097 & 0.055 & 0.051 & 0.044 &  & 0.514 & 0.798 & 0.891 & 0.922 &  & 13.471 & 8.176 & 5.034 & 3.626 \\ 
  X-F  & -0.060 & -0.030 & -0.017 & -0.023 &  & 3.054 & 2.984 & 2.985 & 2.950 &  & 0.171 & 0.064 & 0.049 & 0.040 &  & 0.664 & 0.894 & 0.950 & 0.967 &  & 24.558 & 11.070 & 6.021 & 4.070 \\ 
  X-S  & -0.067 & -0.030 & -0.028 & -0.019 &  & 3.143 & 3.047 & 3.002 & 2.964 &  & 0.323 & 0.106 & 0.060 & 0.045 &  & 0.367 & 0.754 & 0.919 & 0.957 &  & 38.577 & 21.499 & 9.669 & 5.561 \\ 
  X-C  & -0.079 & -0.042 & -0.021 & -0.021 &  & 3.064 & 2.953 & 2.987 & 2.944 &  & 0.209 & 0.066 & 0.049 & 0.037 &  & 0.530 & 0.852 & 0.946 & 0.966 &  & 79.900 & 27.087 & 10.181 & 5.644 \\ 
  DR-F  & -0.106 & -0.073 & -0.034 & -0.022 &  & 3.915 & 3.381 & 3.081 & 2.982 &  & 0.827 & 0.366 & 0.119 & 0.052 &  & 0.317 & 0.770 & 0.912 & 0.948 &  & 13.371 & 10.505 & 6.196 & 4.273 \\ 
  DR-S  & -0.143 & -0.216 & -0.148 & -0.084 &  & 5.350 & 5.320 & 4.033 & 3.317 &  & 1.000 & 0.947 & 0.526 & 0.189 &  & 0.095 & 0.406 & 0.812 & 0.918 &  & 3.696 & 4.807 & 3.992 & 2.940 \\ 
  DR-C  & -0.080 & -0.111 & -0.075 & -0.044 &  & 3.678 & 3.629 & 3.243 & 3.034 &  & 0.960 & 0.668 & 0.243 & 0.085 &  & 0.162 & 0.580 & 0.899 & 0.950 &  & 9.167 & 9.660 & 4.855 & 3.102 \\ 
  R-F  & -0.009 & -0.006 & -0.013 & -0.018 &  & 3.126 & 3.077 & 3.012 & 2.982 &  & 0.233 & 0.128 & 0.063 & 0.049 &  & 0.368 & 0.692 & 0.832 & 0.890 &  & 7.963 & 7.751 & 6.147 & 4.980 \\ 
  R-S  & 0.031 & 0.018 & 0.002 & -0.009 &  & 3.107 & 3.097 & 3.048 & 2.992 &  & 0.207 & 0.151 & 0.082 & 0.054 &  & 0.166 & 0.449 & 0.732 & 0.846 &  & 6.605 & 7.982 & 7.305 & 6.036 \\ 
  R-C  & -0.021 & -0.006 & -0.014 & -0.020 &  & 3.043 & 3.018 & 3.003 & 2.966 &  & 0.088 & 0.063 & 0.052 & 0.042 &  & 0.271 & 0.624 & 0.843 & 0.902 &  & 17.941 & 15.725 & 9.647 & 6.705 \\ 
   \midrule \multicolumn{25}{c}{\parbox{29.0cm}{\footnotesize{\textit{Note:} The results for the $\overline{RMSE}$, $\overline{|BIAS|}$, $\overline{BIAS}$, $\overline{SD}$, $\overline{SKEW}$, and $\overline{KURT}$ show the mean values of the root mean squared error, absolute bias, bias, standard deviation, skewness and kurtosis of all $10'000$ CATE estimates from the validation sample. $SE(\overline{RMSE})$ depicts the standard error of the average RMSE and $JB\%$ presents the share of CATEs for which the Jarque-Bera test has been rejected at the 5\% level. The results for $CORR$ and $VARR$ show the values of the correlation and variance ratio between the true and the estimated CATEs over all replications. Additionally, X-F, DR-F, R-F denote the full-sample versions of the meta-learners, while X-S, DR-S, R-S and X-C, DR-C, R-C denote the sample-splitting and cross-fitting versions, respectively.}}}\\ \bottomrule
\end{tabular}
}
\end{table}

\pagebreak

\subsubsection{Semi-synthetic Simulation}\label{empirical-simulation}

\begin{table}[ht]
\centering
\caption{\label{bigtableplus:CATEEmpirical}CATE Results for Semi-synthetic Simulation} 
\scalebox{0.8}{
\begin{tabular}{lrrrrrrrrrrrrrrrrrrr}
  \toprule
  & \multicolumn{3}{c}{$\overline{RMSE}$} & & \multicolumn{3}{c}{$SE(\overline{RMSE})$} & & \multicolumn{3}{c}{$\overline{|BIAS|}$} & & \multicolumn{3}{c}{$\overline{BIAS}$} & & \multicolumn{3}{c}{$\overline{SD}$} \\
                      \cline{2-4} \cline{6-8} \cline{10-12} \cline{14-16} \cline{18-20}
                      & \textit{500} & \textit{2000} & \textit{8000} & & \textit{500} & \textit{2000} & \textit{8000} & & \textit{500} & \textit{2000} & \textit{8000} & & \textit{500} & \textit{2000} & \textit{8000} & & \textit{500} & \textit{2000} & \textit{8000} \\ \midrule
S & 0.175 & 0.127 & 0.093 &  & 0.025 & 0.015 & 0.009 &  & 0.171 & 0.121 & 0.090 &  & 0.171 & 0.119 & 0.085 &  & 0.035 & 0.035 & 0.023 \\ 
  S-W & 0.131 & 0.109 & 0.078 &  & 0.031 & 0.014 & 0.011 &  & 0.106 & 0.090 & 0.070 &  & -0.011 & -0.050 & -0.043 &  & 0.121 & 0.084 & 0.037 \\ 
  T & 0.150 & 0.111 & 0.079 &  & 0.027 & 0.014 & 0.011 &  & 0.122 & 0.092 & 0.071 &  & -0.063 & -0.053 & -0.044 &  & 0.127 & 0.084 & 0.037 \\ 
  X-F & 0.112 & 0.082 & 0.056 &  & 0.029 & 0.014 & 0.011 &  & 0.092 & 0.069 & 0.052 &  & -0.056 & -0.045 & -0.036 &  & 0.089 & 0.056 & 0.021 \\ 
  X-S & 0.129 & 0.093 & 0.069 &  & 0.041 & 0.019 & 0.011 &  & 0.105 & 0.078 & 0.060 &  & -0.065 & -0.054 & -0.043 &  & 0.104 & 0.067 & 0.040 \\ 
  X-C & 0.103 & 0.077 & 0.055 &  & 0.035 & 0.018 & 0.013 &  & 0.087 & 0.067 & 0.052 &  & -0.065 & -0.054 & -0.043 &  & 0.072 & 0.044 & 0.017 \\ 
  DR-F & 0.147 & 0.105 & 0.070 &  & 0.026 & 0.014 & 0.010 &  & 0.119 & 0.087 & 0.063 &  & -0.061 & -0.051 & -0.042 &  & 0.125 & 0.078 & 0.033 \\ 
  DR-S & 0.256 & 0.180 & 0.123 &  & 0.055 & 0.023 & 0.011 &  & 0.201 & 0.143 & 0.101 &  & -0.066 & -0.056 & -0.047 &  & 0.242 & 0.162 & 0.097 \\ 
  DR-C & 0.159 & 0.116 & 0.078 &  & 0.031 & 0.015 & 0.011 &  & 0.128 & 0.096 & 0.071 &  & -0.068 & -0.057 & -0.046 &  & 0.135 & 0.088 & 0.037 \\ 
  R-F & 0.183 & 0.131 & 0.089 &  & 0.022 & 0.011 & 0.009 &  & 0.146 & 0.107 & 0.078 &  & -0.051 & -0.043 & -0.034 &  & 0.167 & 0.109 & 0.051 \\ 
  R-S & 0.237 & 0.174 & 0.123 &  & 0.046 & 0.021 & 0.011 &  & 0.189 & 0.140 & 0.100 &  & -0.058 & -0.049 & -0.042 &  & 0.224 & 0.158 & 0.099 \\ 
  R-C & 0.144 & 0.109 & 0.076 &  & 0.026 & 0.013 & 0.010 &  & 0.117 & 0.091 & 0.068 &  & -0.058 & -0.050 & -0.040 &  & 0.123 & 0.084 & 0.037 \\ 
    \midrule & \multicolumn{3}{c}{$\overline{SKEW}$} & & \multicolumn{3}{c}{$\overline{KURT}$} & & \multicolumn{3}{c}{$JB\%$} & & \multicolumn{3}{c}{$CORR$} & & \multicolumn{3}{c}{$VARR$} \\
                      \cline{2-4} \cline{6-8} \cline{10-12} \cline{14-16} \cline{18-20}
                      & \textit{500} & \textit{2000} & \textit{8000} & & \textit{500} & \textit{2000} & \textit{8000} & & \textit{500} & \textit{2000} & \textit{8000} & & \textit{500} & \textit{2000} & \textit{8000} & & \textit{500} & \textit{2000} & \textit{8000} \\ \midrule S  & 0.671 & 0.178 & 0.031 &  & 3.310 & 2.998 & 2.988 &  & 1.000 & 0.511 & 0.051 &  & 0.050 & 0.135 & 0.328 &  & 10.481 & 2.315 & 1.666 \\ 
  S-W  & 0.394 & 0.014 & 0.007 &  & 2.838 & 2.979 & 2.984 &  & 1.000 & 0.049 & 0.040 &  & 0.061 & 0.162 & 0.359 &  & 0.562 & 0.331 & 0.448 \\ 
  T  & 0.002 & 0.007 & 0.001 &  & 3.001 & 2.986 & 2.995 &  & 0.055 & 0.056 & 0.058 &  & 0.058 & 0.158 & 0.357 &  & 0.200 & 0.323 & 0.447 \\ 
  X-F  & 0.007 & 0.013 & 0.007 &  & 3.006 & 2.981 & 2.999 &  & 0.053 & 0.049 & 0.055 &  & 0.103 & 0.223 & 0.458 &  & 0.542 & 0.748 & 0.912 \\ 
  X-S  & 0.020 & 0.022 & 0.007 &  & 3.024 & 3.045 & 2.993 &  & 0.063 & 0.100 & 0.045 &  & 0.058 & 0.134 & 0.297 &  & 0.544 & 0.716 & 0.981 \\ 
  X-C  & 0.009 & 0.011 & 0.007 &  & 2.987 & 2.964 & 2.982 &  & 0.028 & 0.027 & 0.049 &  & 0.094 & 0.197 & 0.393 &  & 1.327 & 1.527 & 1.665 \\ 
  DR-F  & 0.012 & 0.008 & 0.001 &  & 3.100 & 3.025 & 2.986 &  & 0.209 & 0.105 & 0.054 &  & 0.059 & 0.139 & 0.372 &  & 0.209 & 0.388 & 0.645 \\ 
  DR-S  & 0.013 & 0.016 & 0.006 &  & 3.678 & 3.476 & 3.130 &  & 0.906 & 0.472 & 0.161 &  & 0.031 & 0.066 & 0.161 &  & 0.065 & 0.110 & 0.220 \\ 
  DR-C  & 0.018 & 0.014 & -0.009 &  & 3.158 & 3.095 & 3.055 &  & 0.318 & 0.190 & 0.098 &  & 0.053 & 0.113 & 0.251 &  & 0.185 & 0.309 & 0.564 \\ 
  R-F  & -0.012 & -0.009 & -0.006 &  & 3.068 & 3.042 & 2.990 &  & 0.179 & 0.125 & 0.047 &  & 0.072 & 0.145 & 0.306 &  & 0.105 & 0.188 & 0.320 \\ 
  R-S  & -0.002 & -0.015 & -0.006 &  & 3.090 & 3.083 & 3.052 &  & 0.166 & 0.154 & 0.109 &  & 0.040 & 0.080 & 0.174 &  & 0.074 & 0.116 & 0.212 \\ 
  R-C  & 0.002 & -0.006 & -0.007 &  & 3.004 & 2.987 & 3.023 &  & 0.053 & 0.060 & 0.062 &  & 0.068 & 0.139 & 0.275 &  & 0.219 & 0.333 & 0.545 \\ 
   \midrule \multicolumn{20}{c}{\parbox{21.75cm}{\footnotesize{\textit{Note:} The results for the $\overline{RMSE}$, $\overline{|BIAS|}$, $\overline{BIAS}$, $\overline{SD}$, $\overline{SKEW}$, and $\overline{KURT}$ show the mean values of the root mean squared error, absolute bias, bias, standard deviation, skewness and kurtosis of all $1'000$ CATE estimates from the validation sample. $SE(\overline{RMSE})$ depicts the standard error of the average RMSE and $JB\%$ presents the share of CATEs for which the Jarque-Bera test has been rejected at the 5\% level. The results for $CORR$ and $VARR$ show the values of the correlation and variance ratio between the true and the estimated CATEs over all replications. Additionally, X-F, DR-F, R-F denote the full-sample versions of the meta-learners, while X-S, DR-S, R-S and X-C, DR-C, R-C denote the sample-splitting and cross-fitting versions, respectively.}}}\\ \bottomrule
\end{tabular}
}
\end{table}

\end{landscape}

\pagebreak

\section{Computation Time}\label{appendix-d-computation-time}

In order to assess the computational trade-offs among different
estimation schemes as well as different meta-learners we evaluate the
computational time for each meta-learner and each estimation scheme for
each sample size over 10 replications of the Main Simulation to illustrate the
performance. The results are summarized in Table \ref{bigtabletime:time6} and Figure \ref{fig:bigplot_time} below.

\subsection{Main Simulation: unbalanced treatment and nonlinear CATE}\label{simulation-6-unbalanced-treatment-and-nonlinear-cate-2}

\begin{table}[ht]
\centering
\caption{\label{bigtabletime:time6}Computation Time Results for Main Simulation} 
\scalebox{0.83}{
%\resizebox{\textwidth}{!}{
\small
\setlength{\tabcolsep}{2.25pt}
\begin{tabular}{lrrrrrrrrrrrrrrrrrrr}
  \toprule
  & \multicolumn{4}{c}{MEAN} & \phantom{..} & \multicolumn{4}{c}{SD} & \phantom{..} & \multicolumn{4}{c}{MIN} & \phantom{..} & \multicolumn{4}{c}{MAX} \\
                      \cline{2-5} \cline{7-10} \cline{12-15} \cline{17-20}
                      & \textit{500} & \textit{2000} & \textit{8000} & \textit{32000} & & \textit{500} & \textit{2000} & \textit{8000} & \textit{32000} & & \textit{500} & \textit{2000} & \textit{8000} & \textit{32000} & & \textit{500} & \textit{2000} & \textit{8000} & \textit{32000} \\ \midrule
S & 1.492 & 8.786 & 53.385 & 252.165 &  & 0.039 & 0.991 & 9.777 & 0.551 &  & 1.440 & 7.110 & 43.000 & 251.050 &  & 1.560 & 9.860 & 67.790 & 252.810 \\ 
  S-W & 1.416 & 6.730 & 39.117 & 263.195 &  & 0.051 & 1.015 & 5.804 & 4.636 &  & 1.330 & 4.890 & 31.920 & 250.630 &  & 1.500 & 8.050 & 49.950 & 267.140 \\ 
  T & 1.203 & 6.168 & 38.933 & 238.932 &  & 0.043 & 0.880 & 5.982 & 26.988 &  & 1.110 & 4.460 & 29.530 & 162.260 &  & 1.260 & 7.540 & 47.340 & 249.560 \\ 
  X-F & 2.894 & 16.512 & 92.803 & 658.892 &  & 0.081 & 1.303 & 10.148 & 49.316 &  & 2.770 & 14.070 & 80.950 & 531.960 &  & 2.980 & 17.590 & 106.570 & 687.320 \\ 
  X-S & 0.915 & 4.233 & 28.863 & 185.232 &  & 0.074 & 0.380 & 3.061 & 21.873 &  & 0.760 & 3.500 & 25.830 & 145.250 &  & 1.000 & 4.640 & 35.250 & 215.630 \\ 
  X-C & 3.027 & 14.353 & 89.644 & 627.236 &  & 0.265 & 0.657 & 9.687 & 91.057 &  & 2.790 & 13.300 & 79.280 & 405.540 &  & 3.470 & 15.720 & 103.810 & 721.920 \\ 
  DR-F & 2.262 & 15.998 & 94.615 & 576.230 &  & 0.129 & 0.696 & 16.618 & 120.102 &  & 2.080 & 14.920 & 57.130 & 323.300 &  & 2.490 & 17.040 & 113.820 & 676.650 \\ 
  DR-S & 0.836 & 4.292 & 30.218 & 214.072 &  & 0.189 & 0.345 & 5.321 & 36.595 &  & 0.580 & 3.780 & 26.640 & 160.380 &  & 1.200 & 4.940 & 42.280 & 277.320 \\ 
  DR-C & 2.684 & 17.272 & 105.261 & 664.728 &  & 0.596 & 3.941 & 14.058 & 60.375 &  & 2.150 & 12.770 & 88.690 & 572.030 &  & 4.300 & 22.850 & 128.400 & 744.670 \\ 
  R-F & 2.058 & 10.588 & 78.830 & 530.529 &  & 0.594 & 1.430 & 18.923 & 78.864 &  & 0.840 & 8.750 & 28.900 & 354.890 &  & 2.450 & 13.020 & 91.270 & 603.520 \\ 
  R-S & 0.919 & 6.514 & 31.234 & 208.910 &  & 0.429 & 2.084 & 7.845 & 49.782 &  & 0.530 & 4.420 & 21.340 & 154.770 &  & 2.100 & 10.440 & 43.750 & 308.480 \\ 
  R-C & 2.177 & 11.934 & 72.912 & 435.450 &  & 0.080 & 1.230 & 23.239 & 140.374 &  & 2.020 & 9.240 & 53.290 & 312.420 &  & 2.250 & 12.970 & 134.670 & 780.560 \\ 
   \midrule \multicolumn{20}{c}{\parbox{18.9cm}{\setstretch{0.95}\footnotesize{\textit{Note:} The results for the MEAN, SD, MIN, and MAX show the values of the mean, standard deviation, minimum and maximum for the computation time in seconds based on 10 simulation replications. The computation time includes both the estimation as well as the prediction task. No multithreading used within the estimation of meta-learners. Additionally, X-F, DR-F, R-F denote the full-sample versions of the meta-learners, while X-S, DR-S, R-S and X-C, DR-C, R-C denote the sample-splitting and cross-fitting versions, respectively.}}}\\ \bottomrule
\end{tabular}
}
\end{table}
\vspace{0.25cm}
\begin{figure}[ht]
\caption{\label{bigfiguretime:time6}Computation Time Results for Main Simulation}\label{fig:bigplot_time}
{\centering \includegraphics[scale=0.8]{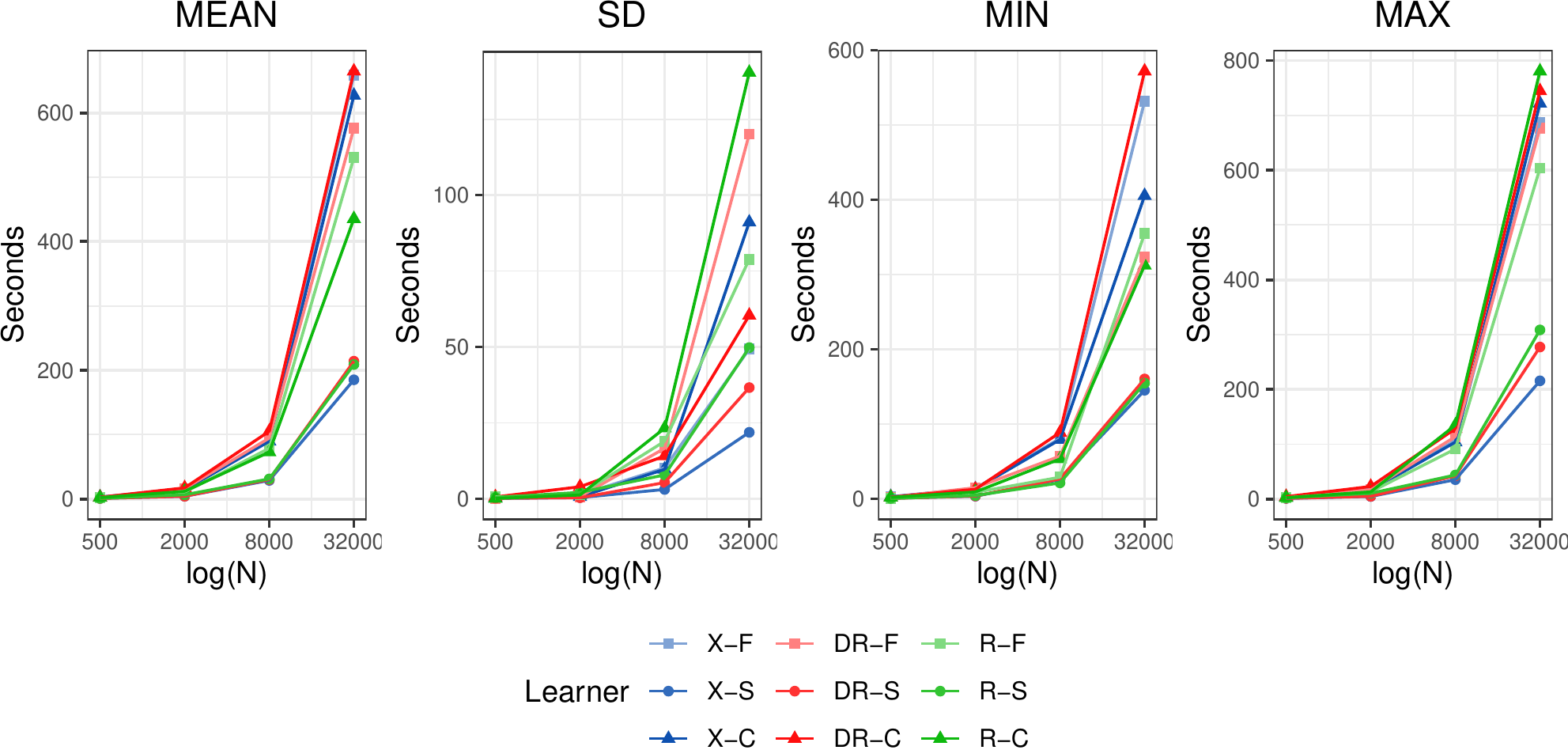} 

}
\vspace{0.25cm}
\caption*{\footnotesize{\textit{Note:} The results for the MEAN, SD, MIN, and MAX show the values of the mean, standard deviation, minimum and maximum for the computation time in seconds based on 10 simulation replications. The figure shows the results based on the increasing training samples of $\{500, 2'000, 8'000, 32'000\}$ observations displayed on the log scale. Additionally, X-F, DR-F, R-F denote the full-sample versions of the meta-learners, while X-S, DR-S, R-S and X-C, DR-C, R-C denote the sample-splitting and cross-fitting versions, respectively.}}
\end{figure}

\end{appendix}

\end{document}